\documentclass[
 reprint,
 superscriptaddress,
 amsmath,amssymb,
 aps,
]{revtex4-2}

\usepackage{graphicx}
\usepackage{dcolumn}
\usepackage{bm}
\usepackage{color}
\usepackage{hyperref}

\begin{document}

\title{First measurement of $\overline{\nu}_{\mu}$ and $\nu_{\mu}$ charged-current inclusive interactions \\ on water using a nuclear emulsion detector}

\author{A.\,Hiramoto}\thanks{E-mail: hiramoto@scphys.kyoto-u.ac.jp}
\affiliation{Kyoto University, Department of Physics, Kyoto, Japan}
\author{Y.\,Suzuki}
\affiliation{Nagoya University, Nagoya, Japan}

\author{A.\,Ali}
\affiliation{Kyoto University, Department of Physics, Kyoto, Japan}
\author{S.\,Aoki}
\affiliation{Kobe University, Kobe, Japan}
\author{L.\,Berns}
\affiliation{Tokyo Institute of Technology, Department of Physics, Tokyo, Japan}
\author{T.\,Fukuda}
\affiliation{Nagoya University, Nagoya, Japan}
\author{Y.\,Hanaoka}
\affiliation{Nihon University, Narashino, Japan}
\author{Y.\,Hayato}
\affiliation{University of Tokyo, Institute for Cosmic Ray Research, Kamioka Observatory, Kamioka, Japan}
\author{A.\,K.\,Ichikawa}
\affiliation{Kyoto University, Department of Physics, Kyoto, Japan}
\author{H.\,Kawahara}
\affiliation{Nagoya University, Nagoya, Japan}
\author{T.\,Kikawa}
\affiliation{Kyoto University, Department of Physics, Kyoto, Japan}
\author{T.\,Koga}\thanks{now at KEK}
\affiliation{University of Tokyo, Department of Physics, Tokyo, Japan}
\author{R.\,Komatani}
\affiliation{Nagoya University, Nagoya, Japan}
\author{M.\,Komatsu}
\affiliation{Nagoya University, Nagoya, Japan}
\author{Y.\,Kosakai}
\affiliation{Toho University, Funabashi, Japan}
\author{T.\,Matsuo}
\affiliation{Toho University, Funabashi, Japan}
\author{S.\,Mikado}
\affiliation{Nihon University, Narashino, Japan}
\author{A.\,Minamino}
\affiliation{Yokohama National University, Yokohama, Japan}
\author{K.\,Mizuno}
\affiliation{Toho University, Funabashi, Japan}
\author{Y.\,Morimoto}
\affiliation{Toho University, Funabashi, Japan}
\author{K.\,Morishima}
\affiliation{Nagoya University, Nagoya, Japan}
\author{N.\,Naganawa}
\affiliation{Nagoya University, Nagoya, Japan}
\author{M.\,Naiki}
\affiliation{Nagoya University, Nagoya, Japan}
\author{M.\,Nakamura}
\affiliation{Nagoya University, Nagoya, Japan}
\author{Y.\,Nakamura}
\affiliation{Nagoya University, Nagoya, Japan}
\author{N.\,Nakano}
\affiliation{Nagoya University, Nagoya, Japan}
\author{T.\,Nakano}
\affiliation{Nagoya University, Nagoya, Japan}
\author{T.\,Nakaya}
\affiliation{Kyoto University, Department of Physics, Kyoto, Japan}
\author{A.\,Nishio}
\affiliation{Nagoya University, Nagoya, Japan}
\author{T.\,Odagawa}
\affiliation{Kyoto University, Department of Physics, Kyoto, Japan}
\author{S.\,Ogawa}
\affiliation{Toho University, Funabashi, Japan}
\author{H.\,Oshima}
\affiliation{Toho University, Funabashi, Japan}
\author{H.\,Rokujo}
\affiliation{Nagoya University, Nagoya, Japan}
\author{I.\,Sanjana}
\affiliation{Kyoto University, Department of Physics, Kyoto, Japan}
\author{O.\,Sato}
\affiliation{Nagoya University, Nagoya, Japan}
\author{H.\,Shibuya}
\affiliation{Toho University, Funabashi, Japan}
\author{K.\,Sugimura}
\affiliation{Nagoya University, Nagoya, Japan}
\author{L.\,Suzui}
\affiliation{Nagoya University, Nagoya, Japan}
\author{H.\,Takagi}
\affiliation{Toho University, Funabashi, Japan}
\author{T.\,Takao}
\affiliation{Nagoya University, Nagoya, Japan}
\author{Y.\,Tanihara}
\affiliation{Yokohama National University, Yokohama, Japan}
\author{K.\,Yasutome}
\affiliation{Kyoto University, Department of Physics, Kyoto, Japan}
\author{M.\,Yokoyama}
\affiliation{University of Tokyo, Department of Physics, Tokyo, Japan}

\collaboration{The NINJA Collaboration}

\date{\today}

\begin{abstract}
This paper reports the track multiplicity and kinematics of muons, charged pions, and protons from charged-current inclusive $\overline{\nu}_{\mu}$ and $\nu_{\mu}$ interactions on a water target, measured using a nuclear emulsion detector in the NINJA experiment. A \mbox{3\,-kg} water target was exposed to the T2K antineutrino-enhanced beam corresponding to $7.1\,\times\,10^{20}$ protons on target with a mean energy of 1.3\,GeV. Owing to the high-granularity of the nuclear emulsion, protons with momenta down to 200\,MeV/$c$ from the neutrino-water interactions were detected. We find good agreement between the observed data and model predictions for all kinematic distributions other than the number of charged pions and the muon kinematics shapes. These results demonstrate the capability of measurements with nuclear emulsion to improve neutrino interaction models.
\end{abstract}

\maketitle

\section{Introduction}
In accelerator-based long-baseline neutrino oscillation experiments, neutrino interactions with the nuclei are essential processes for measuring the neutrino oscillation parameters and searching for CP violation in the lepton sector~\cite{k2k2006,minos2014,t2k2020a,nova2019,hk2015,dune2015}. However, a precise and fully internally coherent model which is able to describe
all the data is a significant challenge ahead of us~\cite{katori2018,alvarez2014}. The charged-current quasi-elastic (CCQE) interactions, which excite one-particle-one-hole states, constitute the dominant interaction process in the energy region of the T2K neutrino oscillation experiment~\cite{t2k2011}. The CCQE interaction has one lepton and one nucleon in the final state. In addition, there are interactions with two-particle-two-hole (2p2h) excitations~\cite{nieves2011,martini2009}. The charged-current 2p2h interaction has one lepton and two nucleons in the final state. The T2K far detector, Super-Kamiokande (SK)~\cite{sk2003}, is insensitive to most neutrons and protons. Events with a single lepton and no other visible particles are selected as the signals, and the incoming neutrino energies are reconstructed from only the outgoing leptons assuming the two-body kinematics of the CCQE interaction. Therefore, the 2p2h interactions involved in the selected events bias the reconstructed neutrino energy. In T2K, neutrino interactions are measured and studied using the near detectors~\mbox{\cite{t2k2015a,t2k2015b,t2k2016,t2k2018a,t2k2018b,t2k2019,t2k2020b,t2k2020c}}. However, at present, the measurement of the 2p2h interaction is poor because the momentum threshold for protons is not sufficiently low to detect all the protons from the neutrino interactions. In addition to the proton measurements, precise measurements of interactions including low-momentum charged pions in the final state are important. They also contaminate the signals at SK when the pions fall short of the Cherenkov threshold in water, although Michel-electron tagging can sometimes be used to veto such events. Measurements of protons and pions from neutrino interactions with low momentum thresholds play an important role in constructing reliable models of the neutrino-nucleus interactions and reducing the systematic uncertainties in T2K.

Low-momentum hadrons produced by neutrino interactions, especially protons with momenta down to 200\,MeV/$c$, have been measured using bubble chambers containing hydrogen or deuterium~\cite{barish1997,baker1981,kitagaki1983} as well as liquid argon time projection chambers~\cite{acciarri2014}. By contrast, recent long-baseline experiments use carbon and oxygen as their targets. The proton momentum thresholds achieved for these nuclei are down to only around 400\,MeV/$c$~\cite{t2k2018b,minerva2018}. Hence, a new experiment using a nuclear emulsion detector was proposed to measure protons from neutrino-water interactions with a momentum threshold as low as 200\,MeV/$c$. A nuclear emulsion detector is a high-granularity three-dimensional tracking device. Emulsion detectors have contributed to advances in fundamental particle physics such as the discovery of the charm particles in cosmic rays~\cite{niu1971}, the direct observation of $\nu_\tau$~\cite{donut2001}, and the discovery of $\nu_\tau$ appearance in neutrino oscillation~\cite{opera2015}. The detection of extremely short tracks was key to these observations. The high granularity allows clear observation of short-range tracks from neutrino interaction vertices. The charged track multiplicity is determined by preparing an alternating structure of emulsion films and thin water-target layers. 

A series of pilot experiments has been carried out by the NINJA collaboration beginning in 2014~\cite{ninja2017a,ninja2017b}. This paper reports the results of a pilot run with a small-mass water target (J-PARC T68). A \mbox{3\,-kg} water target was exposed to the T2K antineutrino mode beam from 2017 to 2018. The signals are the charged-current (CC) inclusive $\overline{\nu}_{\mu}$ and $\nu_\mu$ interactions on water, and muons, charged pions, and protons are detected as the outgoing particles. We measure the distributions of multiplicity, angle, and momentum of the outgoing particles. In particular, we focused on the measurement of protons in the 200--400\,MeV/$c$ range from neutrino-water interactions.

The remainder of this paper is organized as follows. Section~II describes the experimental apparatus. Section~III discusses the Monte Carlo (MC) simulation. Section~IV describes the event reconstruction. Section~V addresses the momentum reconstruction and particle identification (PID). Section~VI describes the selection of neutrino events. Section~VII discusses the estimation of the systematic uncertainties. Section~VIII presents the results. Finally, Section~IX concludes the paper.

\section{Detector configuration and data samples}

\begin{figure}
\includegraphics[width=8.6cm,pagebox=cropbox]{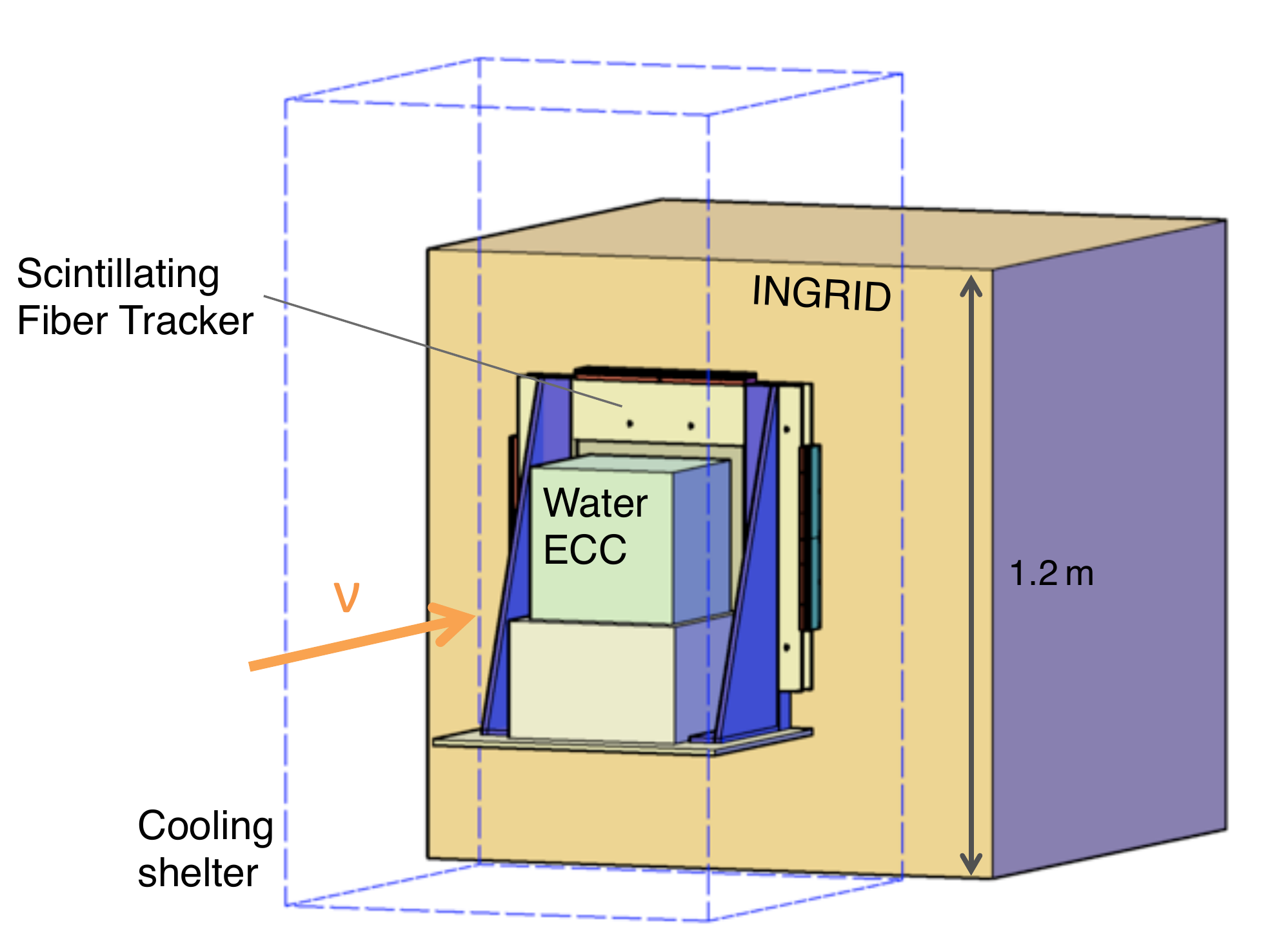}
\caption{\label{fig:SFTinstall_fig}Schematic view of the detectors. The ECC and SFT are installed in a cooling shelter, which is placed in front of an INGRID module.}
\end{figure}

Three detectors were installed in the T2K neutrino near-detector hall. Figure~\ref{fig:SFTinstall_fig} shows a schematic view of the detectors. The main detector that records all charged particles from neutrino interactions is the water-target emulsion cloud chamber (ECC). The ECC was installed upstream of one of the modules of INGRID, which is a T2K near detector~\cite{t2k2012}. In this measurement, INGRID is used to detect muons from the neutrino interactions in the ECC. The emulsion accumulates all the tracks after production without timing information, whereas INGRID records the tracks with timing information. The angular and position resolutions of INGRID are not sufficient to identify corresponding tracks between the ECC and INGRID. Therefore, a scintillating fiber tracker (SFT) was newly developed and installed between them. The ECC and SFT were placed in a cooling shelter to maintain the temperature at around 10$^{\circ}$C and the humidity below 60\%. This is done to prevent the emulsion tracks from fading under high temperature and humidity as well as to prevent the films from warping due to fluctuations in the ambient temperature.

\subsection{J-PARC neutrino beam line}
The J-PARC accelerator provides a high-intensity \mbox{30\,-GeV} proton beam. The proton beam spill is delivered to a graphite target every 2.48\,s. The spill has an eight-bunch structure, and each bunch has a full width of around 58\,ns and separation of about 580\,ns. Hadrons produced by the impinging protons are focused into a decay volume by three electromagnetic horns, where they decay mainly into muons and neutrinos. By changing the polarity of the horns, the charge of the focused hadrons and, thus, the production of either a neutrino or an antineutrino beam are selected. This measurement is performed with the antineutrino mode beam created by the decay of negatively charged hadrons, predominantly $\pi^-$. For further details of the neutrino flux prediction, see Ref.~\cite{t2k2013}

\subsection{INGRID}
\label{sec:ingrid}
\begin{figure}
  \includegraphics[width=8.6cm,pagebox=cropbox]{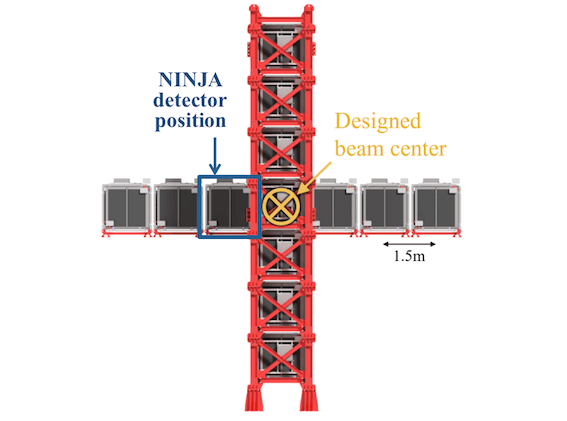}
  \includegraphics[width=8.6cm,pagebox=cropbox]{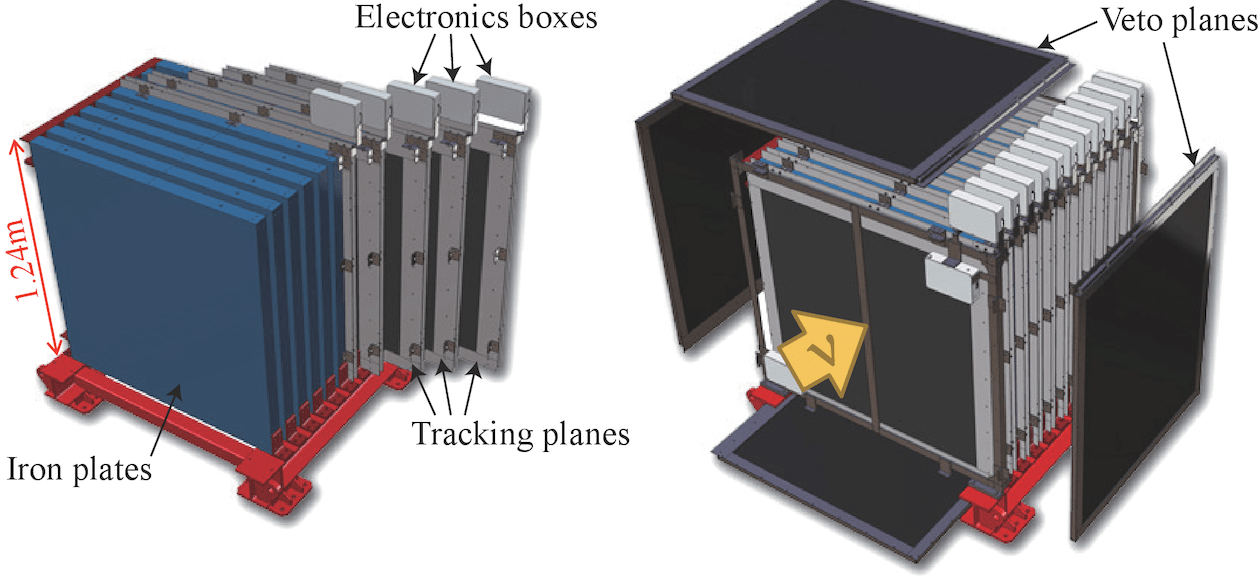}
  \caption{\label{fig:INGRID}INGRID modules (top) and an exploded view of one module (bottom). We use one of these modules behind the NINJA detector as a muon range detector.}
\end{figure}

INGRID is a T2K on-axis near detector located 280\,m downstream of the graphite target. It has 14~modules placed along the vertical and horizontal axes to measure the neutrino event rate and beam profile. We use one horizontal module next to the central module (Fig.~\ref{fig:INGRID} top) as a muon range detector. An INGRID module has a sandwich structure consisting of 9~iron plates and 11~scintillator tracking planes (Fig.~\ref{fig:INGRID} bottom). The thickness of each iron plate is 6.5\,cm. INGRID measures the muon momentum up to around 1\,GeV/$c$ with a resolution of around 10\%. Each scintillator tracking plane consists of 24~plastic scintillator bars aligned horizontally and 24 vertically. Each scintillator bar has dimensions of 120\,cm\,$\times$\,5\,cm$\,\times$\,1\,cm, and photons are collected by a wavelength shifting (WLS) fiber inserted in a hole made along the longitudinal direction of the scintillator. A silicon photomultiplier (SiPM) is attached to one end of the WLS fiber with an optical connector. The angular and position resolutions of the reconstructed tracks are around 2.7\,cm and 3.8$^\circ$, respectively~\cite{t2k2012}.

\subsection{ECC}

\begin{figure}
\includegraphics[width=8.6cm,pagebox=cropbox]{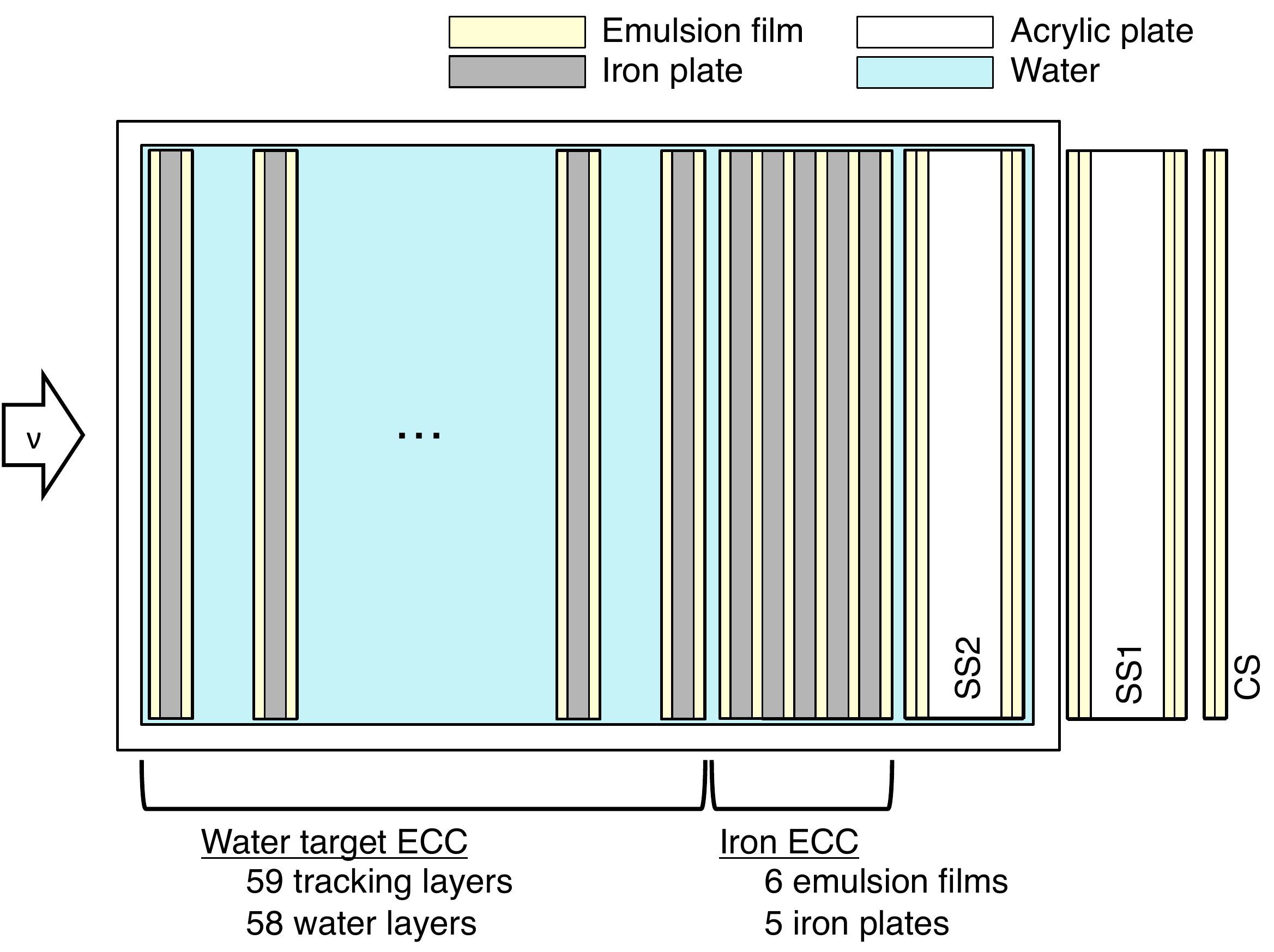}
\caption{\label{fig:detector_ECC_run8}Structure of the water-target ECC. It is an alternating structure of \mbox{2\,-mm} water layers and tracking layers. A tracking layer has an iron plate and two emulsion films. Charged particles from neutrino interactions in the water layers make tracks on the emulsion films.}
\end{figure}

The ECC is an emulsion-based detector composed of alternating layers of emulsion films and target materials. The target materials and their thickness can be selected flexibly. In addition, the alternating structure of emulsion films and thin target layers enables us to achieve a low momentum threshold. Figure~\ref{fig:detector_ECC_run8} shows the structure of the ECC. The components of the ECC are placed in a desiccator. The desiccator is constructed with \mbox{2\,-cm-thick} walls, and it has inner dimensions of \mbox{21\,cm\,$\times$\,21\,cm\,$\times$\,21\,cm}. The structure formed by two emulsion films and a \mbox{500\,-$\mu$m-thick} iron plate vacuum-packed in a \mbox{115\,-$\mu$m-thick} aluminized packing film is referred to as a tracking layer. The iron plate is sandwiched between the two emulsion films, each of which consists of a plastic base film that has been coated with an emulsion gel on both sides. These iron plates are employed as supporting structures for the emulsion films and also used for the momentum measurement described in Section~\ref{sec:mom}. The tracking layers are placed at \mbox{2\,-mm} intervals using acrylic frames with a hollow square shape. The desiccator is filled with water, and \mbox{2\,-mm} water layers are formed inside the acrylic frames. Thus, charged particles from neutrino interactions occurring in the water layers make tracks on the upstream or downstream emulsion films. As the tracks are required to pass through at least one iron plate and two emulsion films, the momentum threshold for proton tracks is around 200\,MeV/$c$, while that for pion tracks is around 50\,MeV/$c$.

An iron ECC is placed downstream of the water region to measure the momentum of the charged particles using multiple Coulomb scattering (MCS) at the iron plates. In addition, two special sheets (SSs) and one changeable sheet (CS) are installed in the most downstream region. SS1 is placed outside the desiccator, while SS2 is placed inside it. Each SS has four emulsion films with a \mbox{2\,-mm-thick} acrylic plate inserted between the emulsion films. This structure enables us to achieve a good angular resolution. The CS contains two emulsion films. They are replaced every month to separate the tracks into several time periods.

\subsection{SFT}
\label{sec:sft}
Although the emulsion detector has excellent angular and position resolutions, it does not provide any time information. For track matching between the ECC and INGRID, another device with time and position resolutions is required because the angular and position resolutions of INGRID are not sufficient to select a track candidate in the ECC to be connected to an INGRID track. In some cases, an emulsion shifter~\cite{ninja2017b, rokujo2013} is used to apply timestamps to tracks in the ECC. However, in this pilot run, the SFT is employed as a timestamper because it can provide more precise time information than the emulsion shifter.

\begin{figure}
\includegraphics[width=8.6cm,pagebox=cropbox]{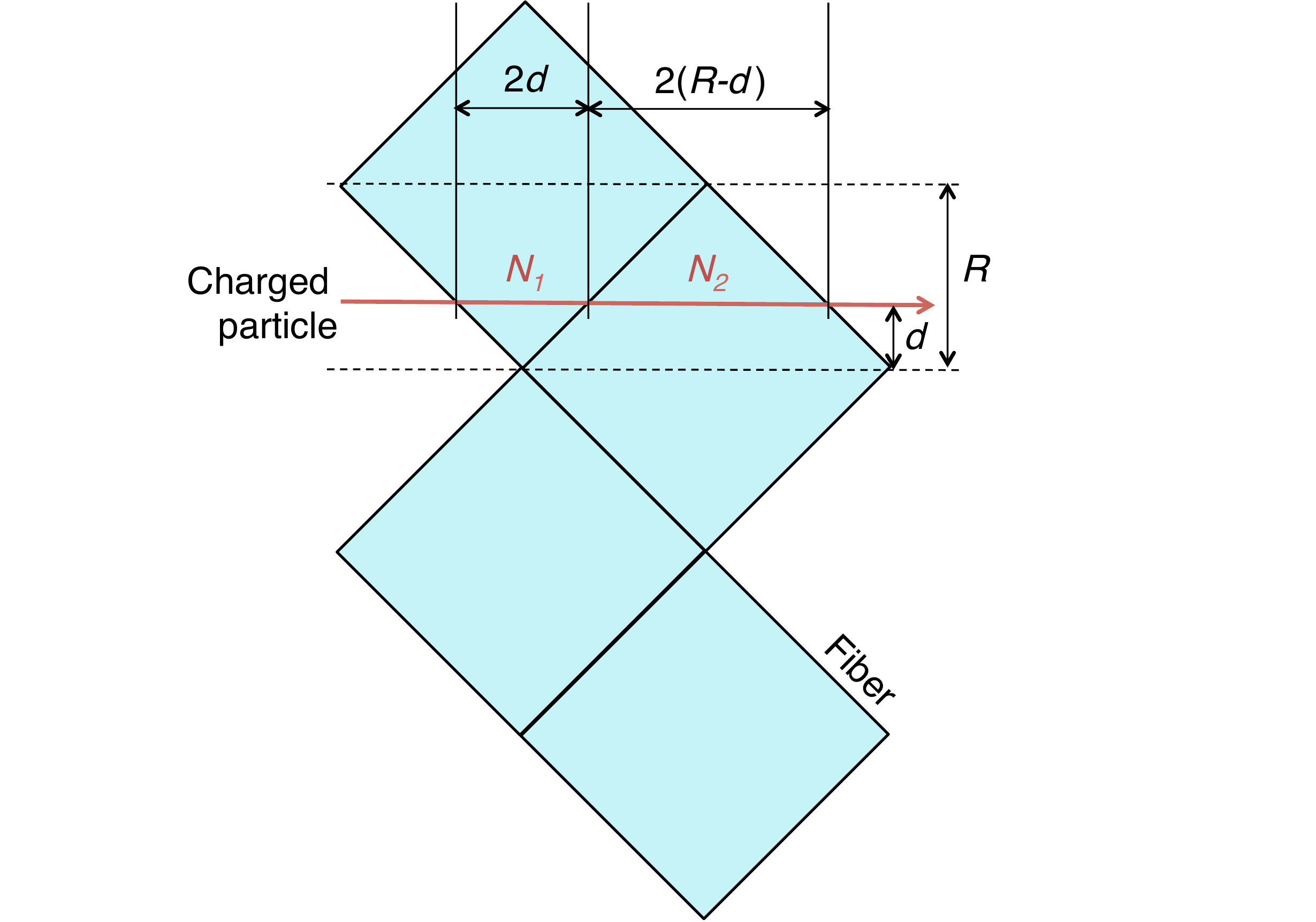}
\caption{\label{fig:sft_idea}Principle of position measurement using square fibers. The ratio of the light yields at neighboring fibers gives precise position information. $N_1$ and $N_2$ are the light yields at each fiber. $R$ and $d$ denote the fiber interval and true hit position, respectively.}
\end{figure}

By arranging square fibers in a slanting lattice pattern as shown in Fig.~\ref{fig:sft_idea}, the ratio of the light yields at neighboring fibers can be used to obtain a precise track position. As the light yield at each fiber is proportional to the path length of a charged particle, the ratio of the light yields changes with the position of the particle. The track position $d$ is calculated as

\begin{eqnarray}
  \label{eq:sftrecon}
  d = \frac{N_1}{N_1+N_2}R
\end{eqnarray}

where $R$ is the fiber interval and $N_1$ and $N_2$ are the light yields from each fiber. The expected position resolution is proportional to $1/\sqrt{N_1+N_2}$ when the ratio of the light yields is used. Thus, with the same number of fibers a position resolution better than the typical $A/\sqrt{12}$, where $A$ is the fiber cross section, can be obtained. Although the position resolution is degraded as the injection angle of the particle increases, this effect is not significant for most muons from the neutrino interactions in the ECC. In this pilot run, \mbox{1\,-mm} square fibers (Kuraray, SCSF-78) are aligned at \mbox{0.725\,-mm} intervals to cover an area of 37\,cm\,$\times$\,37\,cm. A horizontal layer and a vertical layer are constructed, and each layer consists of 512 fibers.

Hamamatsu S13361-3050AE-04 16-channel SiPM arrays are used for the readout of the scintillation light, and NIM EASIROC modules~\cite{nakamura2015} are used as the readout electronics. The total light yield in a layer is around 60\,photoelectrons (p.e.). In this pilot run, the SFT recorded only one event per spill without timing information inside the spill. Therefore, only the first hit is recorded even if there are several hits in a spill.

To reduce the number of readout channels, one SiPM channel reads out four fiber signals, and the signals are read out from both ends of the fibers. As the combinations of the four fibers at the two ends are different, the hit fibers can be identified. Therefore, the total number of readout channels is 512 and the total number of fibers is 1024. 

\subsection{Data samples}
There are two periods of beam exposure, which correspond to T2K Run~9. The first period (\mbox{Run-a}) is from October to December 2017 and the second period (\mbox{Run-b}) is from March to May 2018. Both \mbox{Run-a} and \mbox{Run-b} are separated into three periods using different CS films, and each period corresponds to roughly one month. Considering periods in which both the SFT and INGRID are collecting data, this analysis is performed with $7.1\times 10^{20}$\,protons on target (POT) of the antineutrino mode beam.

\section{Monte Carlo simulation}

The expected signals and backgrounds are generated by MC simulations. Three software packages are used: JNUBEAM~\cite{t2k2013} for the neutrino flux simulation, NEUT~\cite{hayato2009} for the neutrino-nucleus interactions, and a GEANT4~\cite{agostinelli2003}-based framework for the detector response simulation. In this analysis, $\overline{\nu}_{\mu}$ and $\nu_{\mu}$ interactions on $\rm{H_2O}$ and Fe in the antineutrino mode beam are generated by JNUBEAM and NEUT. As the $\overline{\nu}_e$ and $\nu_e$ components of the flux are less than 1\%, $\overline{\nu}_e$ and $\nu_e$ interactions in the ECC are not simulated. The MC predictions are normalized by POT and corrected by the detector efficiencies estimated using the data and the MC simulations.

\subsection{Neutrino flux}

\begin{figure}
\includegraphics[width=8.6cm,pagebox=cropbox]{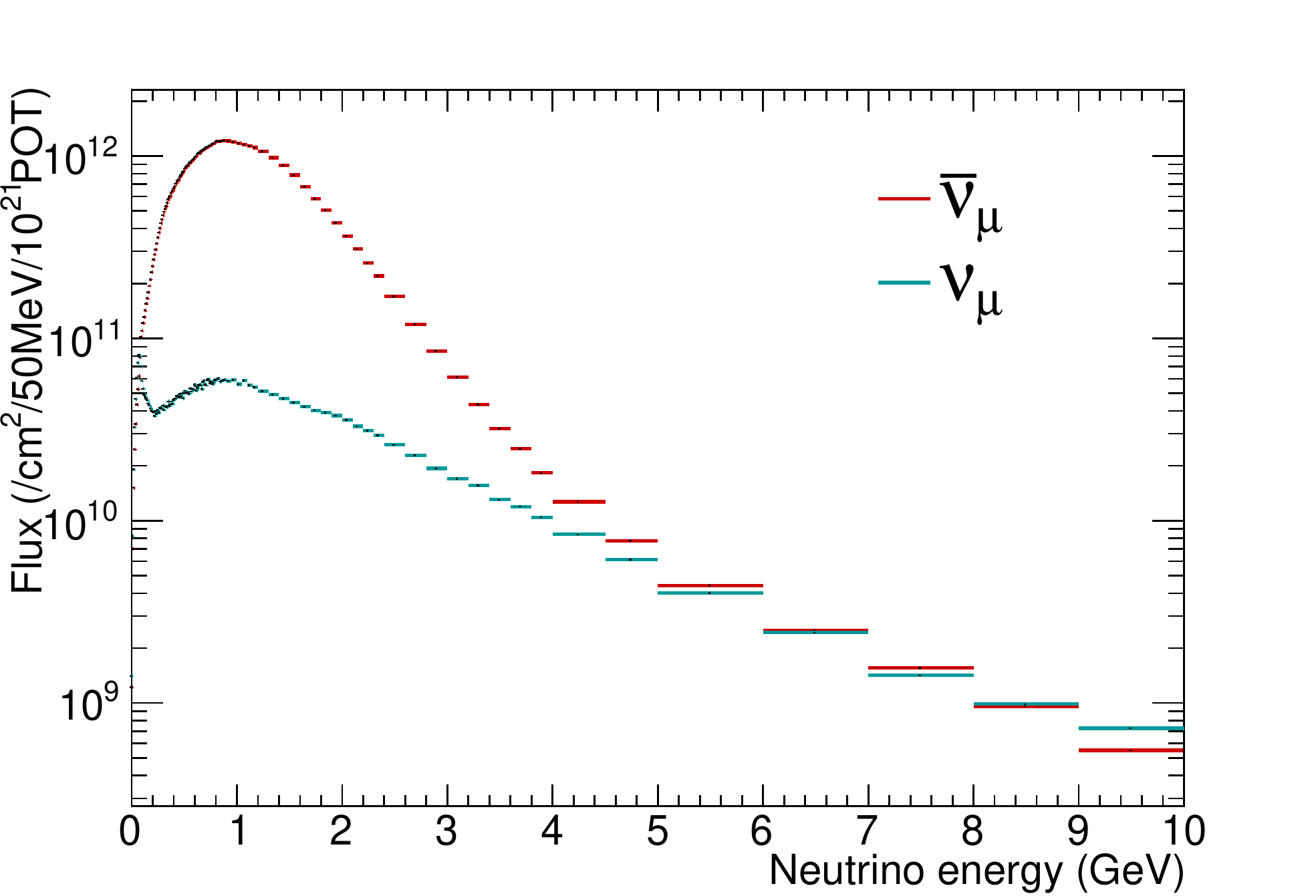}
\caption{\label{fig:Flux}Predicted $\overline{\nu}_\mu$ and $\nu_\mu$ fluxes in the antineutrino mode beam at the location of the NINJA detector.}
\end{figure}

JNUBEAM is a GEANT3~\cite{brun1994}-based neutrino flux simulator developed by T2K. Interactions of the primary protons from the accelerator and the graphite target are simulated by FLUKA 2011.2~\cite{ferrari2005,bohlen2014}. Secondary particles produced are transferred to JNUBEAM, which simulates the propagation, interaction, and decay of these secondary particles. Neutrinos are generated by the decay of the hadrons. The hadron interactions are tuned by external measurements of hadron production such as CERN NA61/SHINE~\cite{na612011,na612012}, most pions exiting the target in particular are tuned using their 2009 data taken with a T2K replica target~\cite{na612016a,na612016b}. Figure~\ref{fig:Flux} shows the predicted flux of the antineutrino mode beam at the location of the NINJA detector. This flux prediction and the covariance of the flux uncertainty between each neutrino energy bin can be found in our data release~\cite{datarelease}. The mean energy of the $\overline{\nu}_\mu$ components is 1.3\,GeV and that of the $\nu_\mu$ components is 2.0\,GeV. 

\subsection{Neutrino interaction}

Using the neutrino flux calculated by JNUBEAM, $\overline{\nu}_{\mu}$ and $\nu_{\mu}$ interactions on $\rm{H_2O}$ and Fe targets are generated by NEUT. In addition, neutrino interactions in the upstream wall and INGRID are generated as background sources. Table~\ref{tab:intmdl} summarizes the neutrino interaction models used in this analysis. The nominal MC predictions are generated using NEUT version 5.4.0, which uses the 1p1h model by Nieves {\it et al.} with correction by random-phase approximation (RPA)~\cite{nieves2012, nieves2004}, and the axial mass $M\mathrm{_A^{QE}}$ is set to 1.05\,GeV/$c^2$ for the CCQE interactions. The local Fermi gas model (LFG) is used as the nuclear model, while the Spectral Function (SF)~\cite{benhar1994,benhar2000} is prepared as an alternative model. For the 2p2h interactions, the model of Nieves {\it et al.}~\cite{nieves2011} is used. The single pion production is modeled by the Rein--Sehgal model~\cite{rein1981}, and the axial mass $M\mathrm{_A^{RES}}$ is set to 0.95\,GeV/$c^2$. The Berger--Sehgal model~\cite{berger2009} is used for the coherent pion production, and the deep inelastic scattering (DIS) is described by the parton distribution function GRV98 and the cross-section model modified by Bodek and Yang~\cite{bodek2003}. The final state interactions (FSI) in the nuclear medium are simulated using a semi-classical intra-nuclear cascade model~\cite{hayato2009}. Samples with other parameters are studied for comparison to the systematic uncertainties as discussed in Section~\ref{sec:sys}.

\begin{table}
  \caption{\label{tab:intmdl}Neutrino interaction models used in the nominal MC simulation.}
  \begin{ruledtabular}
    \begin{tabular}{ll}
      Mode & Model\\ \hline
      CCQE & 1p1h model by Nieves {\it et al.}~\cite{nieves2012} \\
           & LFG with RPA correction ($M\mathrm{_A^{QE}}$=1.05\,GeV/$c$$^2$)\\
      2p2h & Nieves {\it et al.}~\cite{nieves2011}\\
      1$\pi$ & Rein--Sehgal~\cite{rein1981} ($M\mathrm{_A^{RES}}$=0.95\,GeV/$c$$^2$)\\
      Coherent & Berger--Sehgal~\cite{berger2009} \\
      DIS & GRV98 PDF with Bodek--Yang modifications~\cite{bodek2003}\\
      FSI & Semi-classical intra-nuclear cascade model~\cite{hayato2009}
    \end{tabular}
  \end{ruledtabular}
\end{table}

\subsection{Detector response}
The behavior of the particles from neutrino interactions is simulated by a GEANT4-based detector MC framework. The detectors and the wall of the detector hall are modeled. QGSP BERT~\cite{qgsp} is used as the default physics list, and muons, charged pions, and protons from the neutrino events generated by NEUT and their secondary particles are simulated. In addition to the neutrino interactions in the ECC, interactions in the INGRID modules and the upstream wall of the detector hall are generated for the background study. The background from cosmic rays is evaluated using the off-beam timing track data instead of the MC simulation.

\section{Reconstruction}
This section describes the track reconstructions in the ECC and INGRID, the hit position reconstruction at the SFT, and the track matching between all the detectors. 

\subsection{Track reconstruction in ECC}
After the beam exposure, all the emulsion films are developed and several steps of film treatment are performed. Then, they are scanned using Hyper Track Selector (HTS)~\cite{yoshimoto2017} and the tracks are reconstructed automatically for each film~\cite{hamada2012}. The current scanning angle is limited to $|\mathrm{tan}\theta|\lesssim 1.5$, where $\theta$ is the angle of a track with respect to the direction perpendicular to the emulsion films. Tracks satisfying $|\mathrm{tan}\theta|< 1.3$ are used in the analysis. The track density is $\mathcal{O}(10^3)$ per ${\rm cm^2}$ and the detection efficiency of a single emulsion film is 98\%--99\%. The main components of the tracks in the emulsion films are cosmic rays and environmental radiation. Following the track reconstruction in each film, connections between the films are established by the auto-reconstruction process~\cite{hamada2012}. The track connection process is applied not only to adjacent films but also those separated by one or two other films. The angular and position tolerances are defined as functions of the track angle, and they are determined on the basis of the scattering angle of the minimum ionizing particles (MIPs). The connection efficiency between two films for the MIPs is more than 99.8\%. Therefore, by connecting the tracks between both adjacent films and films separated by one or two other films, the connection efficiency becomes greater than 99.99\%.

\subsection{Track reconstruction in INGRID}
Channels with more than 2.5\,p.e. are counted as hits. At least three continuous planes are required to have hits on both horizontal and vertical layers. The tracks are reconstructed using a cellular automaton algorithm~\cite{maesaka2005}, which is the same as that used in the event rate and the profile measurements of the T2K neutrino beam. In our analysis, the tracks are required to start at the most upstream plane of INGRID.

\subsection{Hit reconstruction of SFT}
As described in Section~\ref{sec:sft}, the SFT fiber hits are identified on the basis of the combination of channels at both ends of the fibers. The hit threshold of the SFT is set at 2.5\,p.e. and at least one hit is required in each layer. The hit position is reconstructed from the ratio of the light yields of neighboring fibers. If there is only one hit fiber, the particle is considered to have passed through exactly the center of the fiber because it is likely that the particle penetrated areas that are insensitive due to fiber cladding. To evaluate the track reconstruction efficiency, the effects of accidental noise hits as well as those of missing the second or later hits in a spill by the SFT are calculated using sand muon events which are from neutrino interactions on the upstream wall of the near-detector hall.

\subsection{Track matching}
After reconstructing the tracks and hit positions at each detector, a track matching process is performed. To connect tracks between the ECC and INGRID, matching between INGRID and the SFT is carried out first. Then, using the SFT hit position and INGRID angle, matching between the SFT hits and the ECC tracks is performed.

The track matching between the SFT and INGRID is performed using the position and timing information recorded at each detector. The INGRID tracks are extrapolated to the SFT position. If the extrapolated position is within $\pm 10$\,cm from the reconstructed SFT hit position in the same spill, they are regarded as belonging to the same track. If there are several INGRID track candidates for one SFT hit, the INGRID track in the earliest bunch is selected. This is because the SFT records only the first hit in a spill owing to the limitation of the data acquisition system. By contrast, when there are several INGRID track candidates in the same bunch, or when one INGRID track has several SFT hit candidates, all of them are put forward to the neutrino event selection. The matching efficiency depends on the track angle, but it is higher than 95\% in most regions. The efficiency of the SFT hit reconstruction is included in this matching efficiency.

After matching the INGRID tracks and SFT hits, track matching between the SFT and the ECC is carried out. The tracks recorded on the SS emulsion films are extrapolated to the SFT position. Hits are required to be recorded on at least one of the two films on both sides of SS1. In addition, hits are also required to exist on both CS films. To extrapolate tracks from the SS films, the angle is reconstructed not by a track angle on one film, but by two recorded tracks on the films over the \mbox{2\,-mm-thick} acrylic plate as they give a better angular resolution of around 1\,mrad. The angular resolution of a track reconstructed in one emulsion film is typically 2\,mrad. If the difference between the SFT hit position and the position of the extrapolated SS track is less than 600\,$\mu$m, and the difference of their angles is less than 0.2 in terms of $\tan\theta$ in the horizontal and vertical directions, the track is regarded as a matched track. If there are several candidates, all candidates remain until the neutrino event selection. The angular and position resolutions after the matching are 330\,$\mu$m and 0.05 in terms of $\tan\theta$, respectively.

\begin{figure}
\includegraphics[width=8.6cm,pagebox=cropbox]{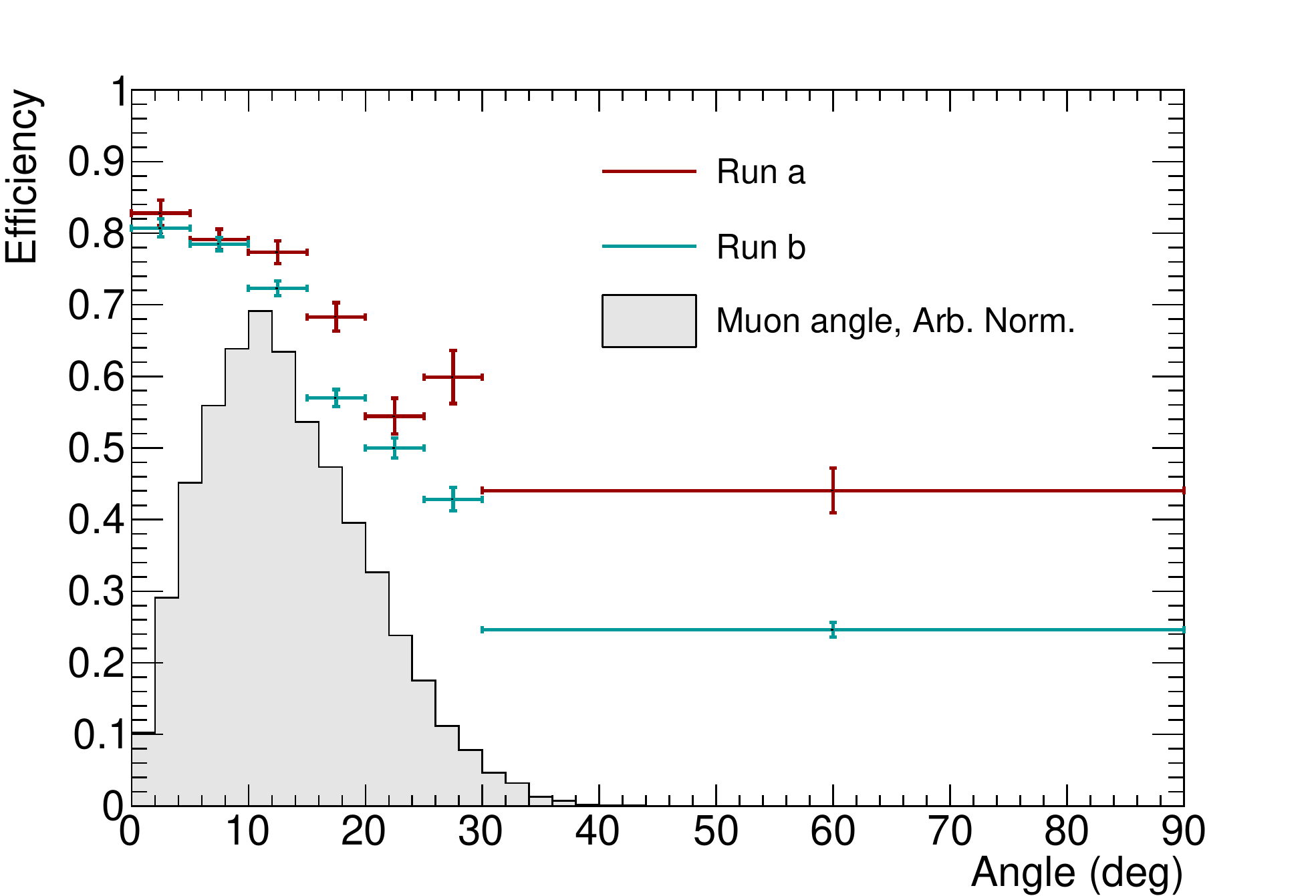}
\caption{\label{fig:mueff}Muon detection efficiency as a function of the angle. The vertical bars denote statistical errors. The CS films were bent in \mbox{Run-b}, which is considered as the reason for the difference between the two periods. The gray histogram represents the expected angle distribution of muons within the INGRID acceptance.}
\end{figure}

Figure~\ref{fig:mueff} shows the total muon detection efficiencies in \mbox{Run-a} and \mbox{Run-b}. The efficiency in \mbox{Run-b} is lower than that in \mbox{Run-a} because the CS films were slightly bent in \mbox{Run-b}. Thus, the distance between the SS and CS films varied depending on the position in the films, and the accuracy of the matching between these films was degraded.

\section{Momentum reconstruction and particle identification}
\label{sec:mom}
\vspace{0.5cm}
The INGRID matched tracks are considered to be muons, while the other tracks recorded in the ECC are considered to be protons or charged pions. This section describes the momentum reconstruction of each particle and separation of protons and pions. In emulsion detectors, the momentum of a charged particle is measured using MCS as $P\beta$, where $P$ is the momentum and $\beta$ is the velocity of the particle. As the angular and position resolutions of the emulsion detectors are sufficient to measure the MCS of the particles, the momentum of the particles can be measured without using a magnetic field. There are two methods for measuring $P\beta$: coordinate method~\cite{kodama2007} and angular method~\cite{agafonova2012}. The coordinate method uses the positional displacement of a track on three films, while the angular method uses the scattering measured by the angular difference of a track between films. In this analysis, the coordinate method is used for the reconstruction of muon momenta, while the momenta of protons and pions are obtained by the angular method, as described later. Moreover, the track range is used to measure the momenta of protons fully-contained in the ECC.

\subsection{Momentum reconstruction of muon tracks}
As described in Section~\ref{sec:ingrid}, INGRID measures muon momentum up to 1\,GeV/$c$. However, most of muons have higher momenta and they penetrate INGRID. Therefore, we adopt the MCS method using the ECC to reconstruct higher momenta. To reconstruct the muon momentum using the coordinate method, three films are used to calculate a positional displacement. The first and second films are used to reconstruct the track angle. Using the reconstructed angle, the track on the second film is extrapolated to the third film. Then, the positional displacement at the third film is measured. The upper limit of the measurable momentum is determined by the measurement error because the scattering angle of a high-momentum particle is small. 
The positional displacement is proportional to $x^{3/2}$ due to the nature of multiple scattering while its measurement error is proportional to $x$, where $x$ is the thickness of the material between the second and third films. Hence, two films placed further apart distance can measure higher momentum compared to adjacent films because the measurement error becomes sufficiently small compared to the scattering angle. In this analysis, the second and third films are separated by five iron plates, which corresponding to around 1.5\,cm and it enables us to measure momentum up to around 5\,GeV/$c$. Films separated by a single water gap are used as the first and second films to reconstruct the track which is extrapolated to a film placed over five iron plates away. This is applied for all available combination of three films. Then, the positional displacement from the predicted position at each combination, $y_{i}$ ($i$=1,\,2,\,3...), is measured. The quadrature sum of $y_{i}$ is taken as $y^2_{\mathrm{meas}}$, which includes both the positional displacement by MCS ($y_0$) and the measurement error ($y_{\mathrm{err}}$):

\begin{eqnarray}
  \label{eq:corrres}
  y^2_{\mathrm{meas}} = y_0^2+y_{\mathrm{err}}^2.
\end{eqnarray}

Therefore, the measurement error needs to be subtracted. The $y_{\mathrm{err}}$ is mainly caused by the alignment error and estimated to be less than 10\,$\mu$m depending on the distance between the segments of the track on the second and third films.
 
Finally, $P\beta$ is calculated from the following relation~\cite{lynch1991}:

\begin{eqnarray}
  \label{eq:coormcs}
  y_0 = \frac{t}{\sqrt{3}}z\frac{13.6\,\mathrm{MeV}}{P\beta}\sqrt{\frac{x}{X_0}}\left(1+0.038\,\ln\left(\frac{x}{X_0}\right)\right),
\end{eqnarray}

where $z$ is the distance between the second and third films, $x$ is the thickness of the material, $X_0$ is the radiation length, and $t$ is a correction factor for the effect of passing through several materials. It is assumed that only the iron plates affect the scattering of a particle when we assign values to $x$ and $X_0$. If the ECC has a simple structure of a single target material and emulsion films, and the mass of the emulsion films is sufficiently small in comparison to that of the target material, the scattering in the emulsion films is negligible. However, the water ECC contains several types of layers of different materials such as iron, water, emulsion film, and vacuum packing film. Thus, scattering in each material is considered and $t$ is estimated to be 1.4 using the MC simulation.

\begin{figure}
\includegraphics[width=8.6cm,pagebox=cropbox]{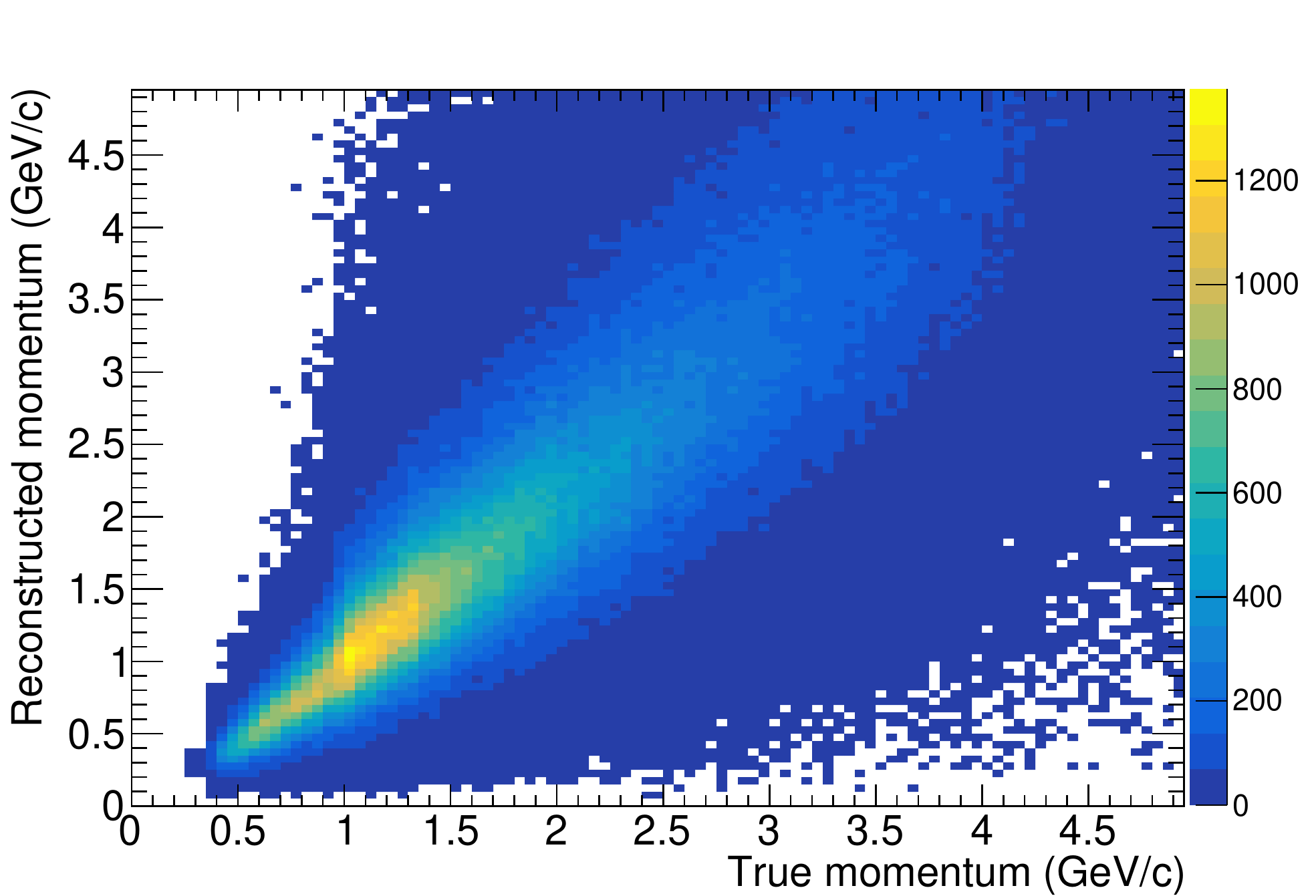}
\caption{\label{fig:Mu_res}Relation between the true and reconstructed momenta of muons from neutrino interactions in the MC simulation.}
\end{figure}

Figure~\ref{fig:Mu_res} shows the relation between the true and reconstructed momenta of muons from the neutrino interactions in the MC simulation. With this method, our detector can reconstruct the muon momentum with a resolution of 30\%--40\%.

\subsection{Momentum reconstruction of non-muon tracks}
Another way to reconstruct momentum is the angular method. The angular difference between two segments of a track on different films is measured instead of the positional displacement. This method enables us to increase the statistic of the combination of films and to reconstruct the momentum of short tracks. However, the angular method cannot be used for the muon momentum measurement because the measurable momentum is limited by the angular resolution of the films, which is typically 2\,mrad for the forward angle tracks. In this analysis, the maximum $P\beta$ measured by the angular method is around 1.5\,GeV/$c$, while $P\beta$ up to 5\,GeV/$c$ can be measured by the coordinate method.

The root mean square of the scattering angle is denoted as $\theta_0$ and it is related to $P\beta$ as follows~\cite{lynch1991}:

\begin{eqnarray}
  \label{eq:angmcs}
  \theta_0 = \frac{13.6\,\mathrm{MeV}}{P\beta}\sqrt{\frac{x}{X_0}}\left(1+0.038\,\ln\left(\frac{x}{X_0}\right)\right).
\end{eqnarray}

As discussed in the coordinate method, the ECC has a complex structure of several materials. The total scattering angle is considered as the quadrature sum of the scattering angle in each material. The measurement error is also considered and subtracted from the measured scattering angles. The angular method reconstructs the momentum of protons and pions with a resolution of 30\%--40\%.

In addition, the momenta of protons stopping in the ECC are measured by the track range in the ECC. This method is used only for the tracks identified as protons. In this method, the momentum is reconstructed with a resolution of $5\%$. Most of the low-momentum protons (typically below 400\,MeV/$c$) are measured by the range.

\subsection{PID}
Muon-like tracks are identified by the track matching with INGRID. This section describes the PID of the other tracks. After the $P\beta$ estimation, all the tracks are separated into nine angle and nine $P\beta$ bins, and separation of the proton-like and pion-like particles in each bin is performed on the basis of the energy deposit in the emulsion films. The emulsion can measure the energy deposit by the blackness of the track, referred to as the volume pulse height (VPH)~\cite{toshito2004,toshito2006}. Tracks with sufficiently large energy deposits, such as the proton tracks, have large VPH, while MIPs have small VPH. Therefore, the distribution of the VPH shows two peaks. The smaller peak is called the MIP peak and the larger one is called the black peak. Each peak is fitted by a Gaussian function to obtain the mean $(\mu_{\mathrm{MIP}}, \mu_{\mathrm{black}})$ and the deviation $(\sigma_{\mathrm{MIP}}, \sigma_{\mathrm{black}})$. Then, the proton-like likelihood $L_{\mathrm{proton}}$ and pion-like likelihood $L_{\mathrm{pion}}$ are defined as follows:
\begin{eqnarray}
  \label{eq:pid1}
  L_{\mathrm{proton}} \equiv \frac{1}{\sigma_{\mathrm{black}}}\exp\left(\frac{-(v-\mu_{\mathrm{black}})^2}{2\sigma_{\mathrm{black}}^2}\right),
\end{eqnarray}
\begin{eqnarray}
  \label{eq:pid2}
  L_{\mathrm{pion}} \equiv \frac{1}{\sigma_{\mathrm{MIP}}}\exp\left(\frac{-(v-\mu_{\mathrm{MIP}})^2}{2\sigma_{\mathrm{MIP}}^2}\right),
\end{eqnarray}

where $v$ is the VPH of the track. The pion-like likelihood ratio $\mathcal{R}$ is defined as
\begin{eqnarray}
    \label{eq:pid}
  \mathcal{R} \equiv \frac{L_{\mathrm{pion}}}{L_{\mathrm{proton}}+L_{\mathrm{pion}}}.
\end{eqnarray}

Particles with $\mathcal{R}$ greater than 0.5 are regarded as pions and those with $\mathcal{R}$ less than 0.5 are regarded as protons. The proton selection efficiency is evaluated as 76.0\% with 98.5\% purity, while the pion selection efficiency is evaluated as 98.7\% with 78.8\% purity for the tracks from the neutrino interactions.

The VPH decreases over time due to the fading of emulsion. The degree of fading in each film is measured using the sand muon samples, and the correction is applied to the beam timing events in the data.

\section{Event selection}
Our signals are CC $\overline{\nu}_\mu$ and $\nu_\mu$ inclusive interactions on the water target. Muon-like tracks are reconstructed in INGRID and the CC interactions in the ECC are selected via track matching between the ECC and INGRID. This section describes the event selection and the method for determining the track multiplicity for measurements of the kinematics distributions of protons and pions.

\subsection{INGRID matching}
  Track matching between the ECC and INGRID is performed using the SFT. After the selection, a total of 14495 events remain as CC interaction candidates.

\subsection{Fiducial volume cut}
  Most of the tracks identified as muons are sand muons. To select neutrino interactions occurring in the ECC, a fiducial volume (FV) is defined as an area of 16\,cm\,$\times$\,17\,cm from the water-gap next to the most upstream one to the most downstream one. The starting points of the muon candidates are required to be in the FV. After the FV cut, 350 events remain as candidates for the interactions in the ECC.

\subsection{Viewer check}
  Track segments might fail to be connected or wrong track segments might be connected due to the inefficiency of track detection on the emulsion films or the failure of the auto-reconstruction process. To find the misconnections and properly determine the muon starting position, all the event candidates are checked by the viewer from the event display. Track segments near the muon starting point are checked whether they are connected to the starting point of the muon track candidate. 
  
  There is also a possibility of misidentifying large-kink sand muons as neutrino events having a backward-going pion. Additional selections are applied to such kink event candidates found in the viewer check based on the angle, momentum, and VPH of tracks upstream and downstream of the kink position. Backgrounds from kink events are evaluated using sand muons in the MC simulation.

\subsection{Manual check with a microscope}
  After the viewer check, the interaction position is confirmed using a microscope manually. The vertex position of the multiple track events can be determined by extrapolating the track data in an emulsion film obtained by HTS. By contrast, the starting positions of single-track events are not determined by the data. Moreover, the selected events are contaminated by the interactions on the emulsion films and packing films because the scanned data do not contain track segments starting in the middle of the emulsion films. To exclude interactions on the emulsion films, the upstream emulsion film of the vertex position is manually checked. If a track starts in the emulsion film, it is considered as an interaction in the emulsion film and excluded. However, the interactions on the packing films cannot be excluded by this check. Therefore, the background from the interactions on the packing films is evaluated by the MC simulation. 
  
  After the viewer and manual checks, 97 events remain as interactions on water, 182 events remain as interactions on iron, and 71 events are excluded as misconnected tracks or interactions in the emulsion films. In the MC simulation, the efficiencies for the viewer and manual checks are assumed to be 100\%.

\subsection{Momentum consistency check}
  Cosmic rays coming from the downstream region may stop in the ECC, and could be connected to the INGRID tracks induced by the neutrino interactions by chance. The protons and pions from the neutrino interactions also contaminate the muon candidates. To exclude such tracks, the consistencies of the muon momentum measured by MCS in the ECC and that measured by the INGRID range are compared event by event. As cosmic rays stopping in the ECC have low momenta, the momentum measured by the INGRID range becomes larger than that measured by MCS. By contrast, in the case of protons and pions, the INGRID range becomes shorter than that expected by MCS. If the momentum measured by MCS is greater (smaller) than 175\% (25\%) of that measured by the INGRID range, these events are excluded. In the case of the INGRID-penetrating tracks, the maximum limit is not set because a momentum above 1\,GeV/$c$ cannot be measured. By this selection, 11 events are excluded from the neutrino-water event candidates, and the contamination of protons and pions from the neutrino interactions is expected to be reduced to around 1\% of the selected event candidates.

\begin{figure}
\includegraphics[width=8.6cm,pagebox=cropbox]{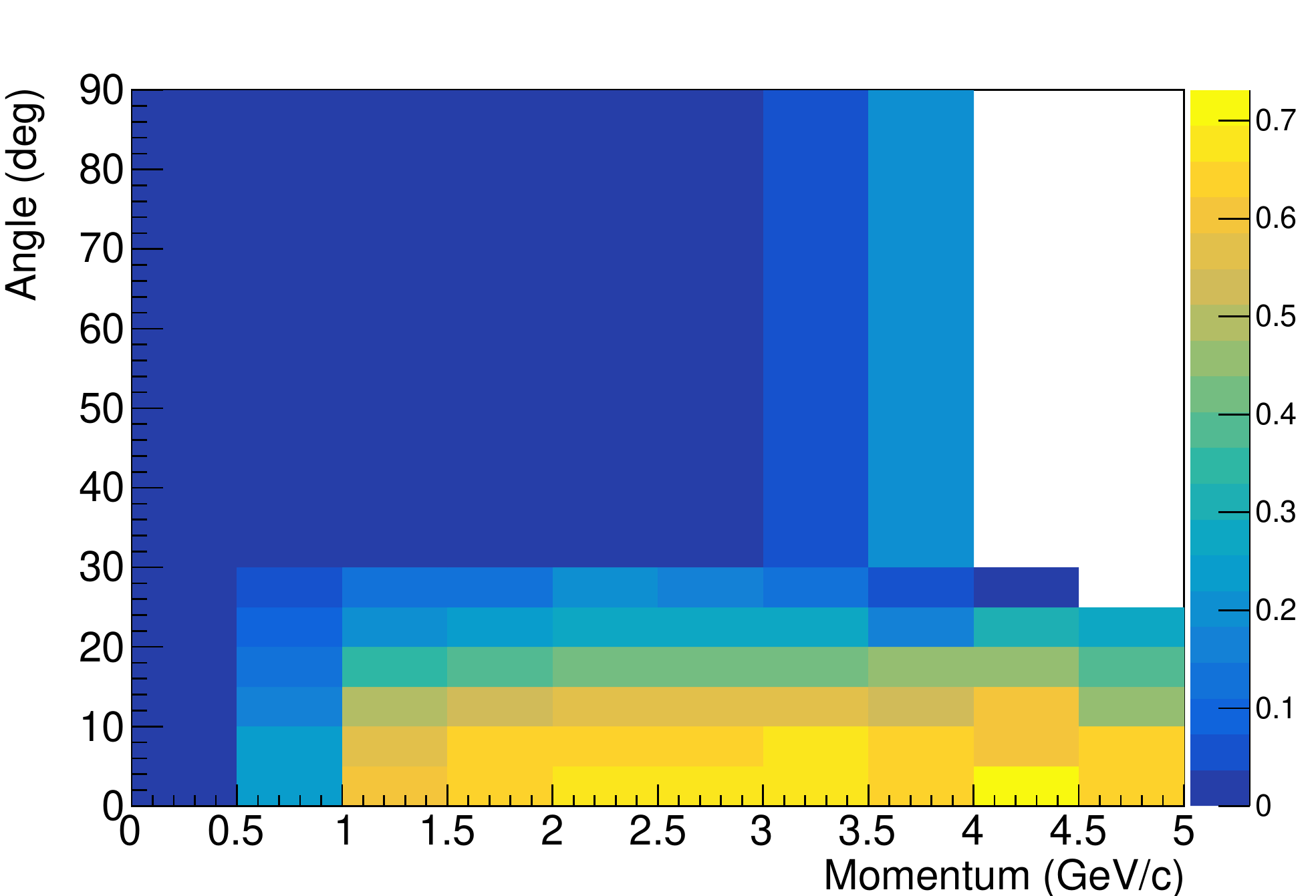}
\caption{\label{fig:mu_eff}Muon detection efficiencies estimated by the MC simulation after all selections.}
\end{figure}

The events selected above are considered as the candidates for muons from the neutrino interactions on the ECC water target. The number of selected events after each step is summarized in Table~\ref{tab:eventsel}. In this pilot run, a total of 86 candidate events of CC interactions on the water target are selected, while the MC prediction is 91.6 events. The observed number of events is consistent with the MC prediction within the statistical uncertainty. In the MC prediction, 58.7\% of the events are $\overline{\nu}_\mu$ interactions, while 18.0\% are $\nu_\mu$ interactions. Although 23.4\% are expected to be background events, cosmic rays are the dominant components and the amount is precisely predicted. The detection efficiency of the CC neutrino interactions within the acceptance of INGRID matching is 63.2\%. It corresponds to a detection efficiency of 26.8\% with respect to all the CC neutrino interactions on water target in the ECC FV and the two-dimensional detection efficiency is shown in Fig.~\ref{fig:mu_eff}.

\begin{table}
  \begin{ruledtabular}
    \caption{Number of selected events after each step.}
    \label{tab:eventsel}
    \begin{tabular}{lcc}
      Step & MC (Background) & Data  \\ \hline 
      INGRID matching & - & 14495  \\ 
      FV cut & - & 350 \\ 
      Viewer/Manual check & 102.4 (25.3) & 97  \\ 
      Momentum consistency check & 91.6 (21.4) & 86  \\ 
    \end{tabular}
    \end{ruledtabular}
\end{table}

Following the selections above, the vertex position and multiplicity of each event are determined as below. 

\subsection{Determination of the vertex}
  After confirming the muon candidates, the vertex positions are determined. First, tracks that have a minimum distance of less than 600\,$\mu$m are clustered. For each track, the midpoint of the closest point of that track and the muon candidate is calculated. The center of mass of those midpoints becomes a temporary vertex. Then, tracks that have a minimum distance less than 100\,$\mu$m from the temporary vertex are clustered again, and their center of mass is regarded as the reconstructed vertex.

\subsection{Partner track determination}
  Finally, partner tracks that make a vertex with the muon track are selected. To determine the track multiplicity, tracks with a minimum distance less than 50\,$\mu$m from the vertex calculated in the previous step are selected. Track length selections are also applied to exclude very short tracks from nuclear spallations. It is required that the length of tracks with large VPH (black) are two or more layers, and those with small VPH (MIP) are nine or more layers.

\begin{figure}
\includegraphics[width=8.6cm,pagebox=cropbox]{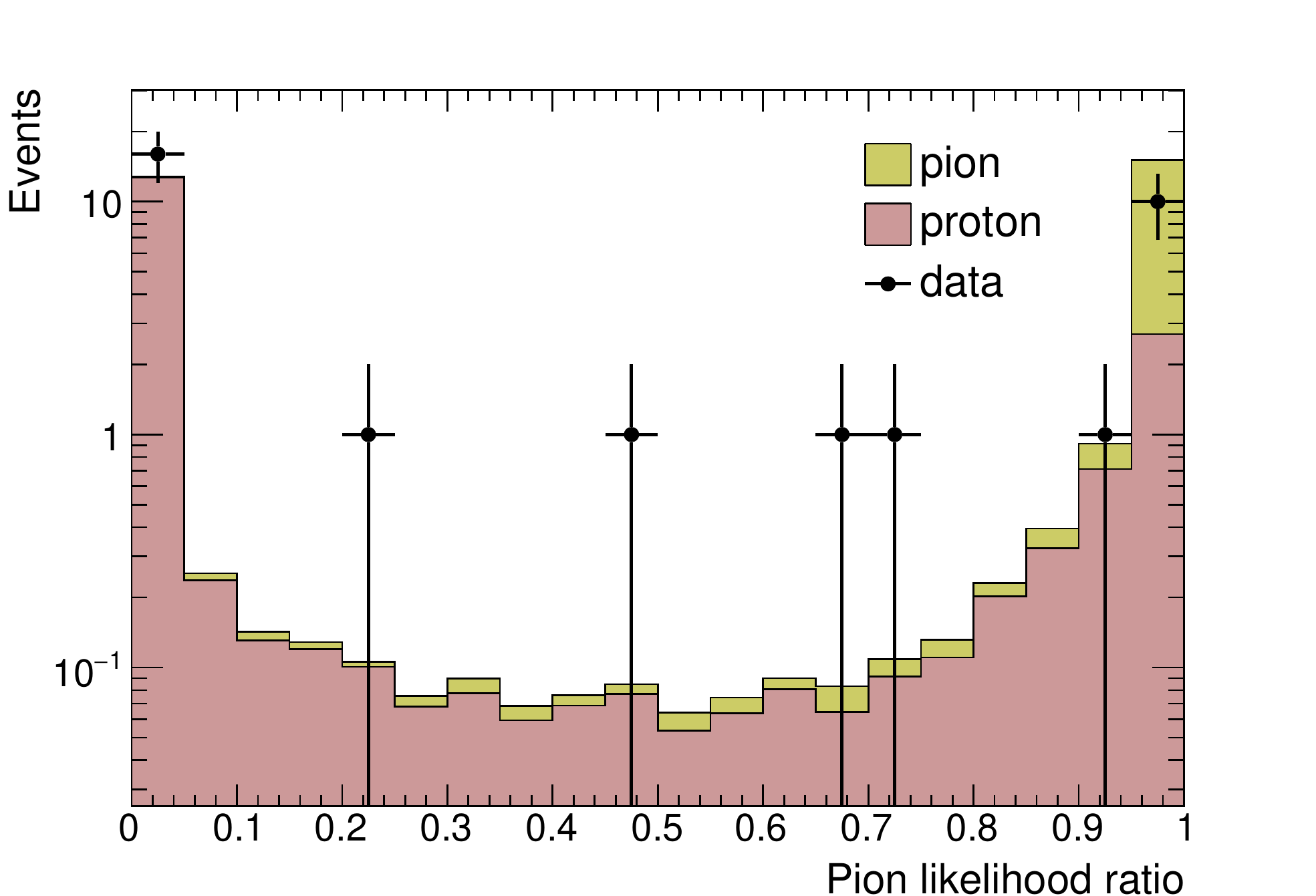}
\caption{\label{fig:pid}Distribution of the pion-like likelihood ratio. Particles with a likelihood ratio more than 0.5 are regarded as pions, and those with less than 0.5 are regarded as protons.}
\end{figure}

After the determination of the multiplicity, the momentum reconstruction and the PID processes are applied. Figure~\ref{fig:pid} shows the distribution of the pion likelihood ratio. The data distributions are consistent with the MC predictions.
  
\begin{figure}
\includegraphics[width=8.6cm,pagebox=cropbox]{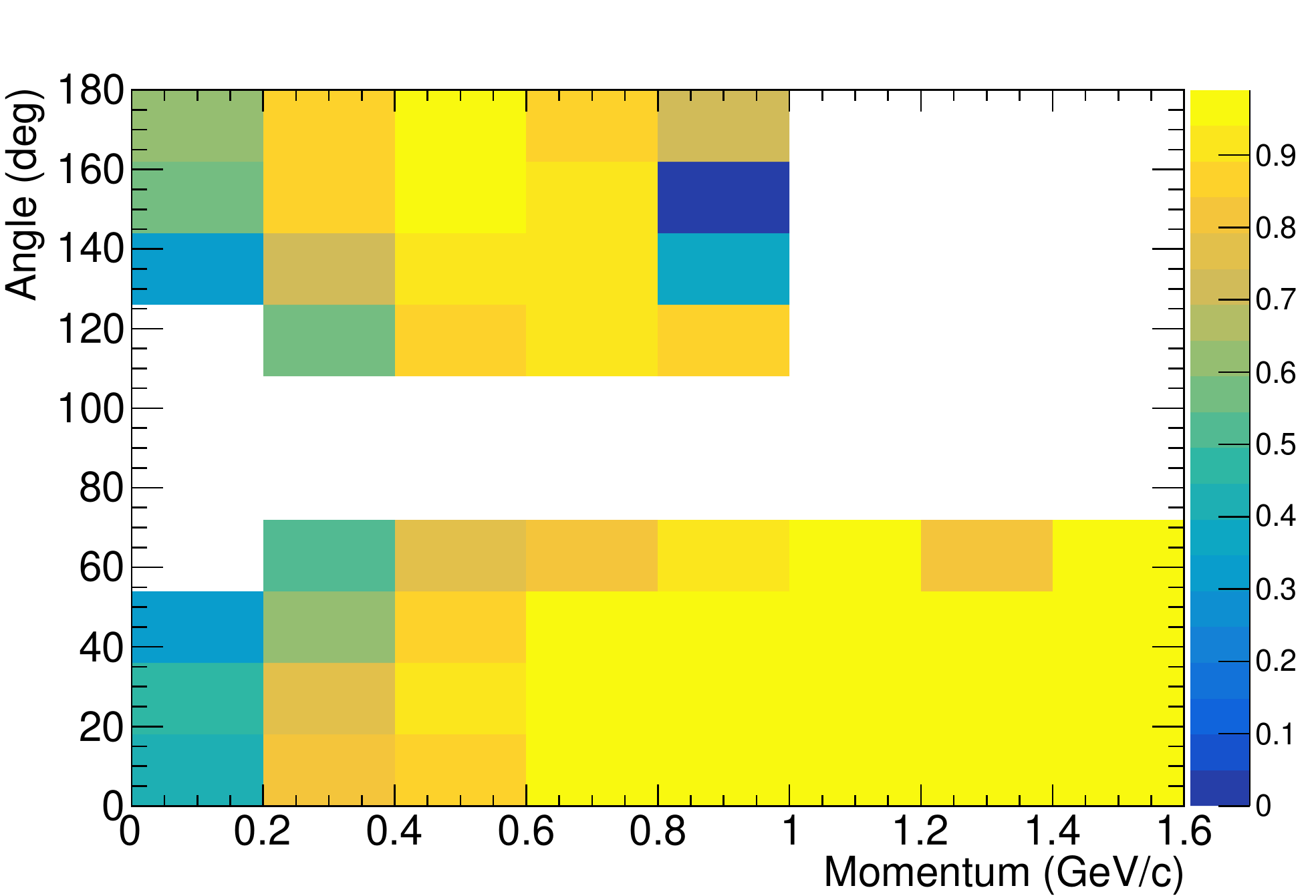}
\includegraphics[width=8.6cm,pagebox=cropbox]{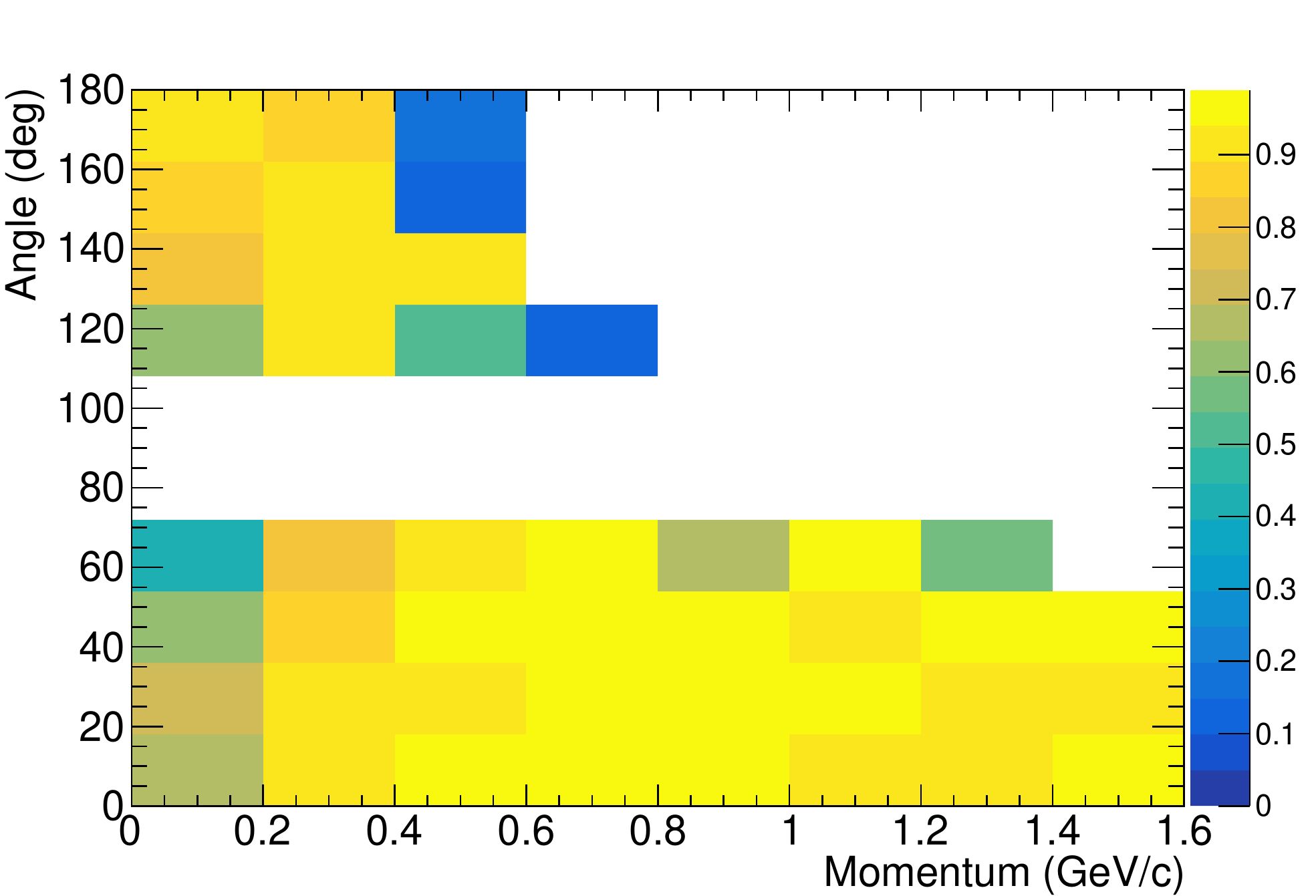}
\caption{\label{fig:ppi_all_2d}Proton (top) and pion (bottom) detection efficiencies estimated by the MC simulation.}
\end{figure}

The selection efficiencies of protons and pions from the neutrino interactions are evaluated by the MC simulation. The efficiency is defined as the number of selected tracks divided by the number of tracks within the scanning angular acceptance. Figure~\ref{fig:ppi_all_2d} shows the result. Blank bins around $90^\circ$ are the region outside the acceptance. More than 50\% of protons are expected to be detected in all angle regions, even in the 200--400\,MeV/$c$ regions. 

\section{Systematic uncertainties}
\label{sec:sys}
The systematic uncertainty sources are classified into three categories: the neutrino flux, the detector response, and the background estimation. In this analysis, comparisons between the data and the MC predictions are shown without unfolding the detector effects. Therefore, the uncertainty of the neutrino interaction modeling does not affect our analysis as a source of the systematic error. The uncertainty from each source is evaluated from the data and the MC simulation as follows.

\subsection{Neutrino flux}
\begin{figure}
    \includegraphics[width=8.6cm,pagebox=cropbox]{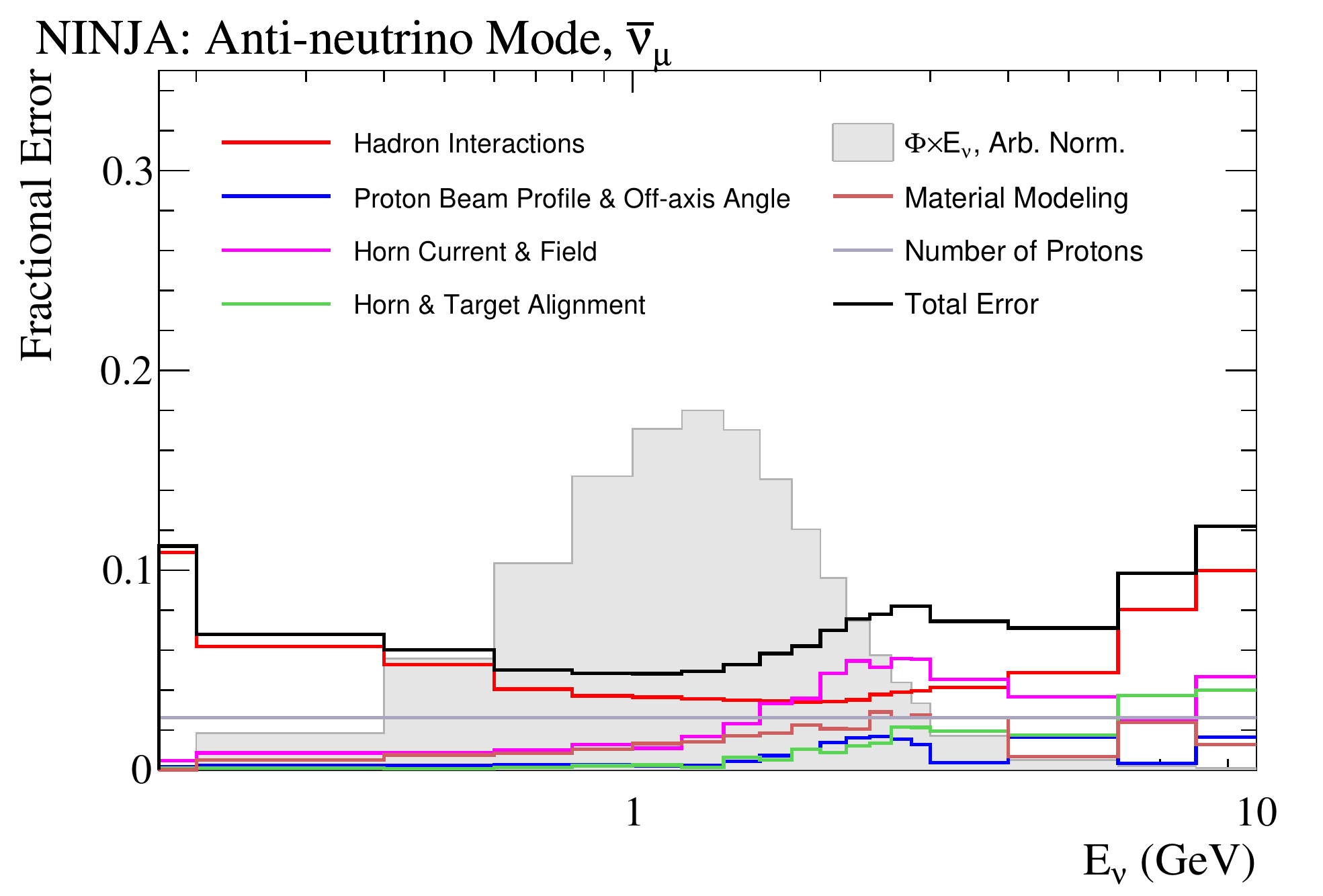}
    \includegraphics[width=8.6cm,pagebox=cropbox]{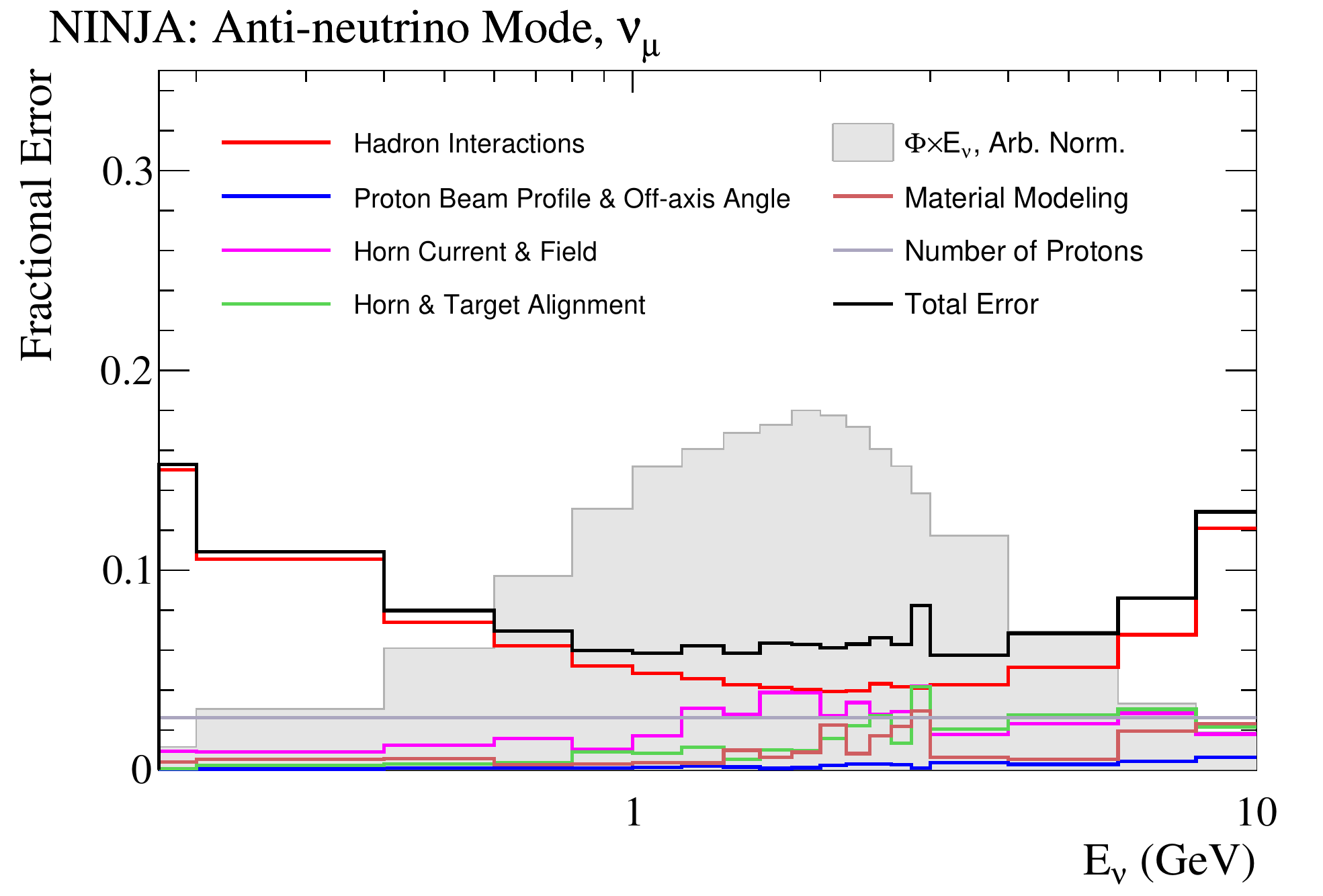}
    \caption{\label{fig:flux_err}Flux uncertainty in the NINJA detector. The gray histograms correspond to fluxes multiplied by the neutrino energy in an arbitrary normalization. $\overline{\nu}_{\mu}$ (top) and $\nu_{\mu}$ (bottom) components in anti-neutrino beam mode are shown.}
  \end{figure}

 The neutrino flux uncertainty and correlations between each neutrino energy bin of both $\overline{\nu}_\mu$ and $\nu_\mu$ components at the detector position are evaluated as a covariance matrix. The matrix is obtained from the uncertainty of the hadron interaction and the various configurations of the J-PARC neutrino beamline. Figure~\ref{fig:flux_err} shows the total flux uncertainty of $\overline{\nu}_\mu$ and $\nu_\mu$ components in the antineutrino mode beam. Systematic uncertainties from the neutrino flux are calculated using a set of toy MC simulations. Weighting factors on flux bins are thrown according to the flux covariance matrix. Then, the change in the number of predicted neutrino interactions from the nominal value is estimated at each bin. This process is repeated $10^5$ times, and the 68\% range of the distribution is regarded as the size of the systematic uncertainty.

\subsection{Detector response}
  The uncertainties from the detector response are evaluated by the sand muons and the MC simulation. The reconstruction efficiency and matching efficiency between the detectors are evaluated using the sand muons, and its statistical error is taken as the uncertainty. The uncertainty of the muon momentum reconstruction is evaluated in the MC simulation by varying the measured position within the position error ($\sim1.5\,\mu$m) which can be obtained from the alignment accuracy of the auto-reconstruction process. The uncertainty from the efficiency of the partner track search, momentum reconstruction and PID performance are checked by varying the selection criteria in the MC simulation on the basis of the emulsion angular resolution ($\sim2$\,mrad). For the PID performance, the difference in the VPH distribution between the data and the MC simulation is also taken into account. The dip positions of the MIP peak and the black peak are checked. There is around 11\% difference at maximum between the data and the MC prediction, thus the VPH distribution in the MC simulation is varied to see how the PID efficiency and purity change. The uncertainty due to the GEANT4 physics list is evaluated by trying various different physics lists, and the uncertainty of the detector modeling in the GEANT4 simulation is also checked by varying the thickness of materials on the basis of measured thickness. The uncertainty of the target mass is calculated from the error of the water layer thickness estimated by the auto-reconstruction process. 
  
  Overall the connection between the ECC and the SFT has around 3\% uncertainty, and it is the dominant uncertainty source for muon detection. Around 7\% uncertainty is assigned to the PID performance, which is the dominant uncertainty source for the kinematics measurements of the protons and pions.

\subsection{Background estimation}
  The beam-induced particles from outside the ECC and cosmic rays are considered as background sources for the muon candidates and the partner track candidates. The uncertainty of the beam-induced background mainly originates from the normalization of the sand muons. The number of sand muons in the MC simulation is normalized with the data. There is a difference of around 30\% between the MC prediction and the data; this is taken as the uncertainty. The cosmic background comes from misconnections in the track matching; however, the uncertainty is less than 1\%, as the contamination is precisely estimated using mock data. The mock data are generated by merging the nominal data and shifted data in which the positions of all the tracks are shifted by a few mm. Then the mock data are processed to the event selection, and the number of extra events compared to the nominal data is treated as the cosmic ray background. By repeating this process many times, the uncertainty is reduced.
  
  Besides the contamination to the muon candidates, cosmic rays which stop in the ECC may be selected as partner track candidates by chance. This background and its uncertainty are also evaluated using another set of mock data in which positions of the muon track candidates are shuffled while the other tracks remain in their original positions. This is also repeated many times, resulting in an uncertainty of less than 1\%.
  
  As above, the total uncertainties of these backgrounds are sufficiently small compared to other uncertainties in most regions.

Besides the uncertainties above, effects of model uncertainties of the neutrino interactions are evaluated as follows. They are compared to the systematic uncertainties to see if the total systematic uncertainty is smaller than the effects of the uncertainties in the neutrino interaction model and the FSI.

\subsection{Neutrino interaction}
  The neutrino interaction model and the FSI in NEUT have many uncertainties. Uncertainties from these sources are evaluated by changing parameters in the model based on the current understanding of the neutrino interaction model and the FSI~\cite{t2k2019}. Table~\ref{tab:reweightdial} shows the nominal value and the $1\sigma$ error size of each parameter. In this analysis, the uncertainty of the nuclear binding energy is not evaluated. However, the uncertainty is covered by the comparison with an alternative nuclear model discussed in Section~\ref{sec:result}. After evaluating the uncertainty induced by each parameter, the uncertainty of the normalization of 2p2h interaction is found to be about 8\%.
  
Although the neutrino interaction uncertainty does not directly affect our analysis, it slightly changes the detection efficiency. The change of the detection efficiency by the change of the interaction model is separately estimated. The typical value is around 1--2\% in each bin. It is added to the uncertainty of the detector response.
  
  \begin{table}
  \caption{\label{tab:reweightdial}Summary of the nominal values of the parameters and their 1$\sigma$ uncertainties used in NEUT.}
  \begin{ruledtabular}
    \begin{tabular}{lll}
      Parameter & Nominal value & Uncertainty\\\hline
      $M\mathrm{_A^{QE}}$ & 1.05\,GeV/$c$$^2$ & 0.20\,GeV/$c$$^2$ \\
      $M\mathrm{_A^{RES}}$ & 0.95\,GeV/$c$$^2$ & 0.15\,GeV/$c$$^2$ \\
      $C\mathrm{^A_{5}}$& 1.01 & 0.12 \\ 
      $\mathrm{Isospin_{\frac{1}{2}}}$ background & 1.30 & 0.20 \\
      CCother shape & 0 & 0.40 \\ 
      CCcoh normalization & 100\% & 100\% \\ 
      NCother normalization & 100\% & 30\% \\
      NCCoh normalization & 100\% & 30\% \\
      2p2h normalization & 100\% & 100\% \\
      Fermi momentum & 225\,MeV/$c$ & 31\,MeV/$c$ \\ \hline
      Pion absorption & 1.1 & 50\% \\ 
      Pion charge exchange (low E) & 1.0 & 50\% \\ 
      Pion charge exchange (high E) & 1.8 & 30\% \\
      Pion quasi elastic (low E) & 1.0 & 50\% \\ 
      Pion quasi elastic (high E) & 1.8 & 30\% \\
      Pion inelastic & 1 & 50\% \\ 
    \end{tabular}
  \end{ruledtabular}
\end{table}

\begin{figure*}
  \begin{minipage}{0.48\hsize}
  \includegraphics[width=8.6cm,pagebox=cropbox]{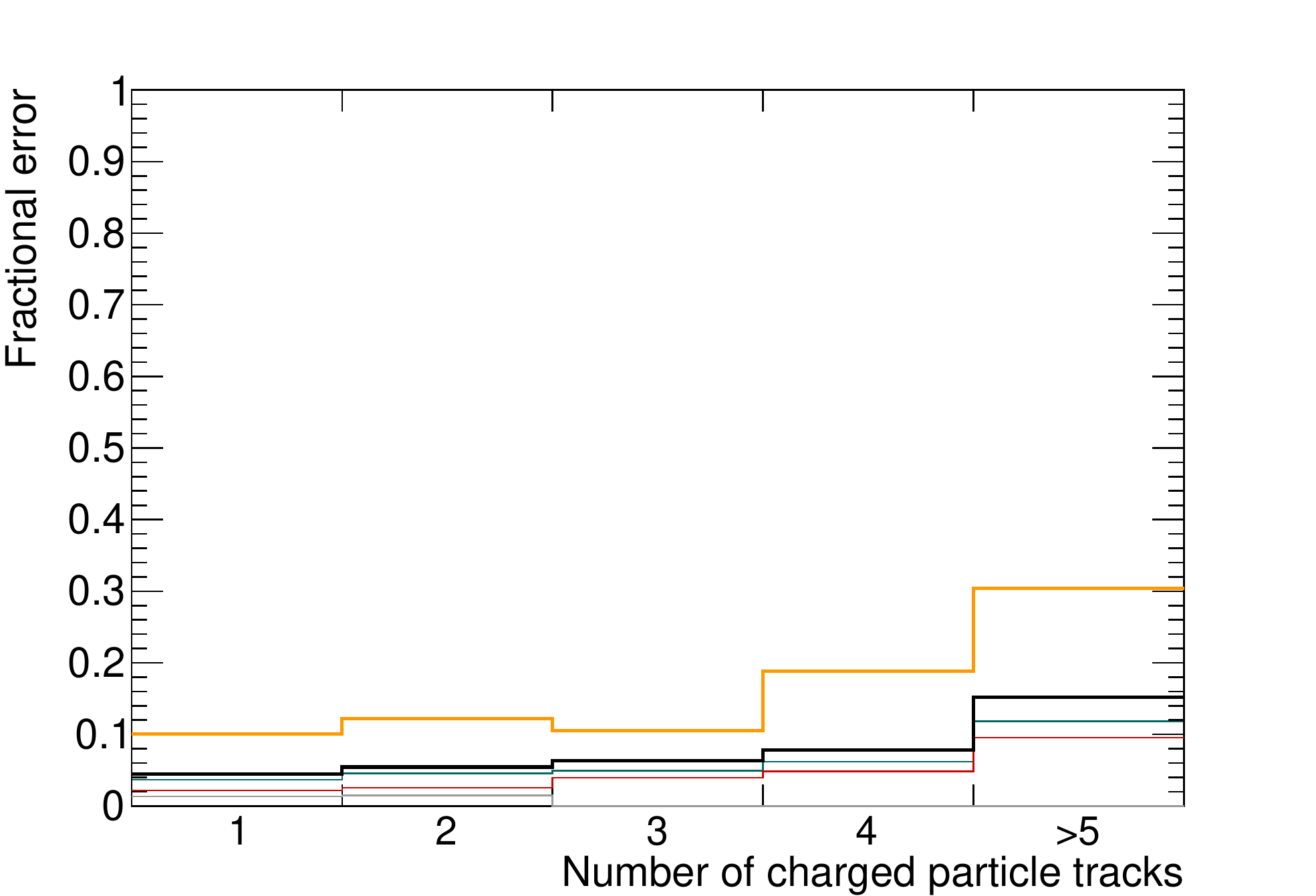}
    \end{minipage}
  \begin{minipage}{0.48\hsize}
  \includegraphics[width=8.6cm,pagebox=cropbox]{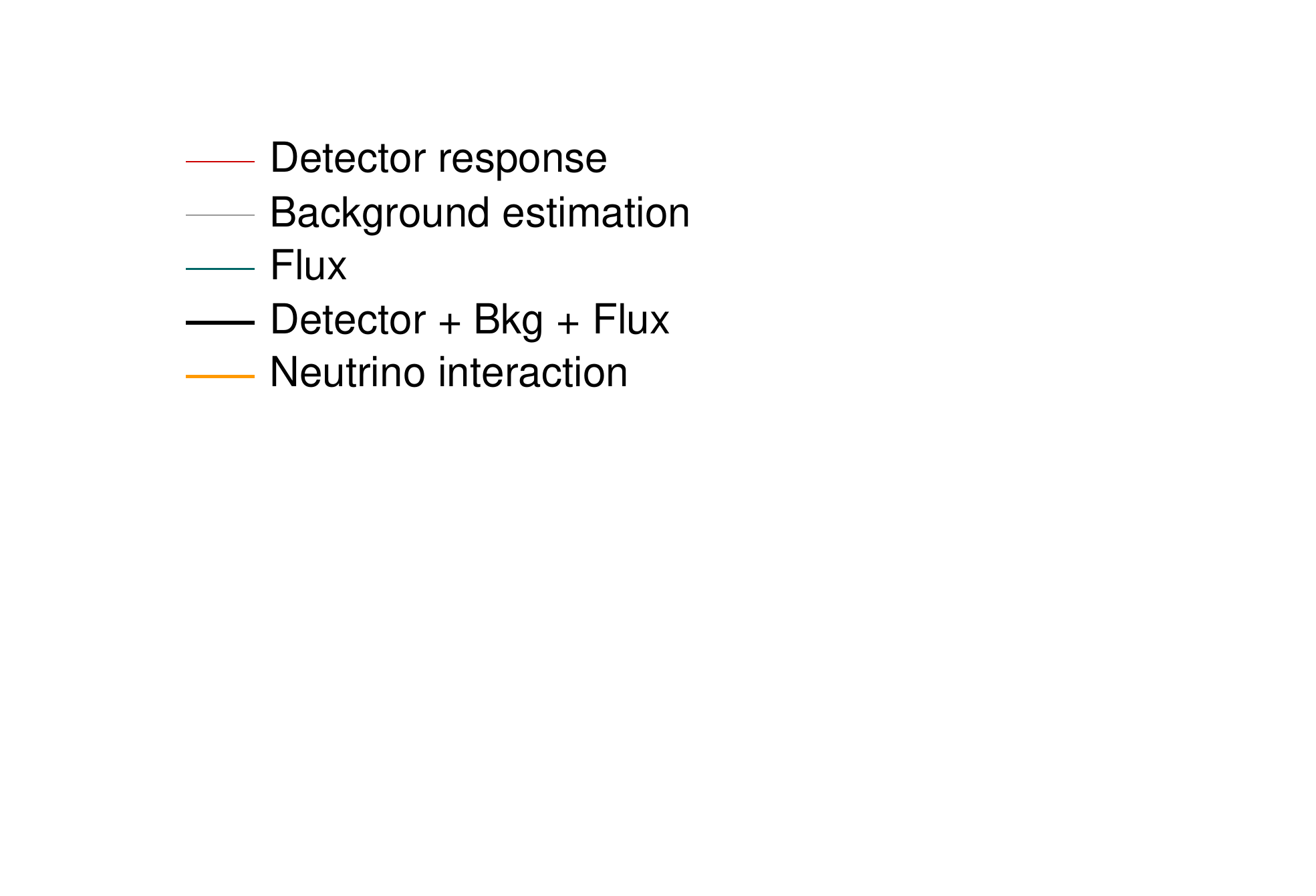}
    \end{minipage}
  \begin{minipage}{0.48\hsize}
  \includegraphics[width=8.6cm,pagebox=cropbox]{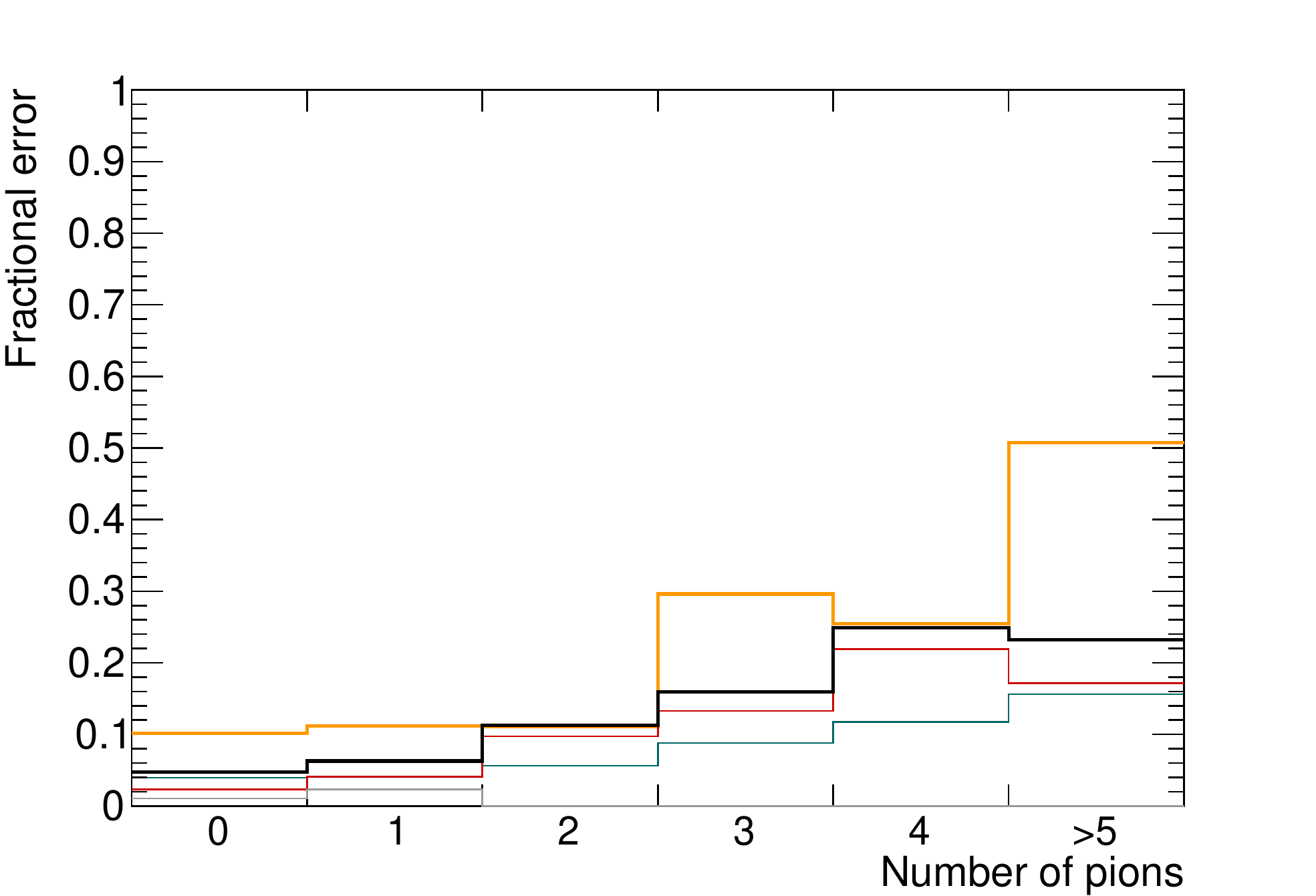}
    \end{minipage}
  \begin{minipage}{0.48\hsize}
  \includegraphics[width=8.6cm,pagebox=cropbox]{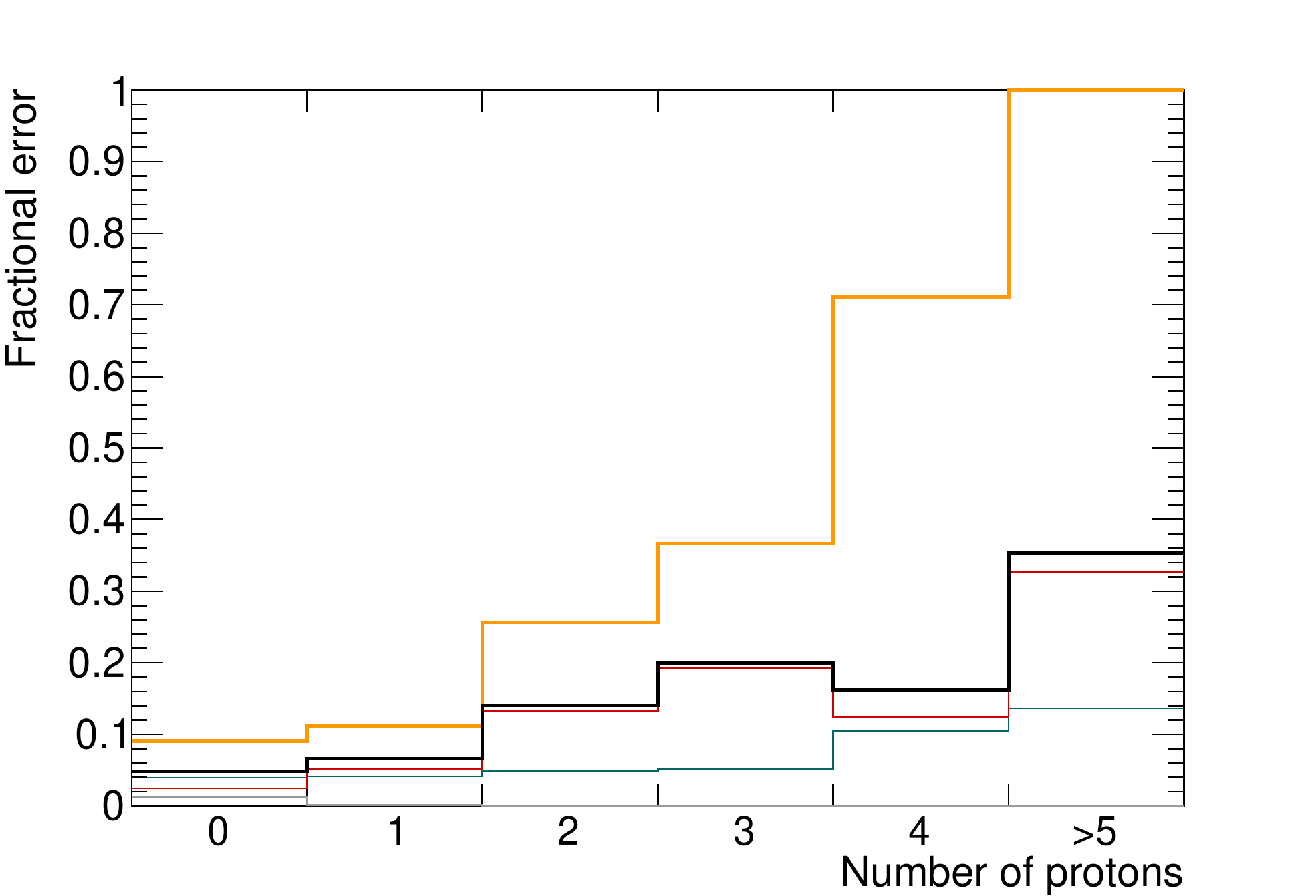}
  \end{minipage}
\caption{\label{fig:ntrk_all}Summary of the fractional uncertainties of charged particle multiplicity (top), the number of pions (bottom left), and the number of protons (bottom right) with a breakdown by the uncertainties from the neutrino flux, detector response, and background estimation. The uncertainty of neutrino interaction modeling is compared to the other uncertainties.}
\end{figure*}

\begin{figure*}
  \begin{minipage}{0.48\hsize}
  \includegraphics[width=8.6cm,pagebox=cropbox]{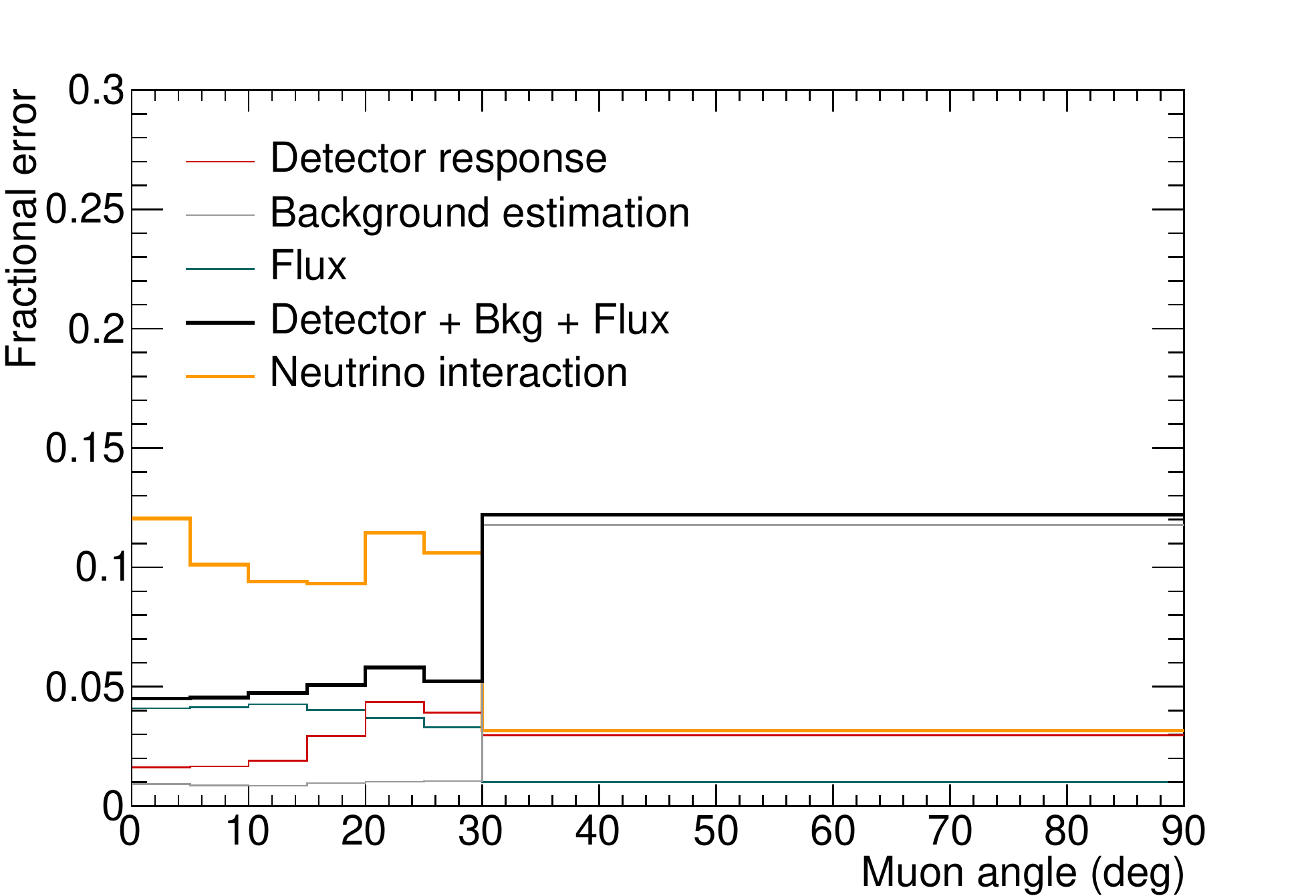}
  \end{minipage}
  \begin{minipage}{0.48\hsize}
    \includegraphics[width=8.6cm,pagebox=cropbox]{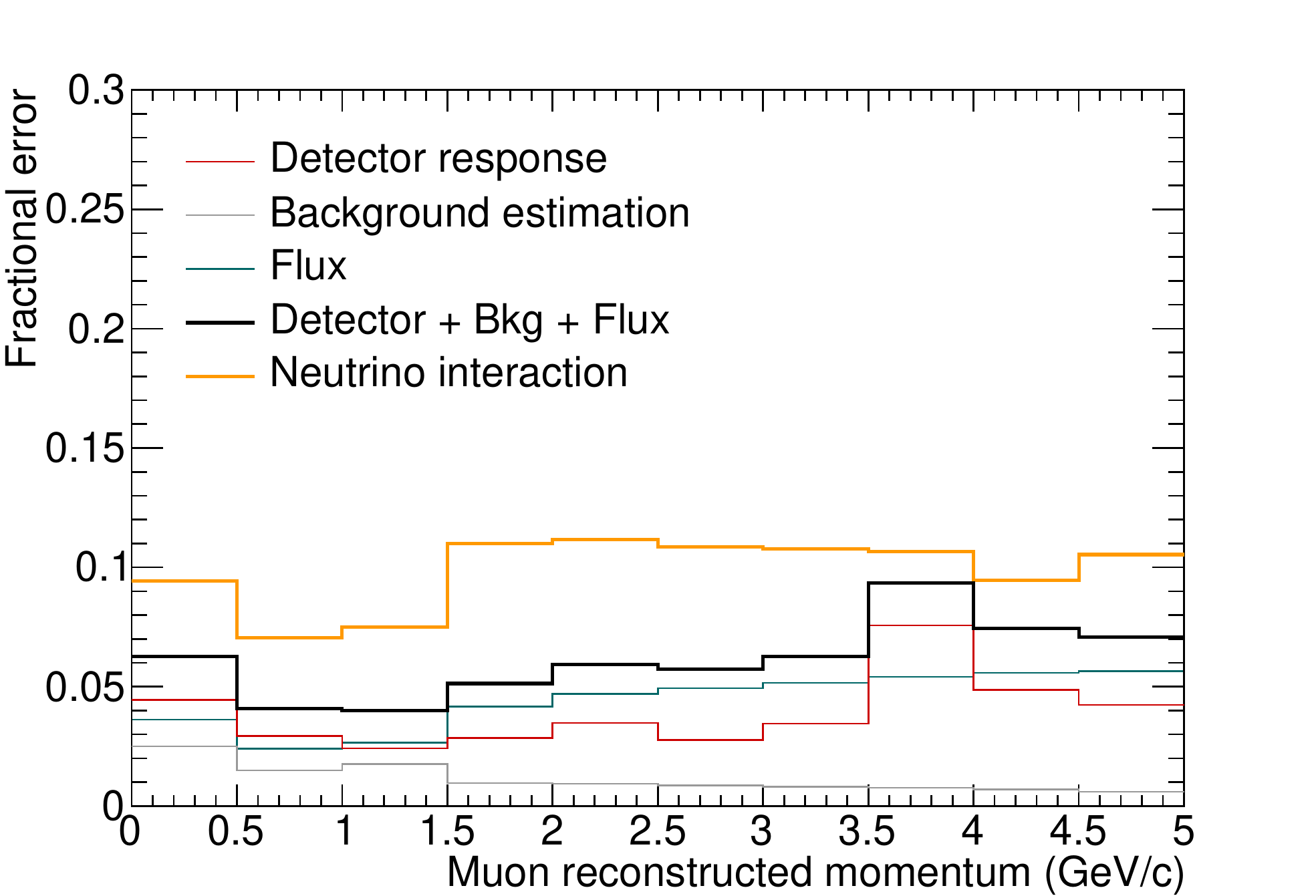}
  \end{minipage}
    \begin{minipage}{0.48\hsize}
  \includegraphics[width=8.6cm,pagebox=cropbox]{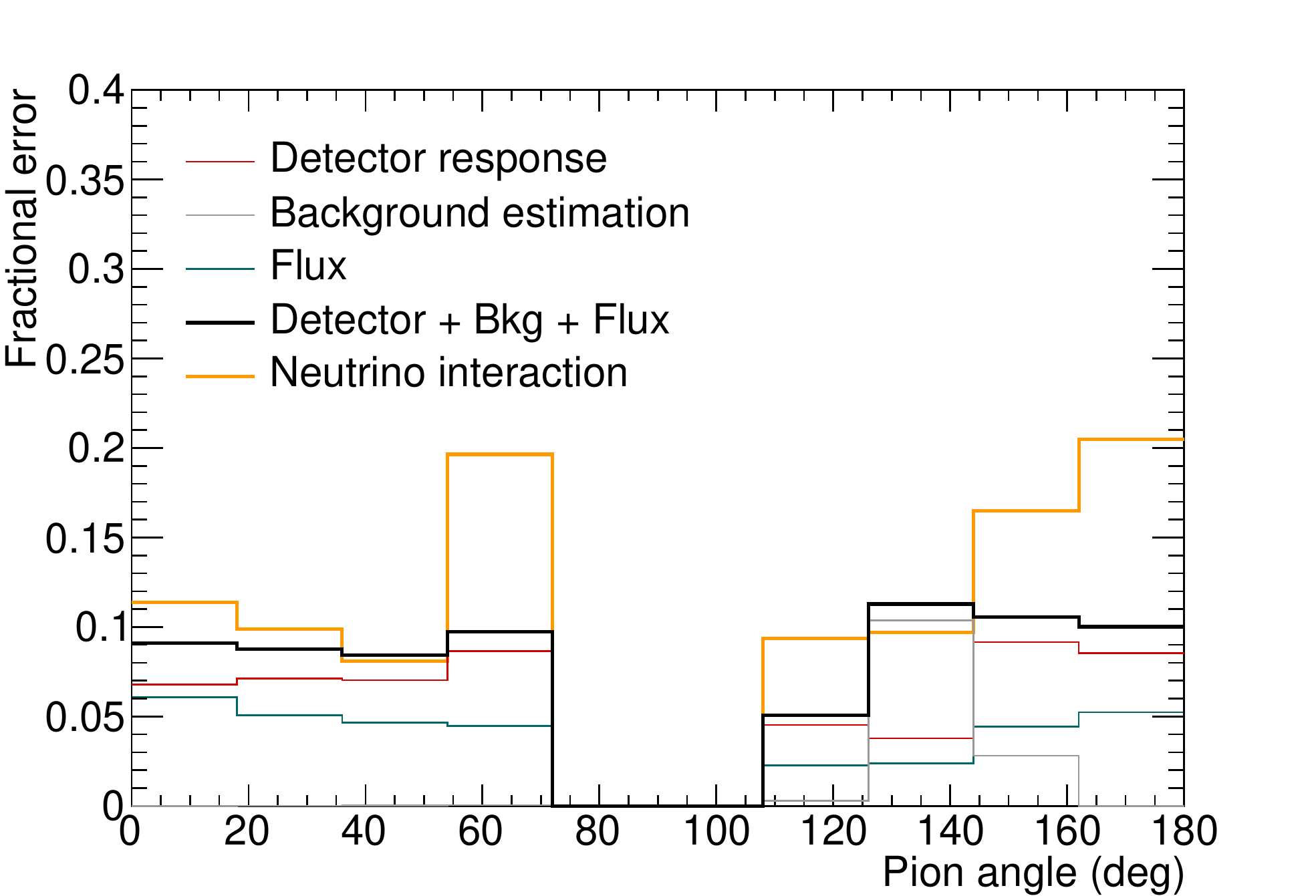}
  \end{minipage}
    \begin{minipage}{0.48\hsize}
  \includegraphics[width=8.6cm,pagebox=cropbox]{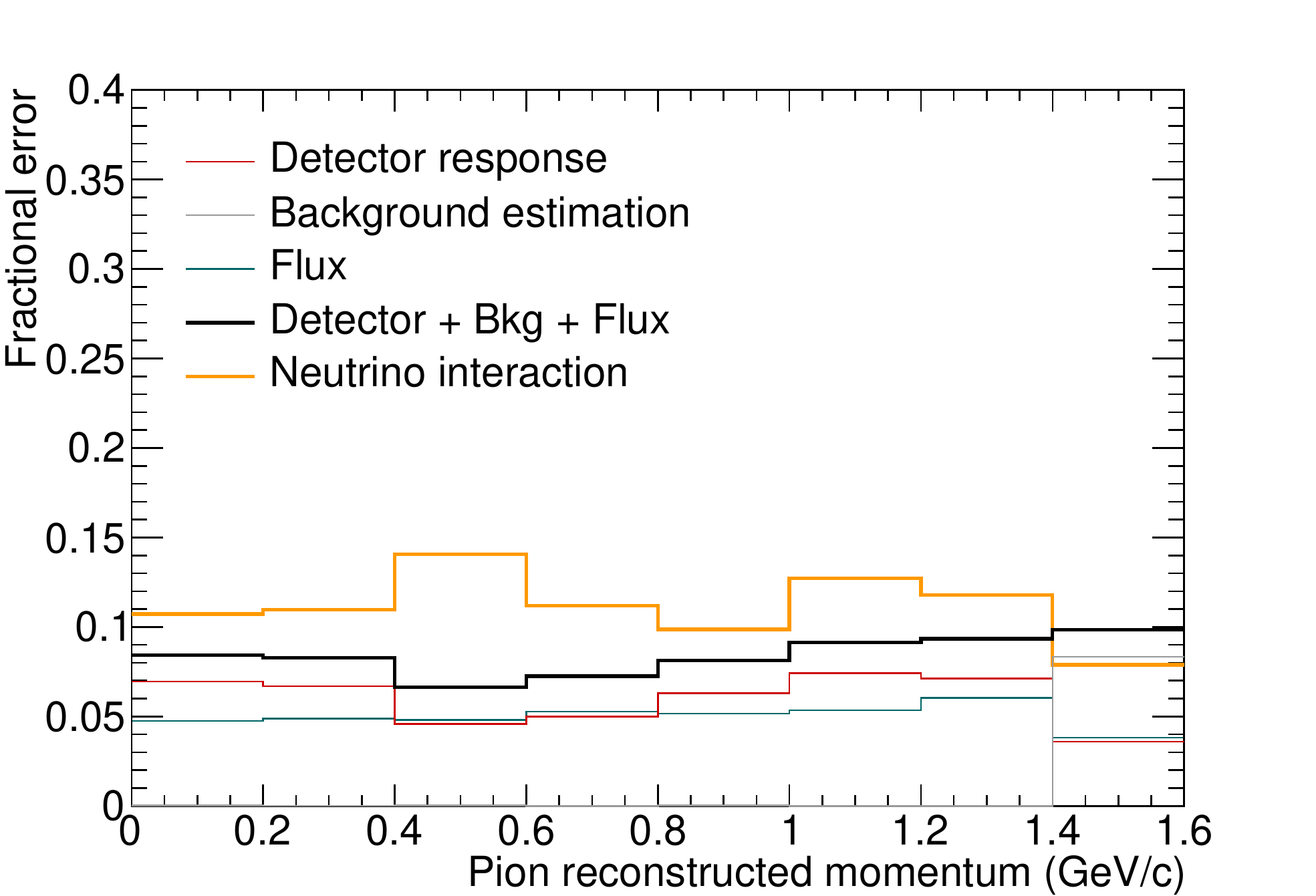}
  \end{minipage}
    \begin{minipage}{0.48\hsize}
    \includegraphics[width=8.6cm,pagebox=cropbox]{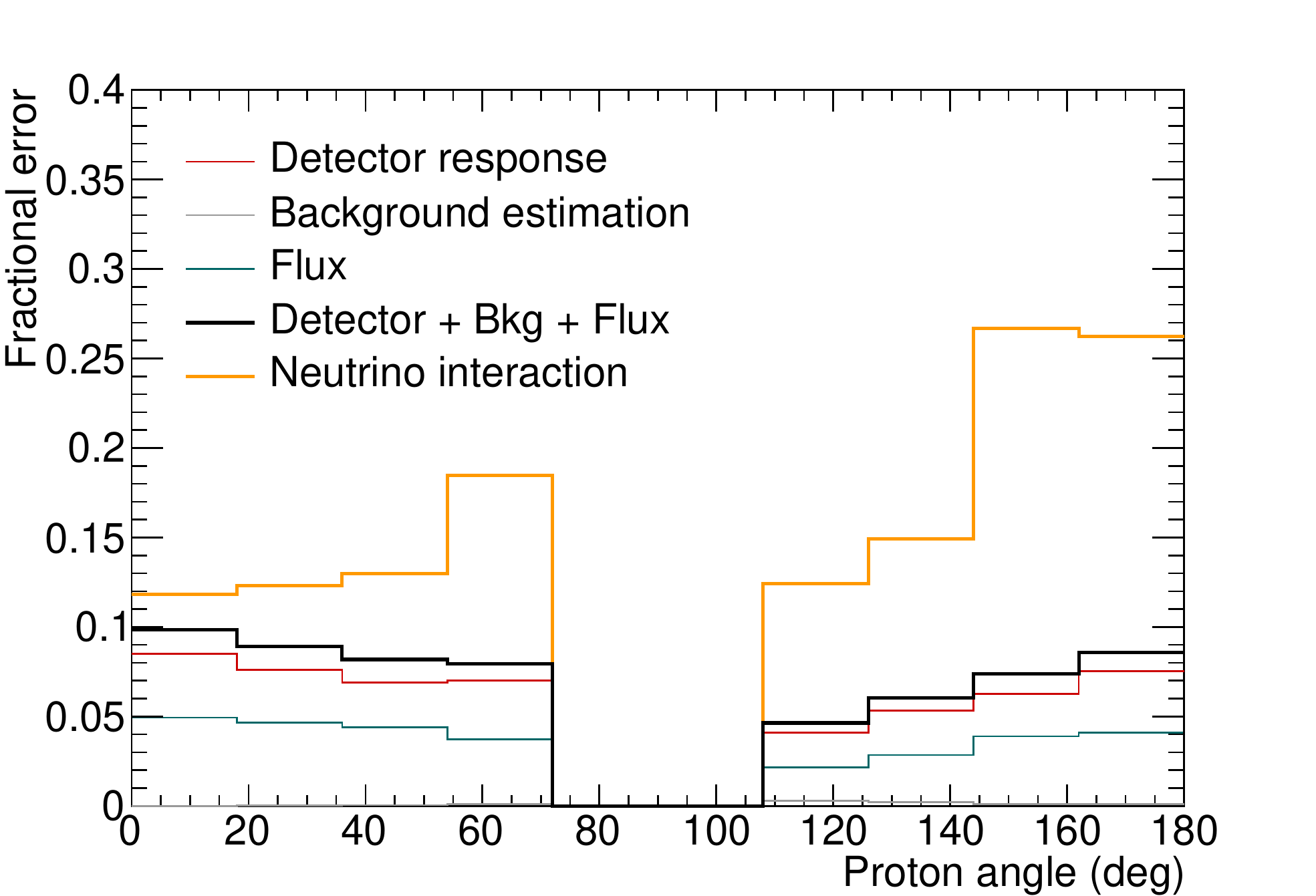}
  \end{minipage}
  \begin{minipage}{0.48\hsize}
    \includegraphics[width=8.6cm,pagebox=cropbox]{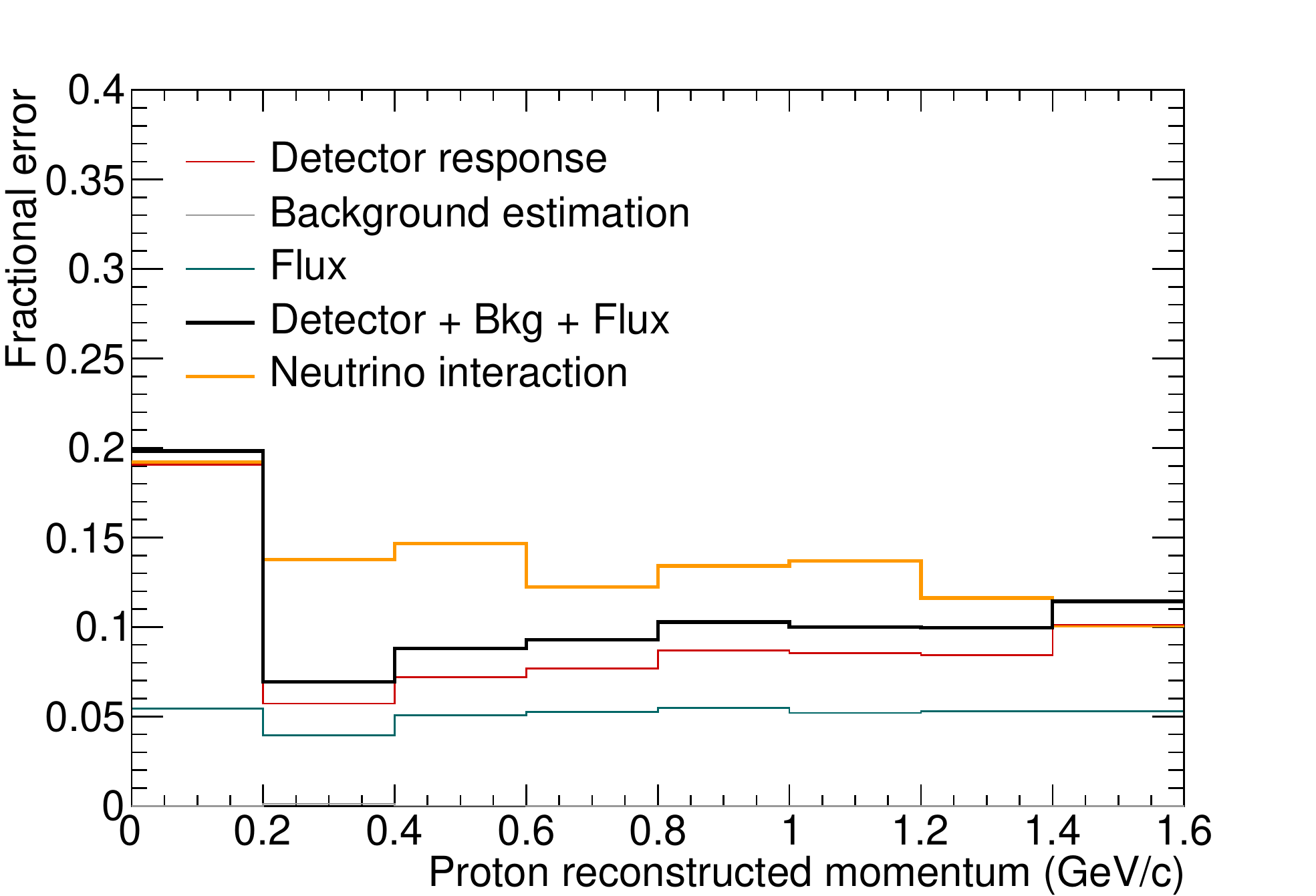}
  \end{minipage}
  \caption{\label{fig:systematics}Summary of the fractional uncertainties of muon, pion, and proton kinematics with a breakdown by the uncertainties from the neutrino flux, detector response, and background estimation. The uncertainty of neutrino interaction modeling is compared to the other uncertainties. The left column corresponds to the fractional uncertainties of angle distribution, while the right column corresponds to those of momentum distribution.}
\end{figure*}

Figures~\ref{fig:ntrk_all} and \ref{fig:systematics} show the systematic uncertainties of each measurement with a breakdown by category. The fractional uncertainty of the expected number of selected events in each bin is plotted. The quadrature sums of the uncertainties from the neutrino flux, the detector response, and the background estimation are smaller than the uncertainty of the neutrino interaction model in almost all bins. This shows that our future measurements will give constraints on the models. An uncertainty of only 5\%--8\% is derived from the flux uncertainty owing to the great improvement in the hadron interaction modeling by NA61/SHINE. The current detector uncertainty is slightly larger than the flux uncertainty, and desired to be improved in future analysis. 

\vspace{0.5cm}
\section{Results}
\label{sec:result}
\vspace{0.5cm}
In this measurement, the statistics is not sufficient to precisely extract the cross section from the reconstructed distributions. Thus, distributions of the charged particles from the neutrino interactions on the water are compared with model predictions to get insights into the model validity and to demonstrate the feasibility of the NINJA detector for future measurements. It should be noted that our results include low-momentum charged particles, especially protons with momenta of 200--400\,MeV/$c$, owing to the high granularity of the emulsion films.

\begin{figure*}
  \begin{minipage}{0.48\hsize}
  \includegraphics[width=8.6cm,pagebox=cropbox]{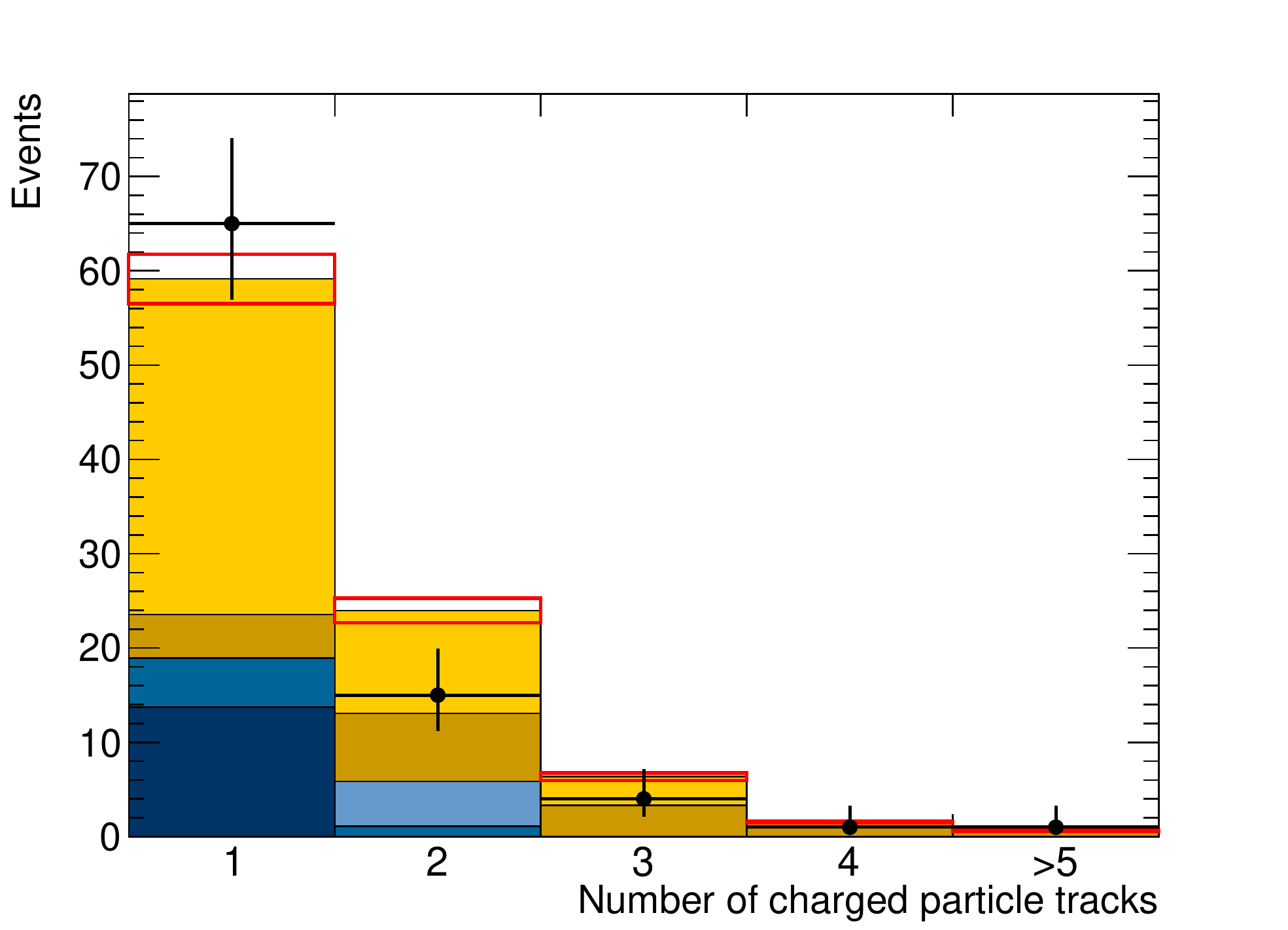}
  \end{minipage}
  \begin{minipage}{0.48\hsize}
    \includegraphics[width=8.6cm,pagebox=cropbox]{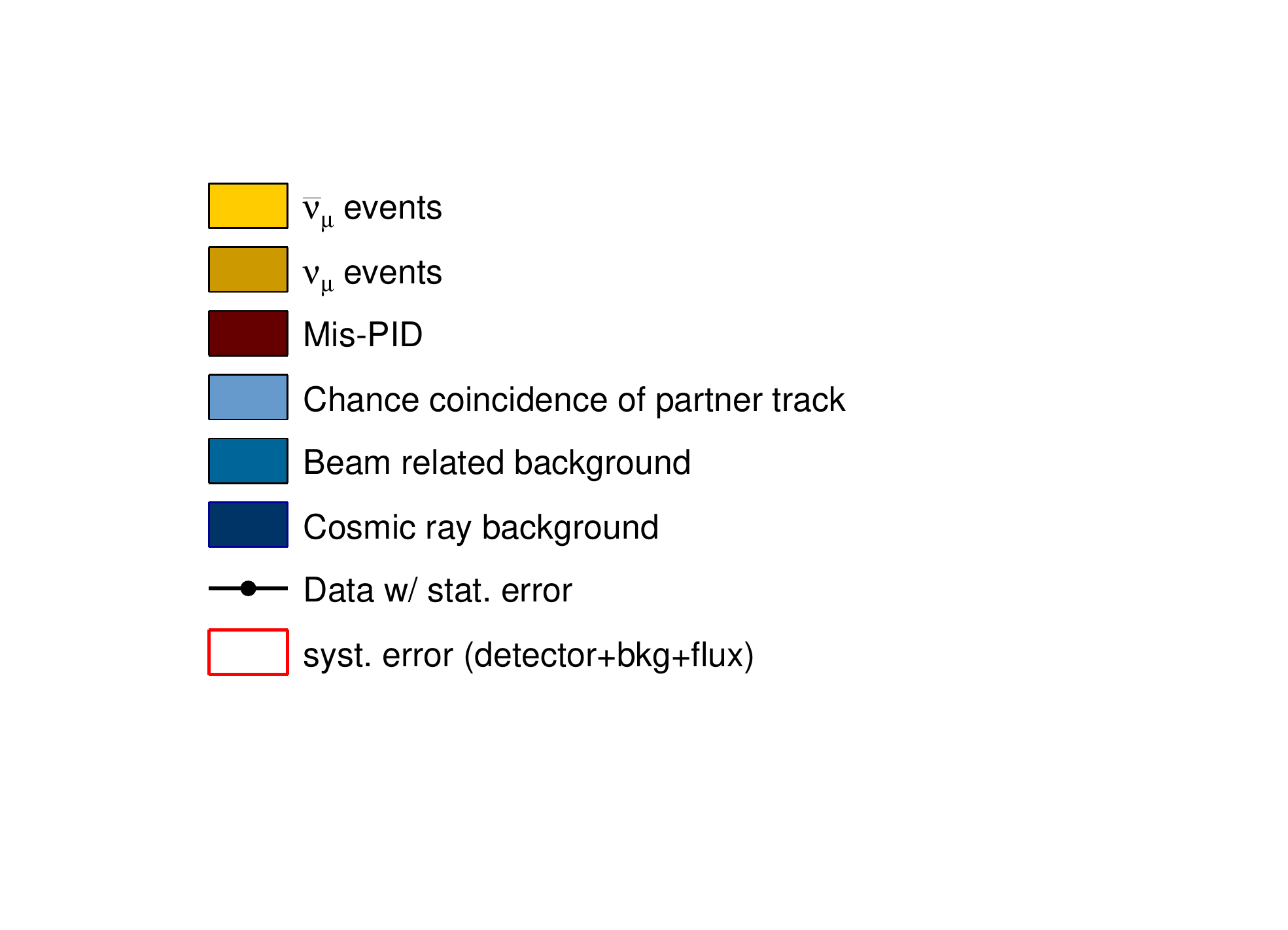}
  \end{minipage}
  \begin{minipage}{0.48\hsize}
  \includegraphics[width=8.6cm,pagebox=cropbox]{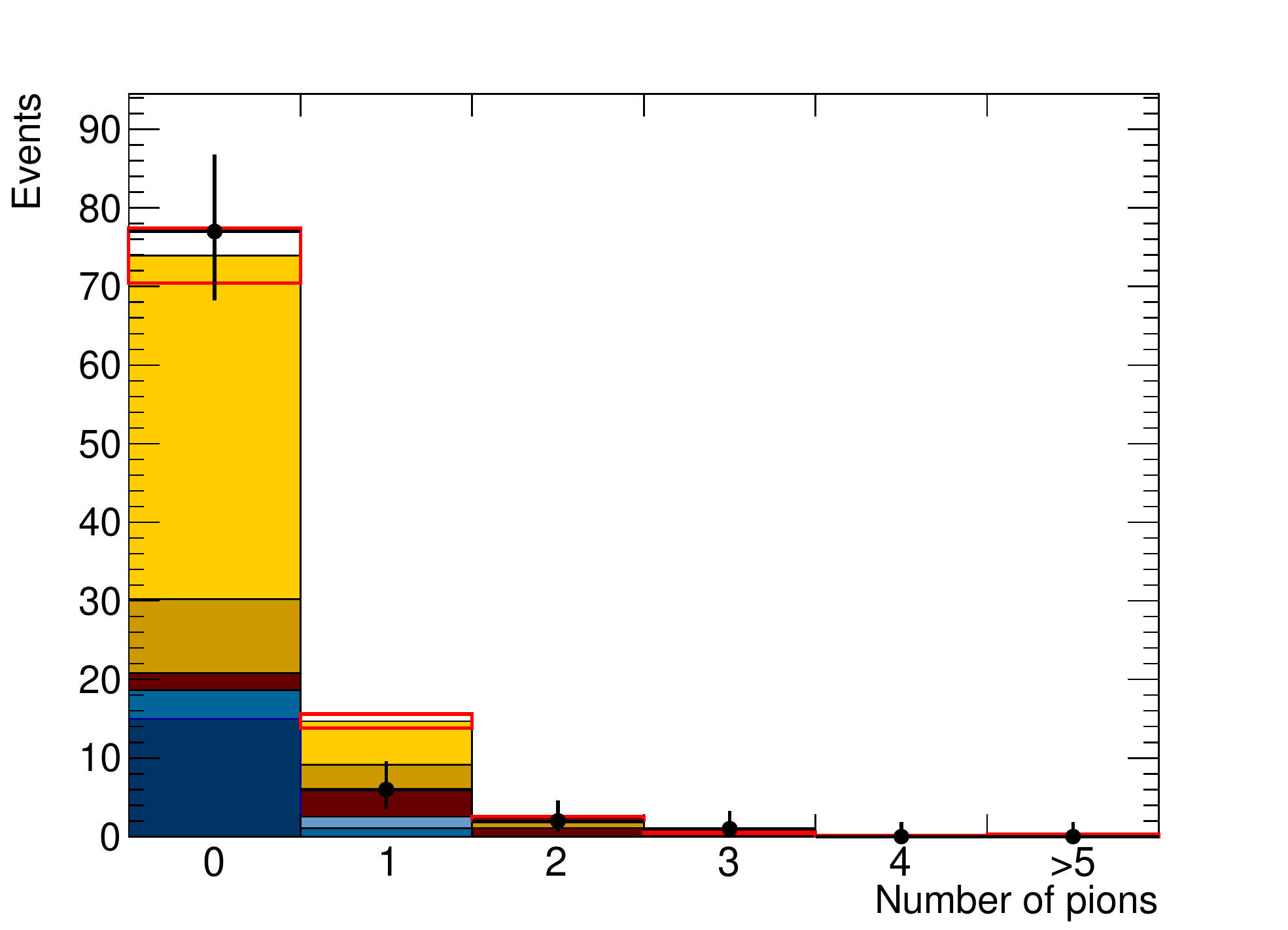}
  \end{minipage}
  \begin{minipage}{0.48\hsize}
    \includegraphics[width=8.6cm,pagebox=cropbox]{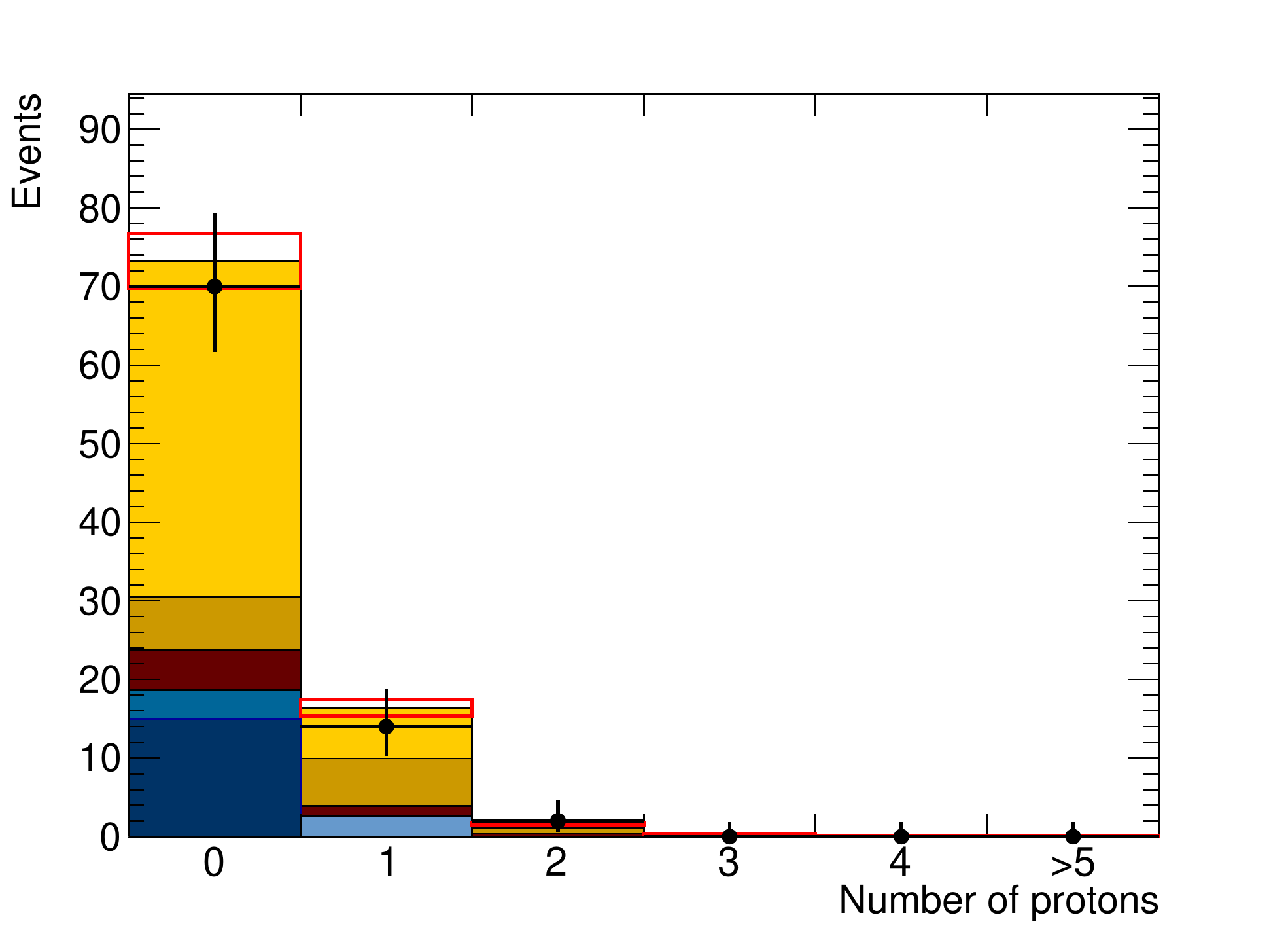}
  \end{minipage}
\caption{\label{fig:ntrk}Multiplicity of charged particles from neutrino-water interactions and backgrounds including muon candidates (top). The bottom plots show the number of pions (left) and protons (right). The data points are shown by marker points with statistical error bars and the predictions are shown by histograms with systematic uncertainties as red boxes, which are the quadrature sum of the uncertainties of neutrino flux, the deector response, and the background estimation.}
\end{figure*}

\begin{figure*}
  \begin{minipage}{0.48\hsize}
  \includegraphics[width=8.6cm,pagebox=cropbox]{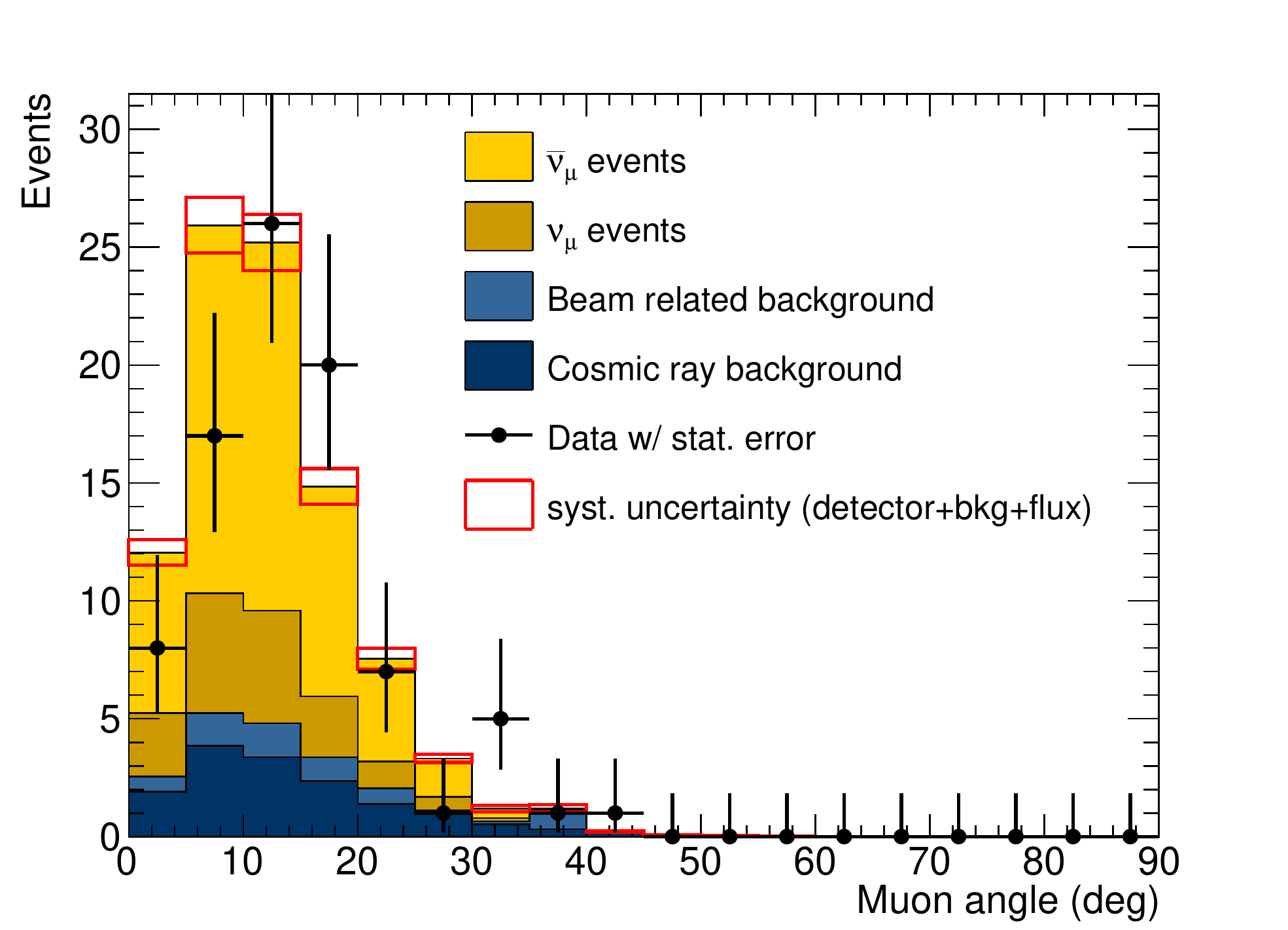}
  \end{minipage}
  \begin{minipage}{0.48\hsize}
  \includegraphics[width=8.6cm,pagebox=cropbox]{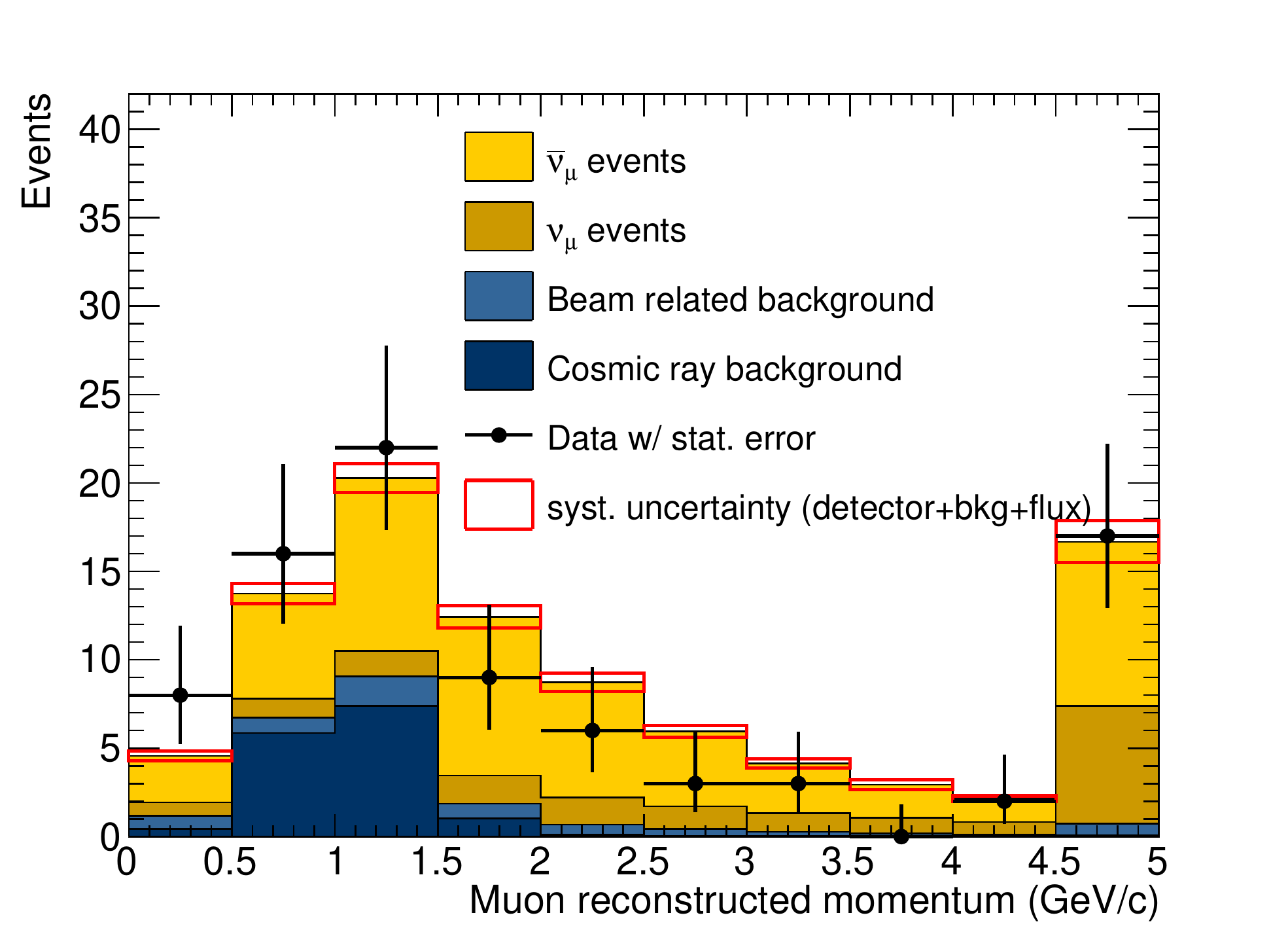}
  \end{minipage}
    \begin{minipage}{0.48\hsize}
  \includegraphics[width=8.6cm,pagebox=cropbox]{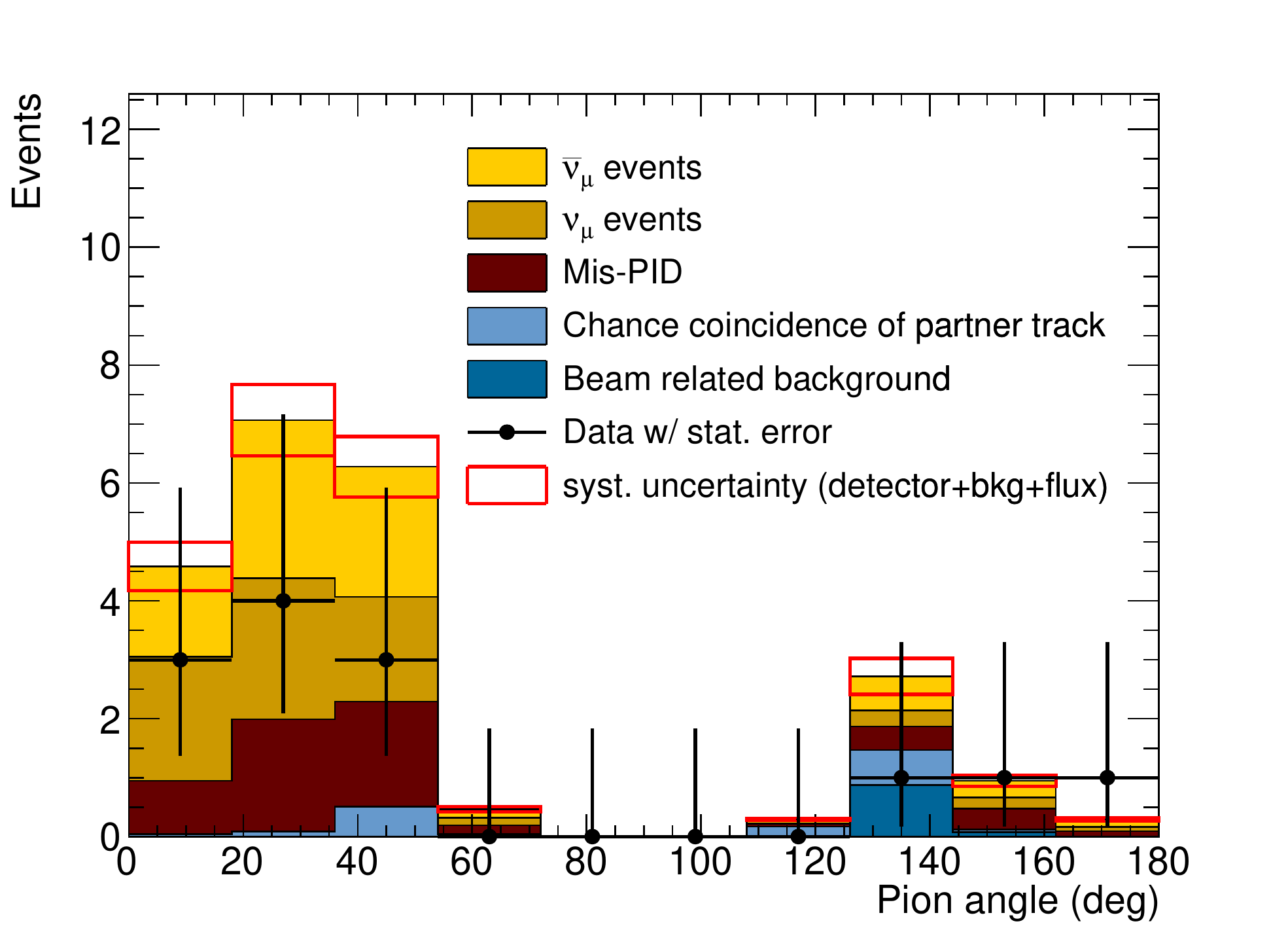}
  \end{minipage}
  \begin{minipage}{0.48\hsize}
  \includegraphics[width=8.6cm,pagebox=cropbox]{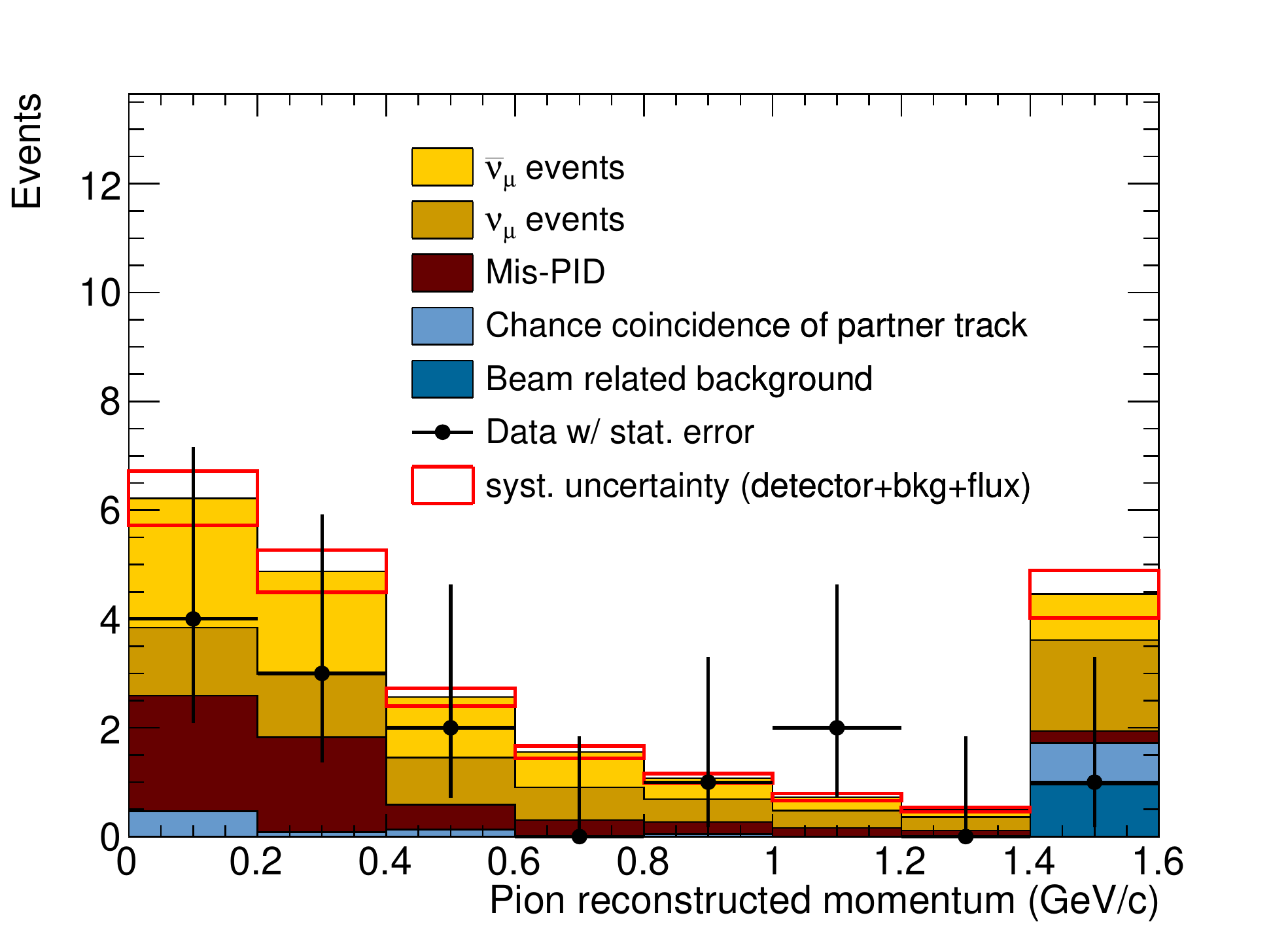}
  \end{minipage}
  \begin{minipage}{0.48\hsize}
    \includegraphics[width=8.6cm,pagebox=cropbox]{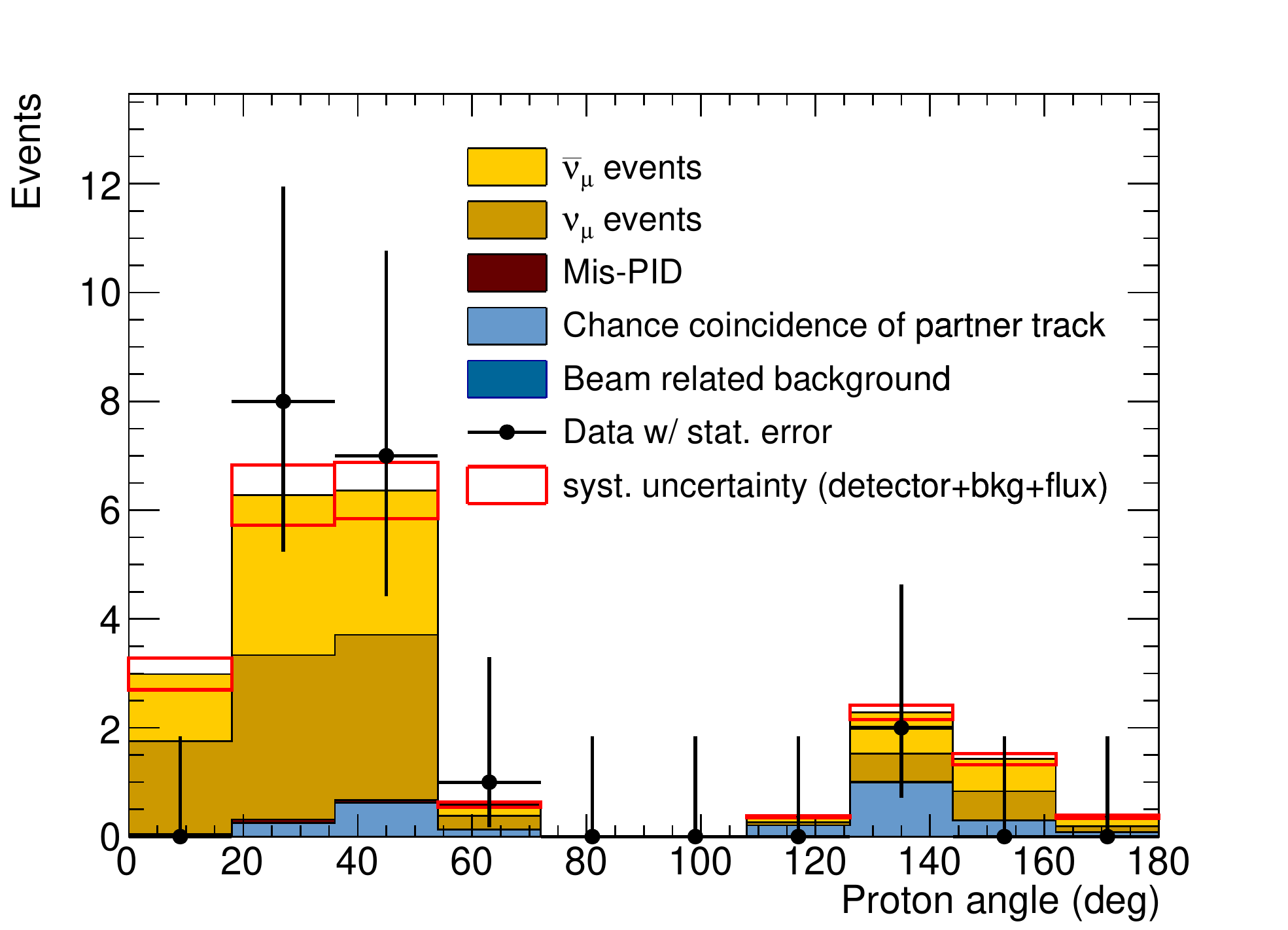}
  \end{minipage}
  \begin{minipage}{0.48\hsize}
    \includegraphics[width=8.6cm,pagebox=cropbox]{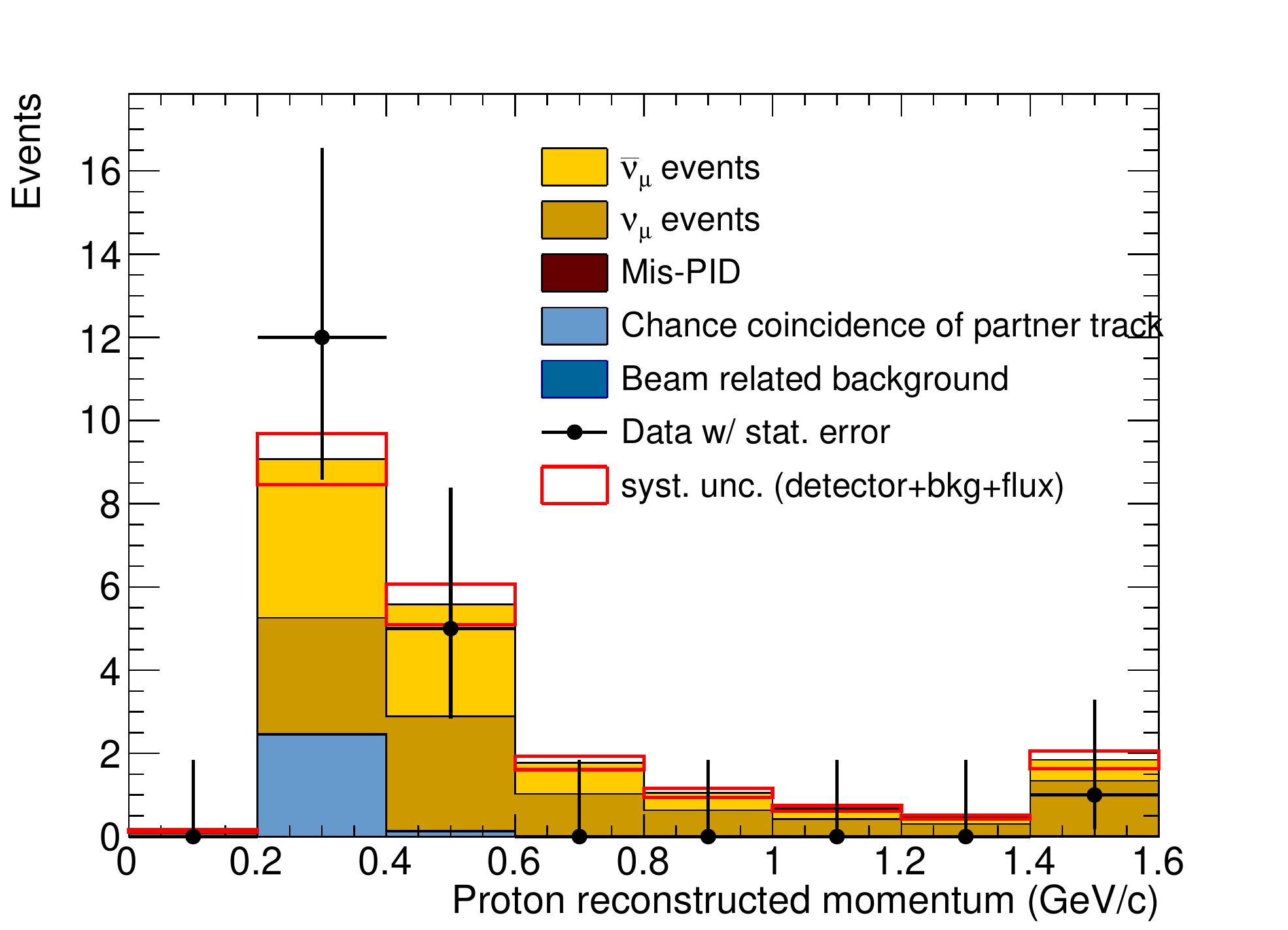}
  \end{minipage}
  \caption{\label{fig:results}Distributions of muon, pion, and proton kinematics from neutrino-water interactions and backgrounds. The left column shows angle distributions, while the right column shows reconstructed momentum distributions. Though the angular resolution for all particles is sufficiently small compared to the bin width in the angle plots, the momentum resolution is typically larger than the momentum binning, especially for high momentum muons. In the right-most bins of the momentum distributions, all the events with momenta above 5\,GeV/$c$ for muons and 1.6\,GeV/$c$ for pions and protons are contained. The data points are shown by marker points with statistical error bars and the predictions are shown by histograms with systematic uncertainties as red boxes, which are the quadrature sum of the uncertainties of neutrino flux, the detector response, and the background estimation.}
\end{figure*}

First, raw data distributions are compared with sum of the neutrino event prediction estimated with the MC simulation and the cosmic-ray background prediction estimated with the off-beam timing data. Figure~\ref{fig:ntrk} shows the multiplicity of the charged particles and the number of pions and protons. The red boxes on the prediction correspond to the quadrature sum of the uncertainties of neutrino flux, the detector response, and the background estimation. Figure~\ref{fig:results} shows distributions of the reconstructed kinematics of muons, pions, and protons. Though the angular resolution for all particles is sufficiently small compared to the bin width in the angle plots, the momentum resolution is typically larger than the momentum binning, especially for high momentum muons. In the proton momentum distribution, protons with momentum 200--400\,MeV/$c$ are successfully detected for the first time in measurements of neutrino-water interactions.

\begin{figure*}
    \begin{minipage}{0.48\hsize}
  \includegraphics[width=8.6cm,pagebox=cropbox]{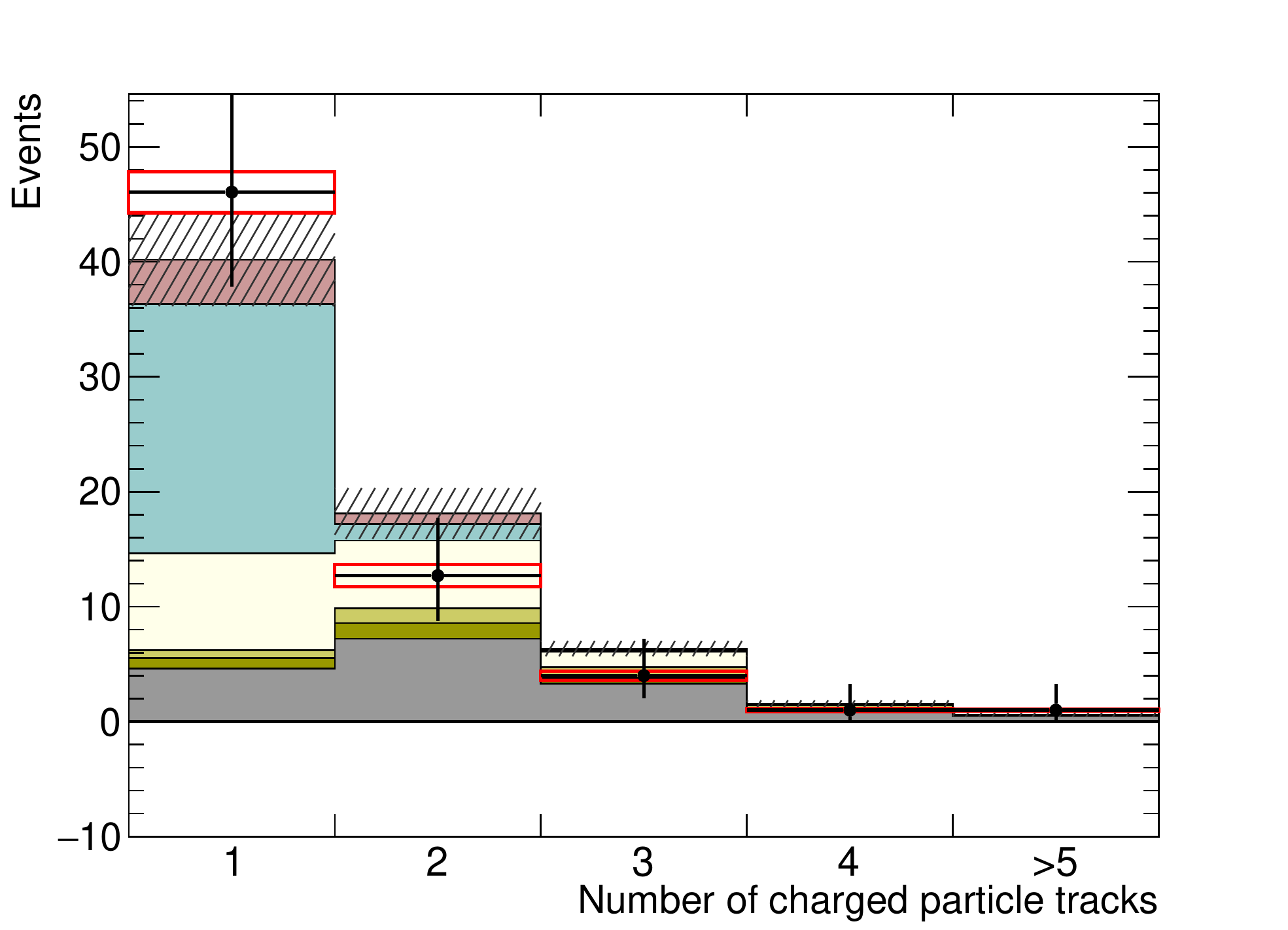}
  \end{minipage}
  \begin{minipage}{0.48\hsize}
    \includegraphics[width=8.6cm,pagebox=cropbox]{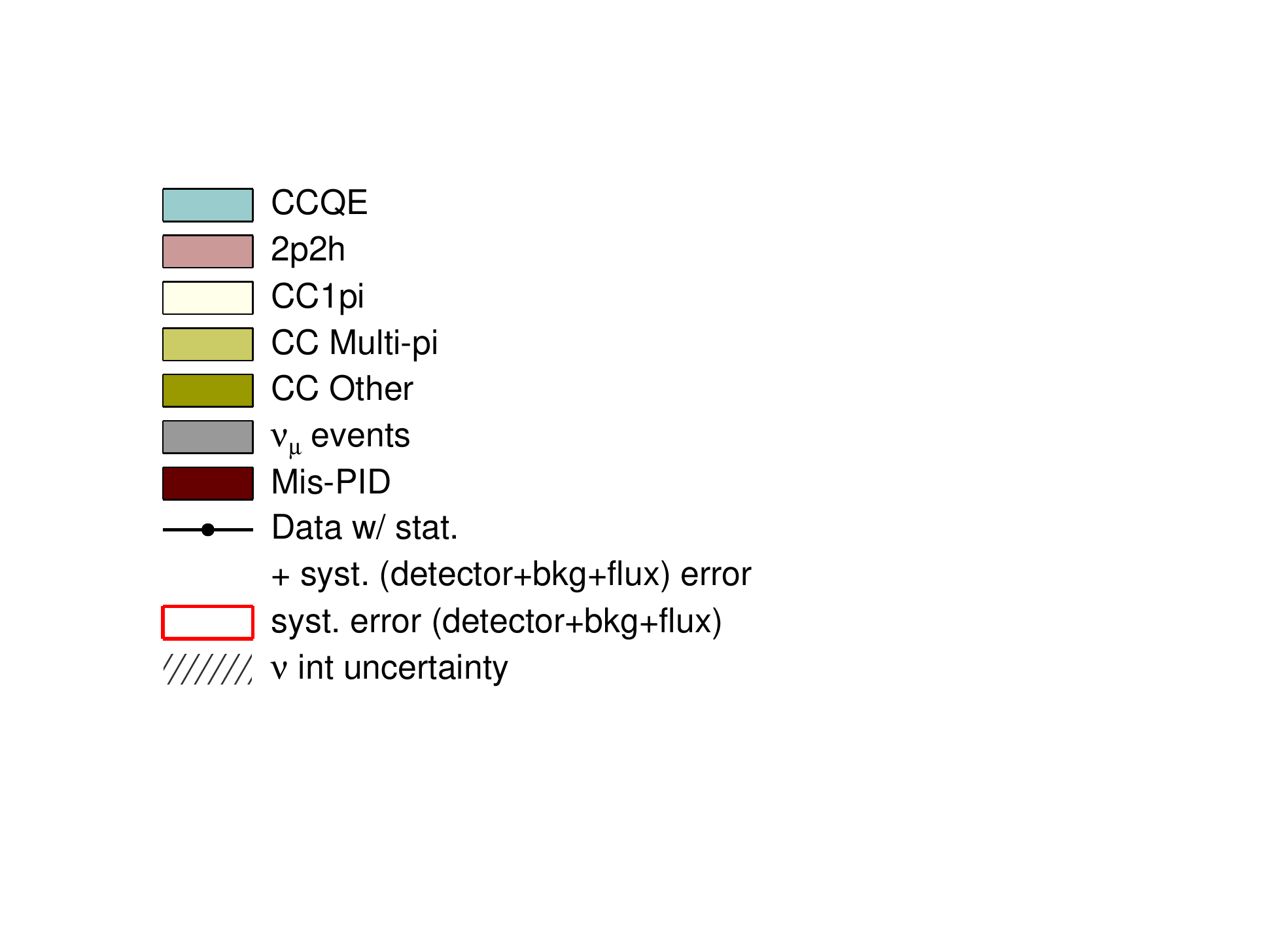}
  \end{minipage}
    \begin{minipage}{0.48\hsize}
  \includegraphics[width=8.6cm,pagebox=cropbox]{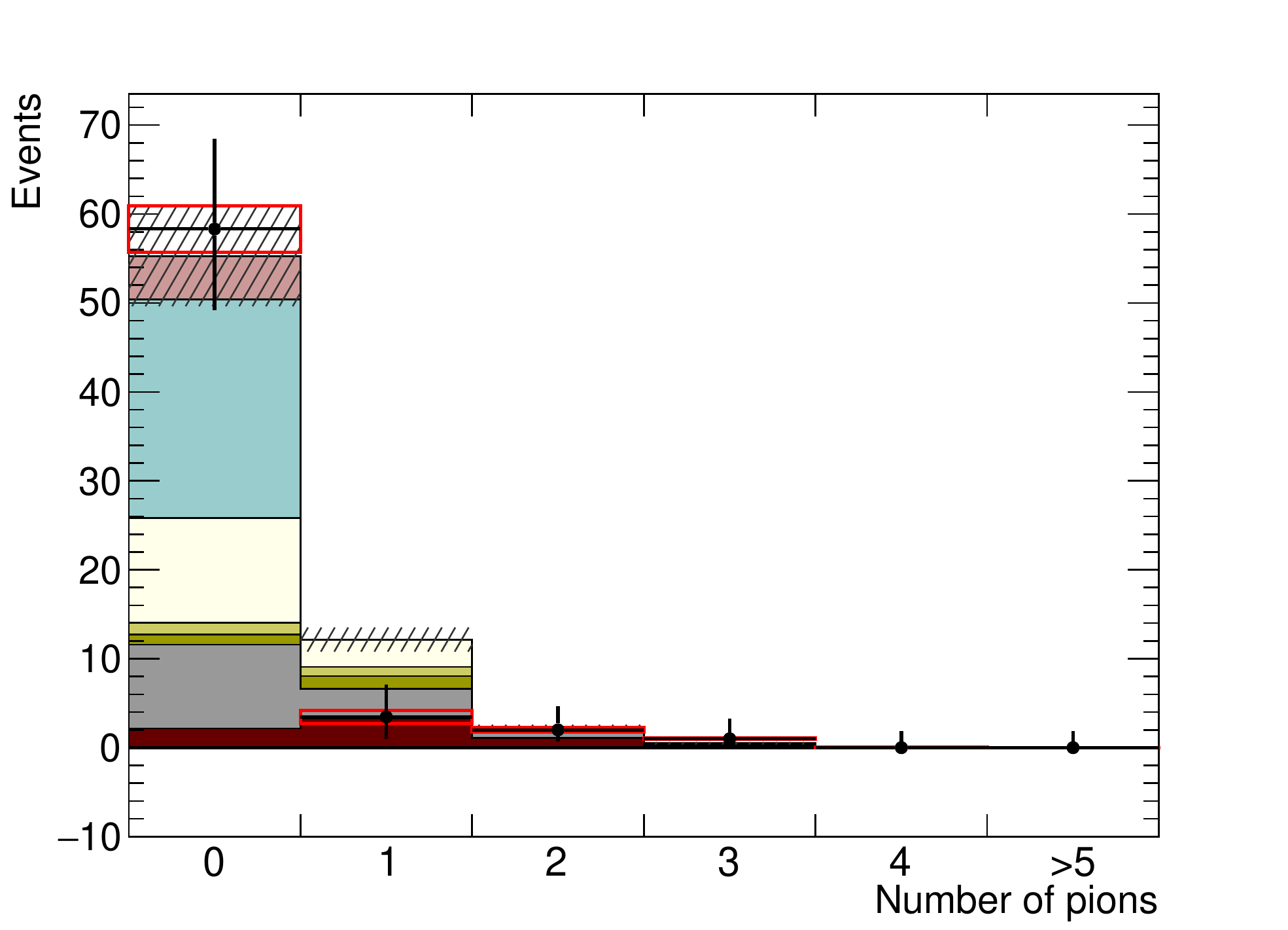}
  \end{minipage}
  \begin{minipage}{0.48\hsize}
    \includegraphics[width=8.6cm,pagebox=cropbox]{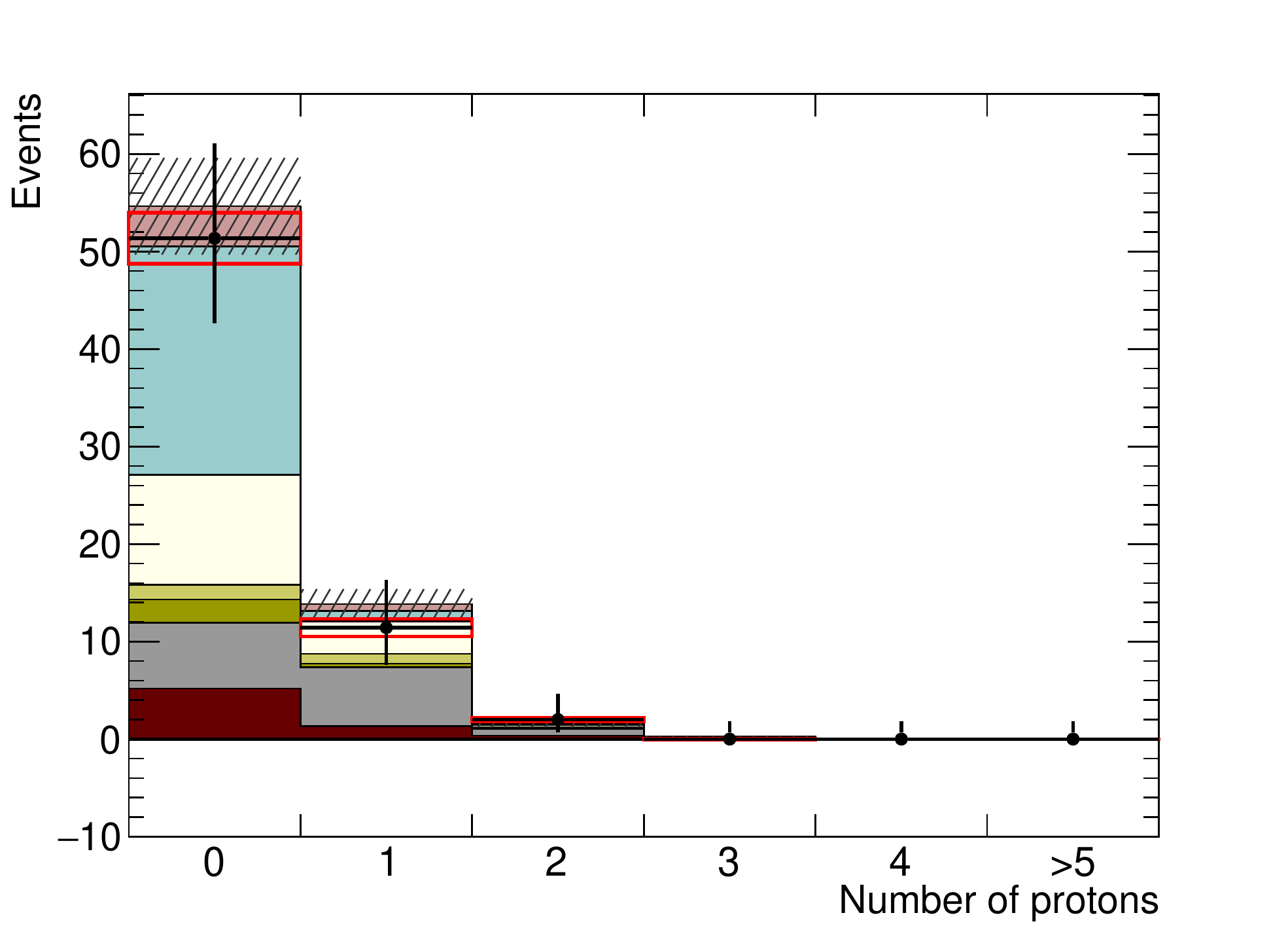}
  \end{minipage}
\caption{\label{fig:ntrk_bkgsubt}Multiplicity of charged particles from neutrino-water interactions  including muon candidates (top). Backgrounds are subtracted from both the data and the prediction. The bottom plots show the number of pions (left) and protons (right). The flux, detector response, and background estimation uncertainties are included in the error bars on the data points, while the hatched regions correspond to uncertainty of the neutrino interaction model.}
\end{figure*}

\begin{figure*}
  \begin{minipage}{0.48\hsize}
  \includegraphics[width=8.6cm,pagebox=cropbox]{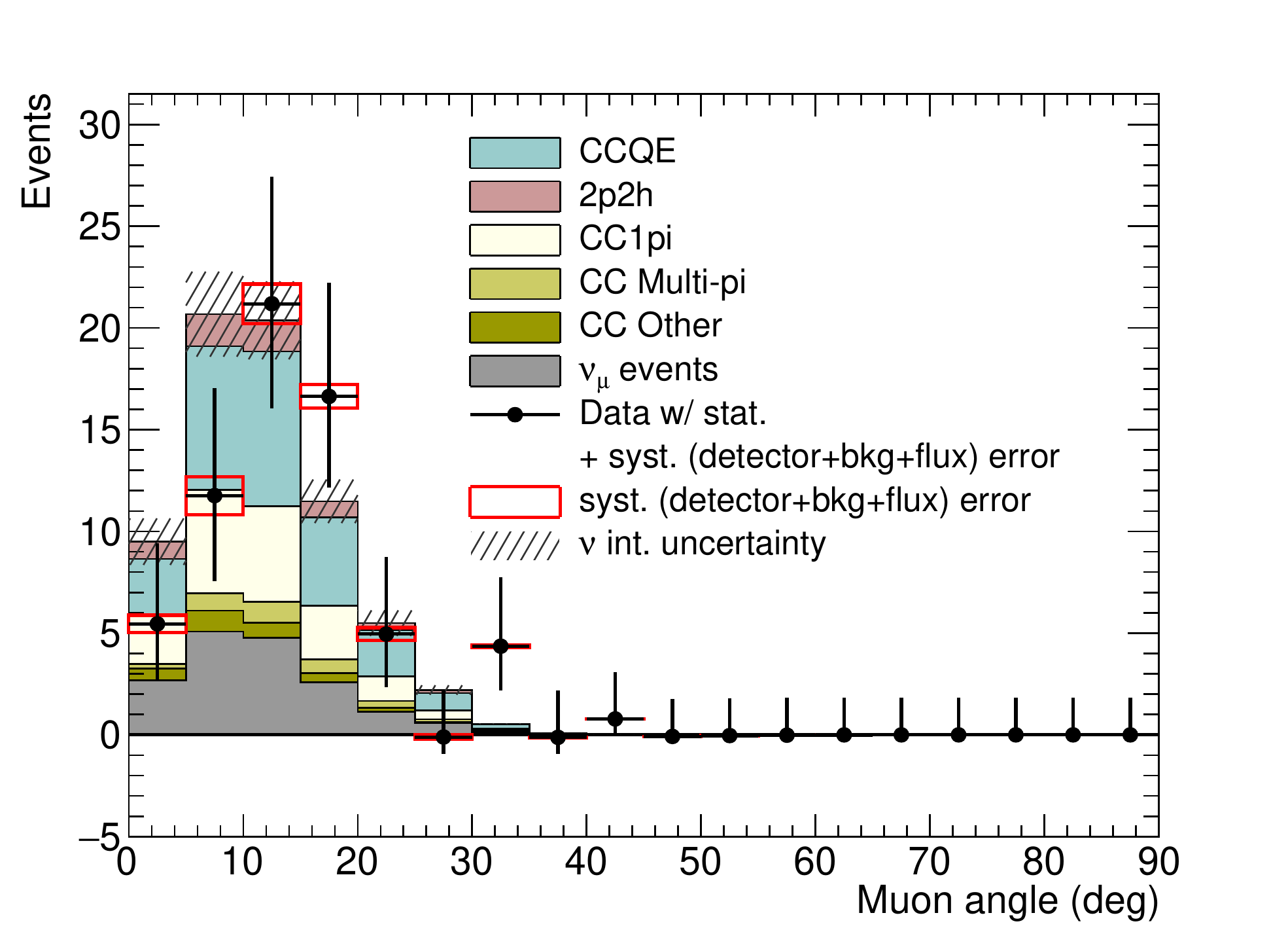}
  \end{minipage}
  \begin{minipage}{0.48\hsize}
    \includegraphics[width=8.6cm,pagebox=cropbox]{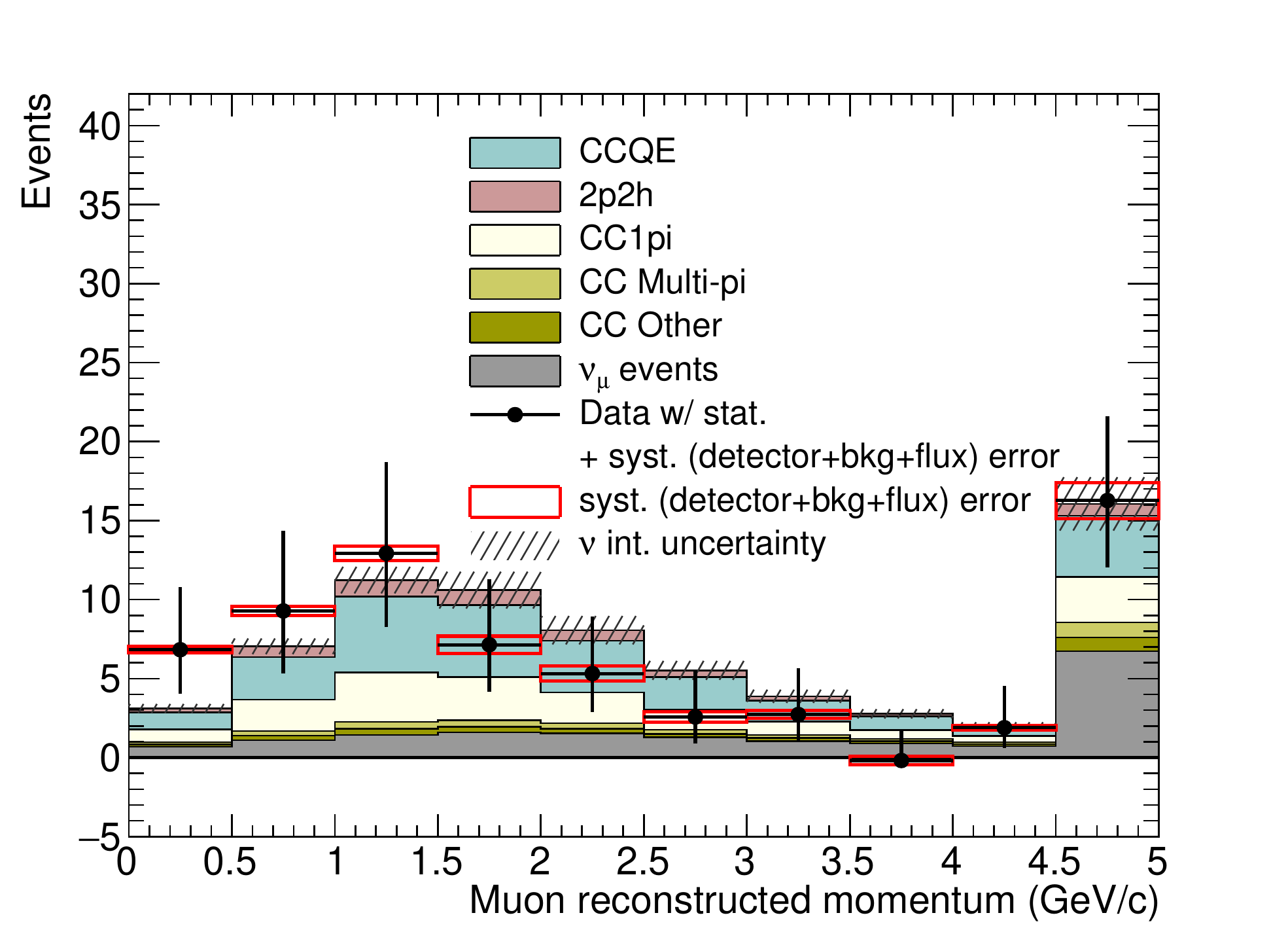}
  \end{minipage}
    \begin{minipage}{0.48\hsize}
  \includegraphics[width=8.6cm,pagebox=cropbox]{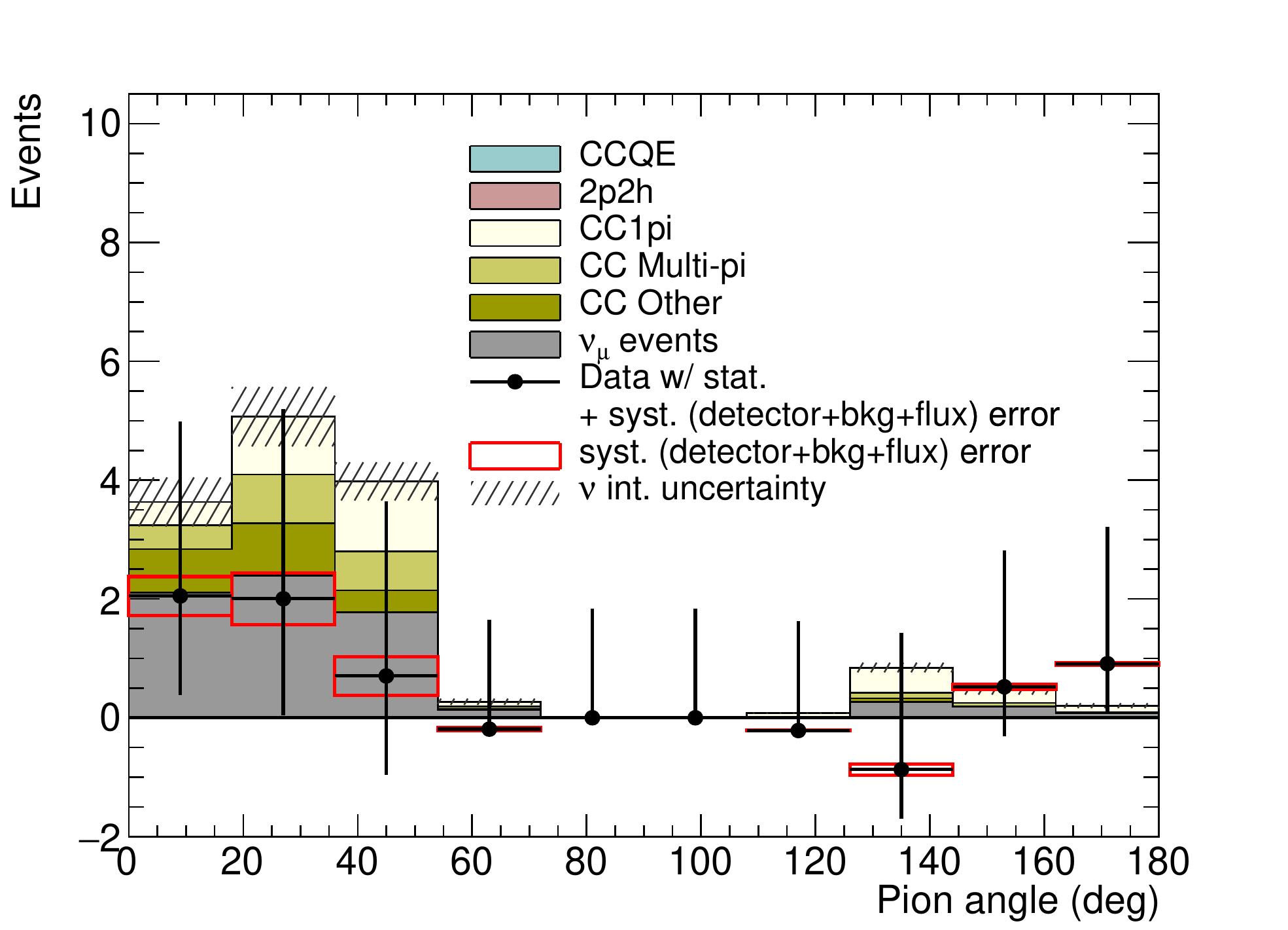}
  \end{minipage}
    \begin{minipage}{0.48\hsize}
  \includegraphics[width=8.6cm,pagebox=cropbox]{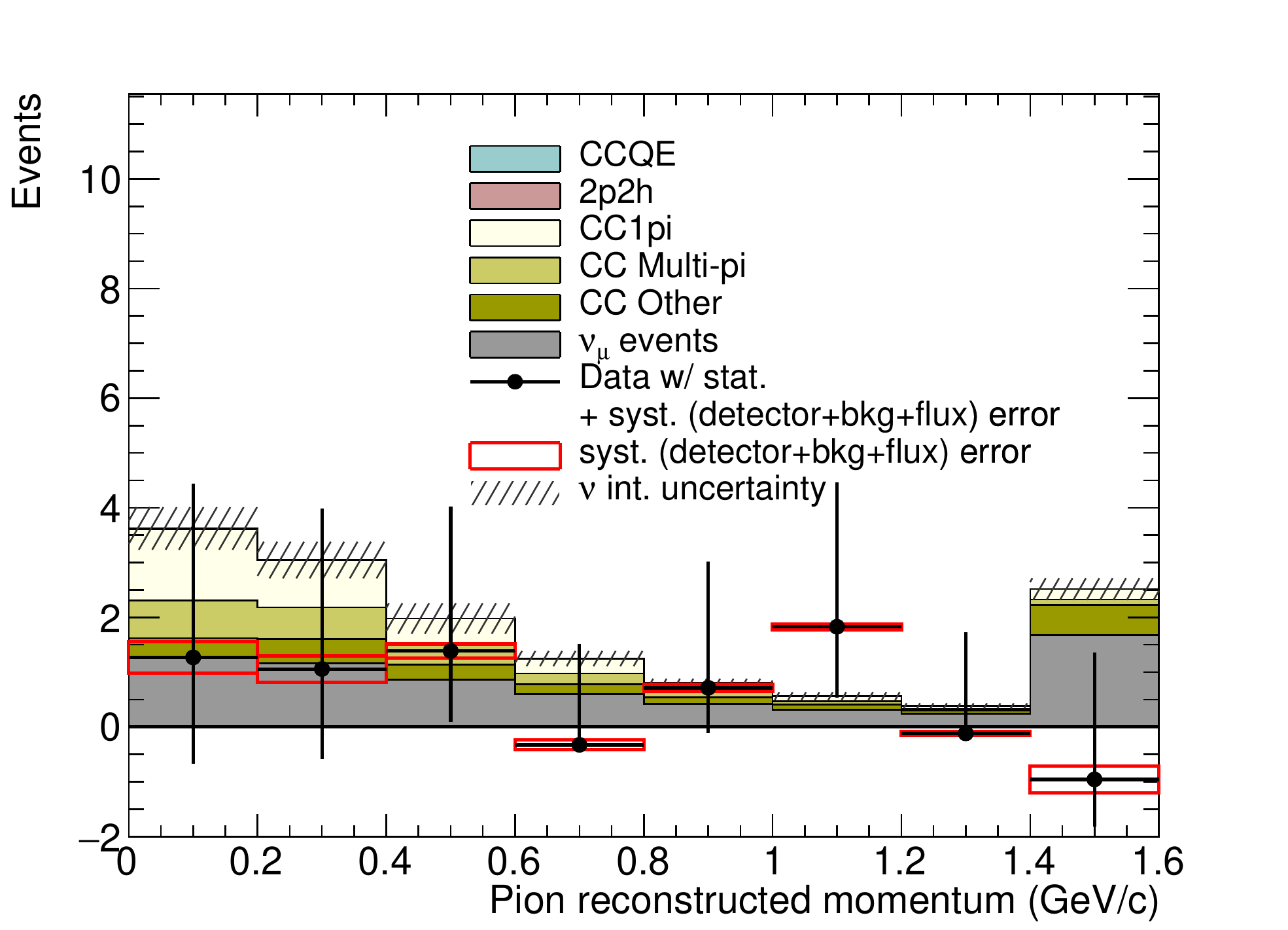}
  \end{minipage}
  \begin{minipage}{0.48\hsize}
    \includegraphics[width=8.6cm,pagebox=cropbox]{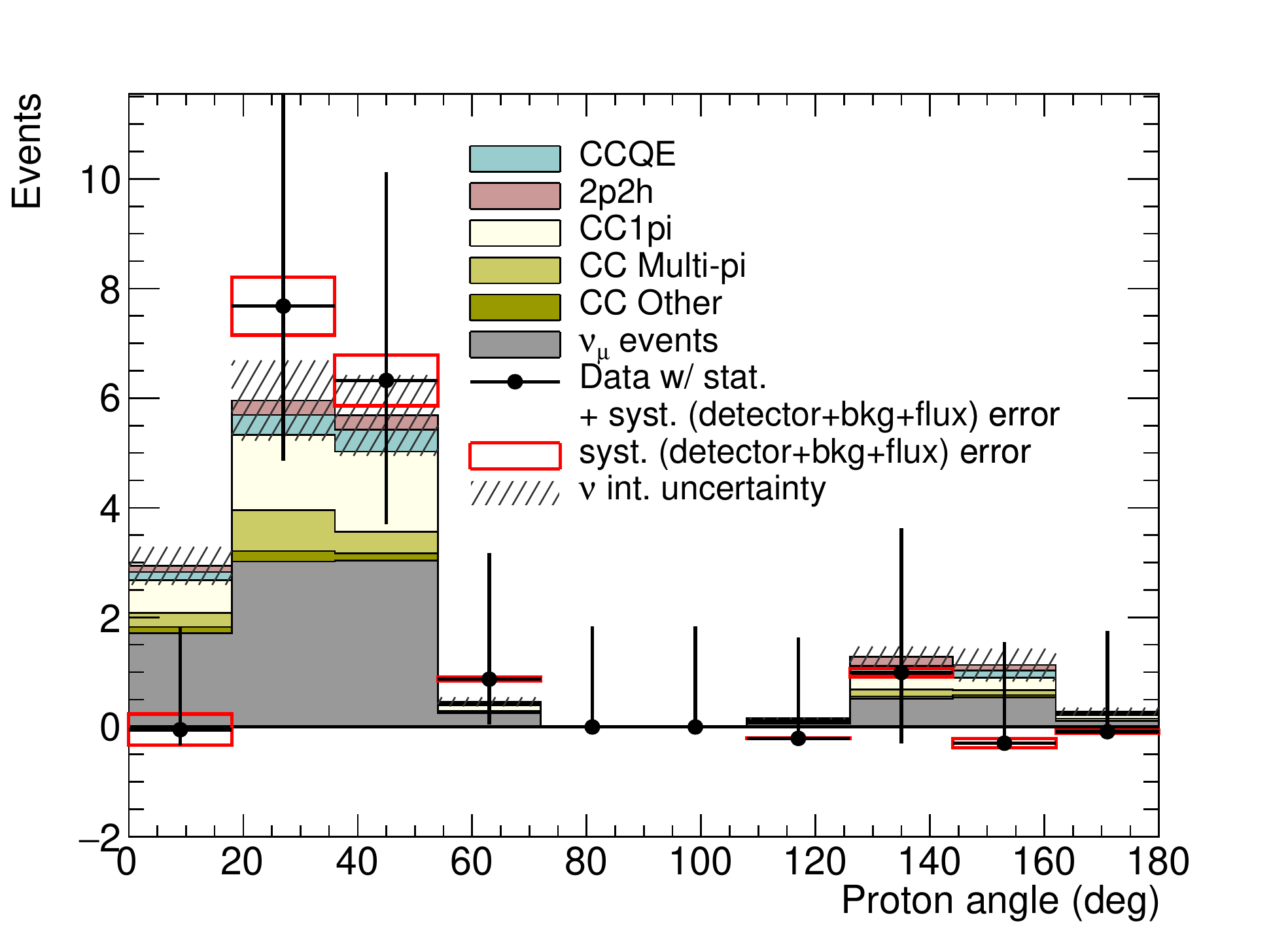}
  \end{minipage}
  \begin{minipage}{0.48\hsize}
    \includegraphics[width=8.6cm,pagebox=cropbox]{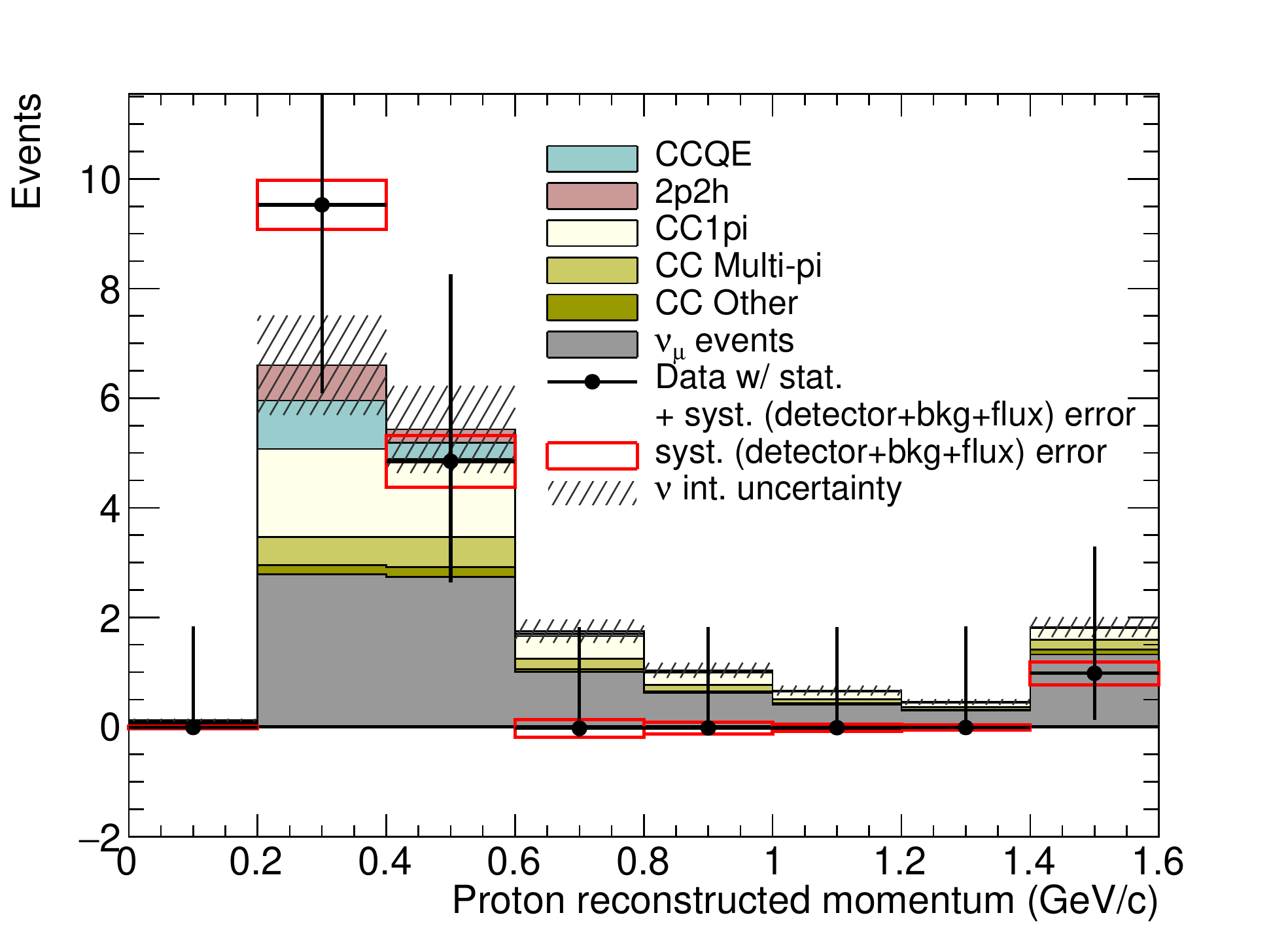}
  \end{minipage}
  \caption{\label{fig:results_bkgsubt}Distributions of muon, pion, and proton kinematics. Backgrounds are subtracted from both the data and the prediction. The left column shows angle distributions, while the right column shows momentum distributions. Though the angular resolution for all particles is sufficiently small compared to the bin width in the angle plots, the momentum resolution is typically larger than the momentum binning, especially for high momentum muons. In the right-most bins of the momentum distributions, all the events with momenta above 5\,GeV/$c$ for muons and 1.6\,GeV/$c$ for pions and protons are contained. The flux, detector response, and background estimation uncertainties are included in the error bars on the data points, while the hatched regions correspond to uncertainty of the neutrino interaction model.}
\end{figure*}

In order to extract the signal distributions from our data and to compare them with the prediction, backgrounds from neutral-current interactions, interactions on the packing films, cosmic rays, and chance coincidence of the off-beam timing tracks are subtracted from the data using the background prediction. Figures~\ref{fig:ntrk_bkgsubt} and \ref{fig:results_bkgsubt} show the results, in which the signal distributions are compared with the predictions. In Fig.~\ref{fig:results_bkgsubt}, contaminations by misidentification between protons and pions are also subtracted. In these plots, the flux, detector response, and background estimation uncertainties are included in the error bars on the data points, while the hatched regions correspond to the uncertainty of the neutrino interaction model. The systematic uncertainties shown in the red boxes were originally applied to the MC predictions in Figs.~\ref{fig:ntrk} and \ref{fig:results}. The absolute values of the uncertainty on the bin-by-bin MC predictions are transferred to the data points to show our measurement errors clearly. The measurement's systematic error is smaller than the current model uncertainty. Hence measurements with the NINJA detector can be expected to constrain neutrino interaction models given more statistics.

Although the statistical uncertainty is large, the measurement result shows a slightly lower multiplicity of charged particles compared to the prediction. As shown in the bottom plots of Fig.~\ref{fig:ntrk_bkgsubt}, there is a tendency for the prediction to overestimate the number of pions. The number of detected pions is 4.9~$\pm$~3.6~(stat.)~$\pm$~0.6~(syst.), while 14.5 are expected in the prediction. By contrast, 15.2~$\pm$~4.2~(stat.)~$\pm$~0.8~(syst.) protons are detected, which is consistent with the prediction of 17.7 protons. The overestimation of charged pions may be induced by the inaccuracy of the modeling of either neutrino interactions or FSI. Besides this overestimation, the muon distributions have slightly higher angle and lower momentum shape than the MC prediction. In the other plots, the predictions explain the data well.

\begin{figure*}
  \begin{minipage}{0.48\hsize}
  \includegraphics[width=8.6cm,pagebox=cropbox]{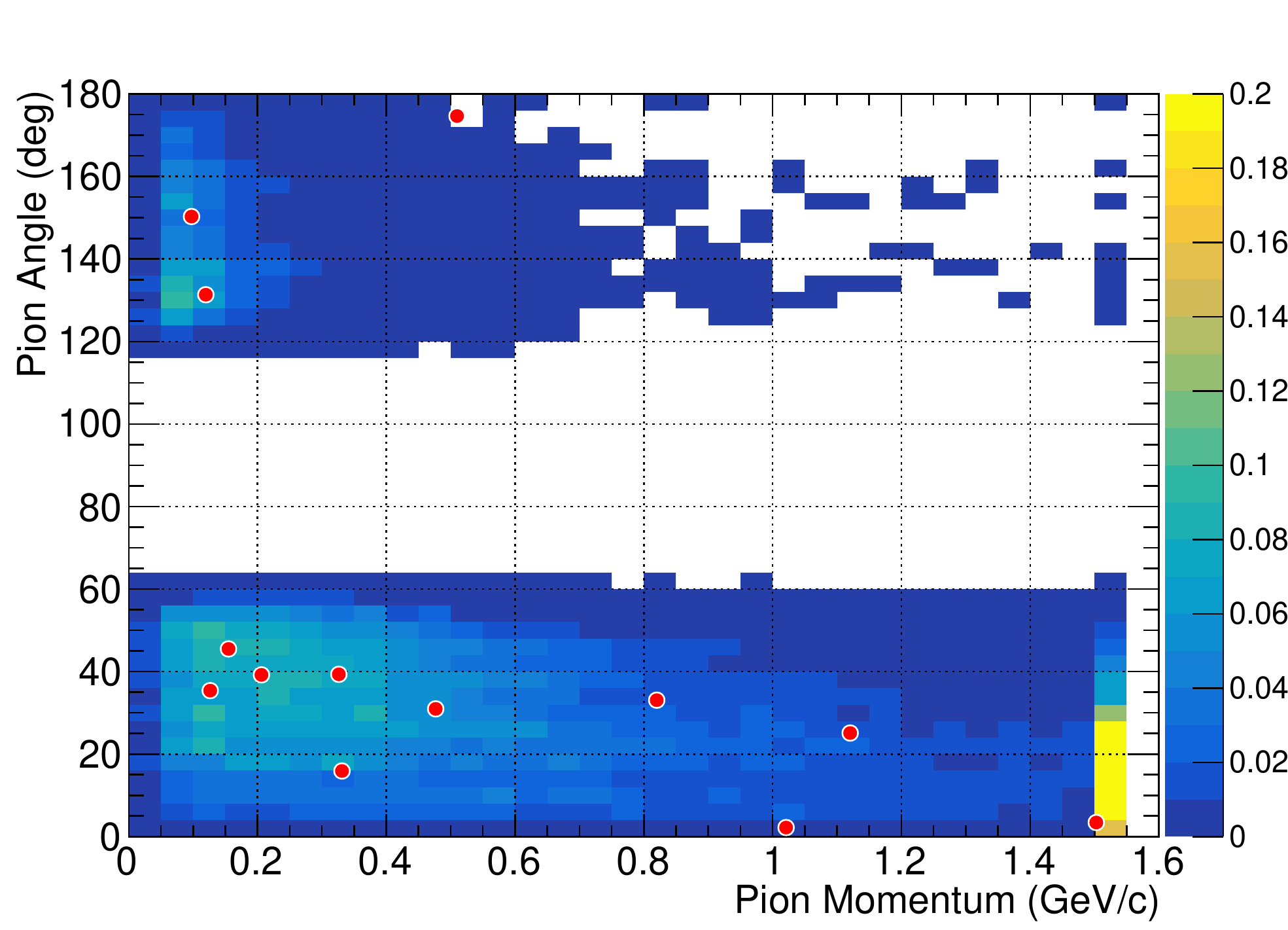}
  \end{minipage}
  \begin{minipage}{0.48\hsize}
    \includegraphics[width=8.6cm,pagebox=cropbox]{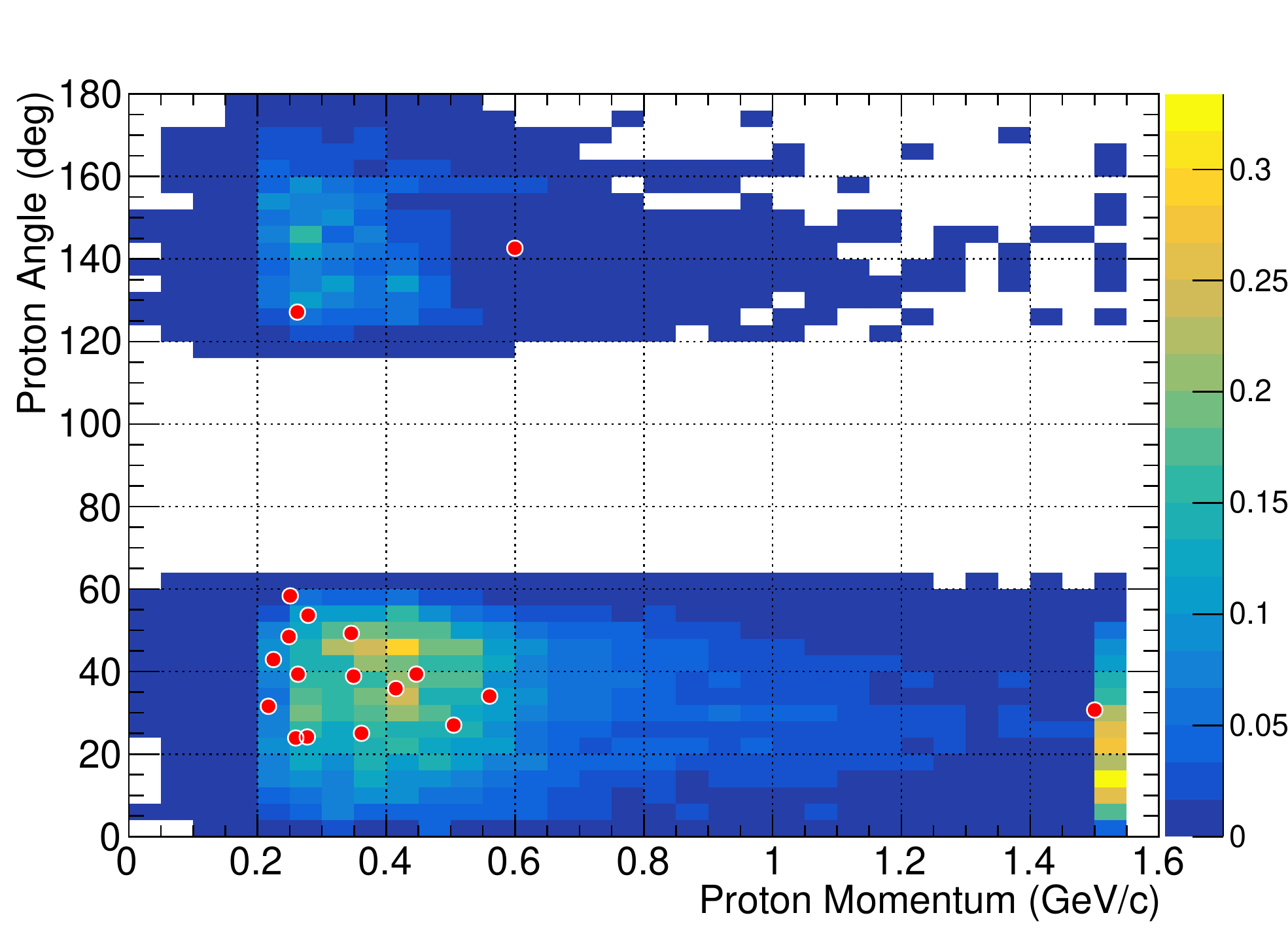}
  \end{minipage}
  \caption{\label{fig:result_ptheta}Two-dimensional kinematics distributions of pions (left) and protons (right) from neutrino-water interactions. The red points show the detected tracks and the colored histograms represent the predictions.}
\end{figure*}

In addition to the one-dimensional kinematics distributions,  Fig.~\ref{fig:result_ptheta} shows the two-dimensional distribution of the angle and the momentum for protons or pions. Although the statistics is limited, these plots show general agreement between the data and the predictions.

\begin{figure*}
    \begin{minipage}{0.48\hsize}
  \includegraphics[width=8.6cm,pagebox=cropbox]{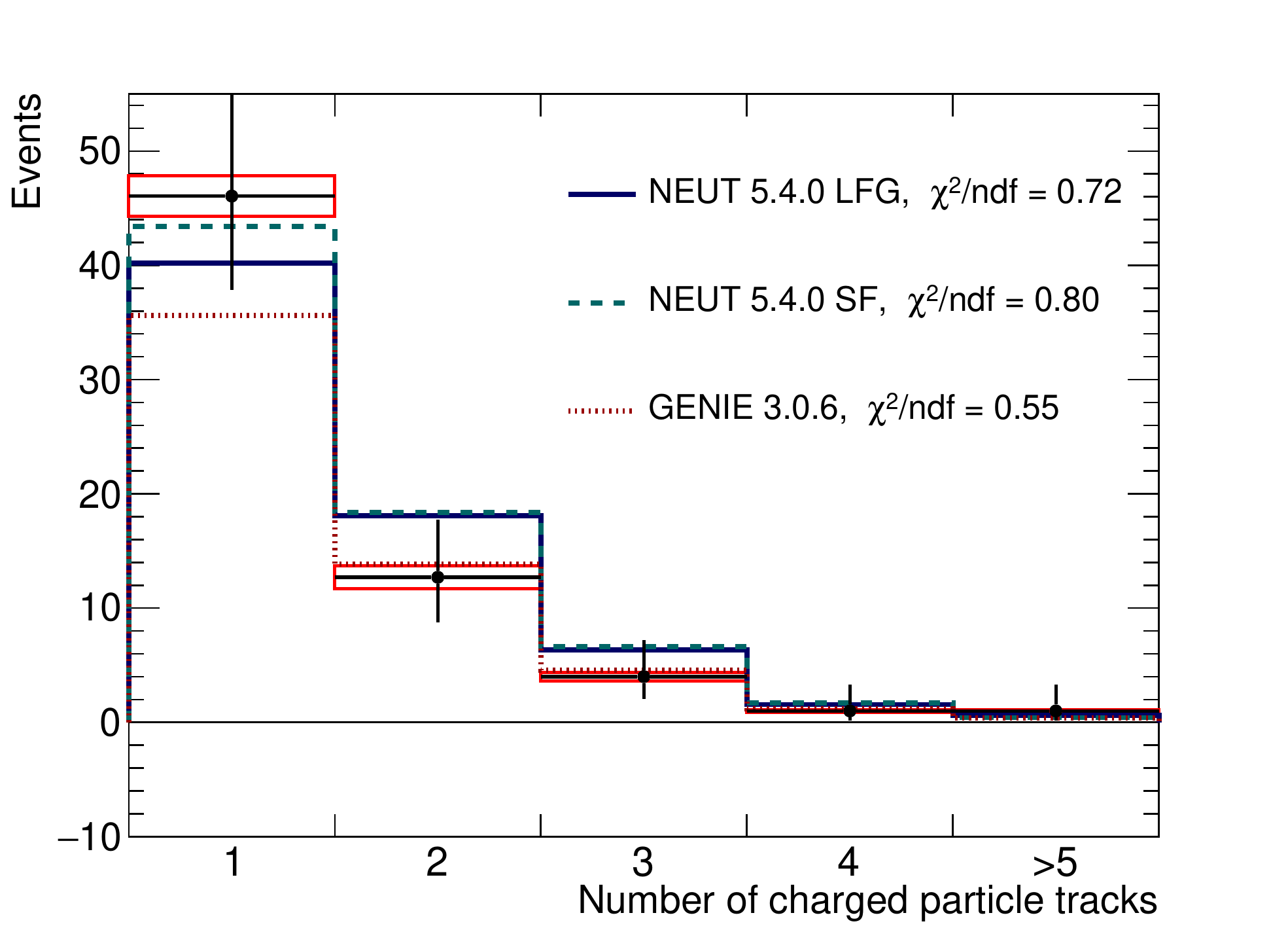}
  \end{minipage}
  \begin{minipage}{0.48\hsize}
    \includegraphics[width=8.6cm,pagebox=cropbox]{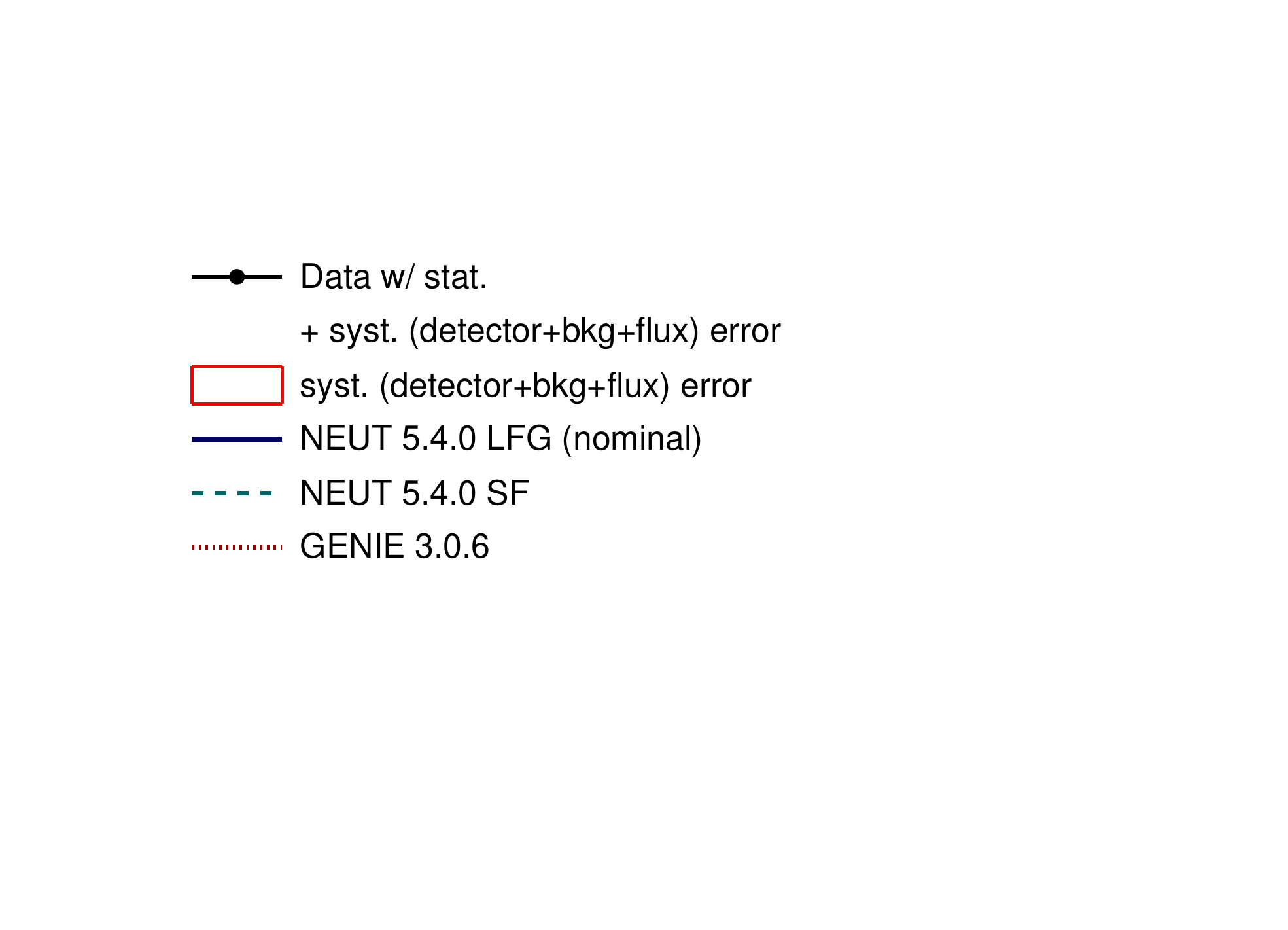}
  \end{minipage}
    \begin{minipage}{0.48\hsize}
  \includegraphics[width=8.6cm,pagebox=cropbox]{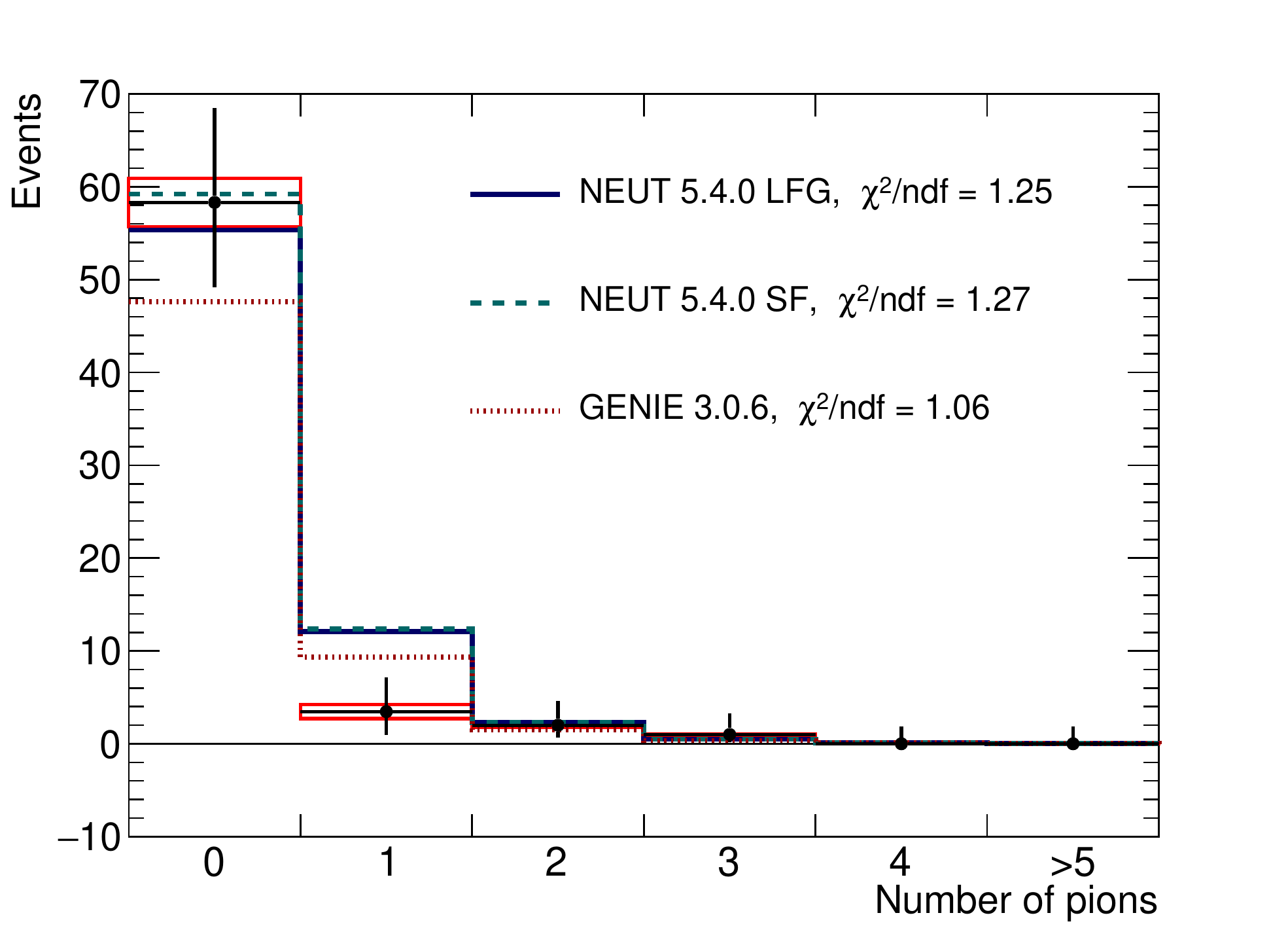}
  \end{minipage}
  \begin{minipage}{0.48\hsize}
    \includegraphics[width=8.6cm,pagebox=cropbox]{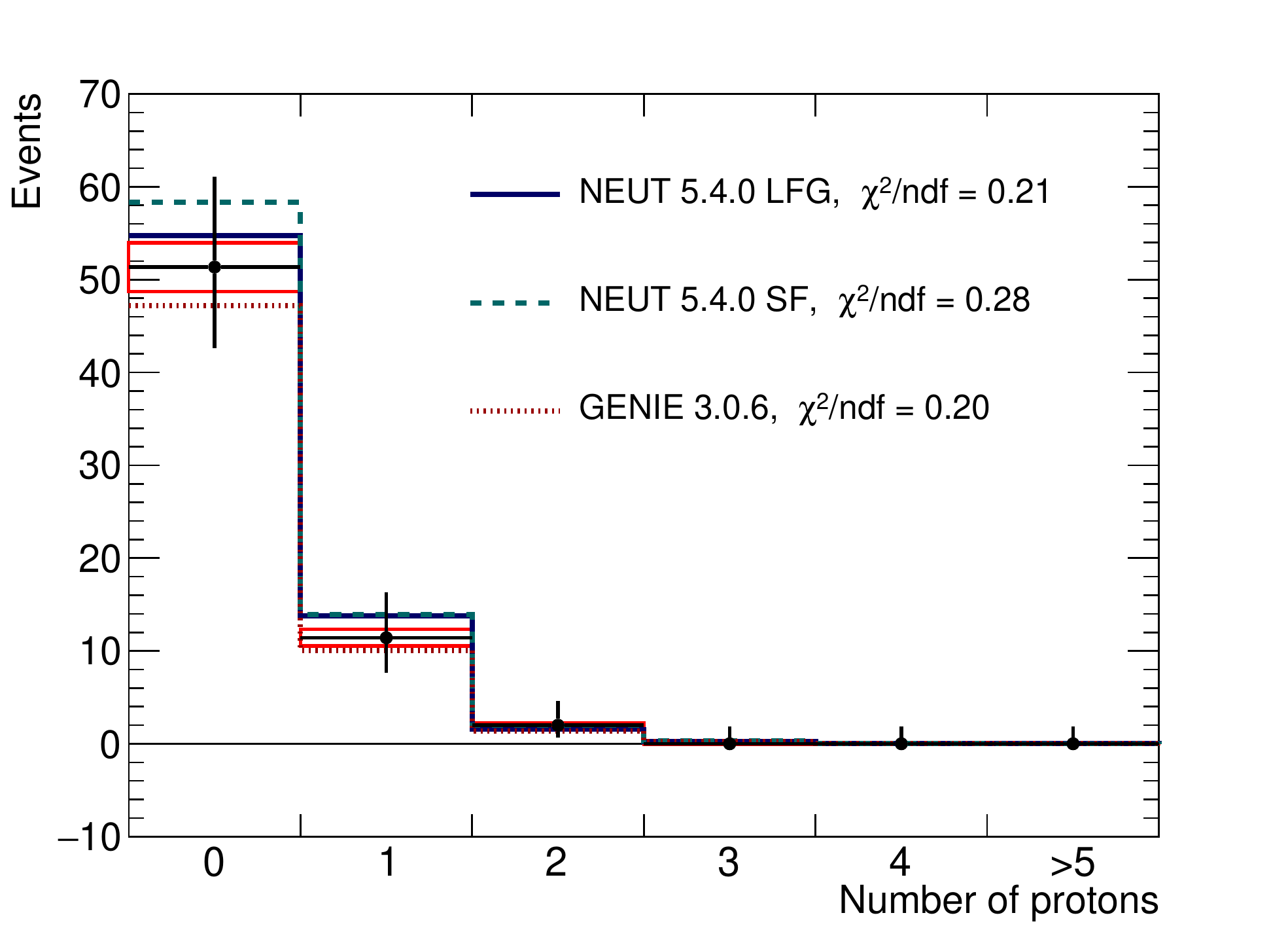}
  \end{minipage}
\caption{\label{fig:ntrk_model}Multiplicity of charged particles from neutrino-water interactions  including muon candidates (top) in comparison with NEUT LFG (nominal), NEUT SF, and GENIE predictions. Backgrounds are subtracted from the data using the nominal prediction. The bottom plots show the number of pions (left) and protons (right). The flux, detector response, and background estimation uncertainties are included in the error bars on the data points. Reduced $\chi^2$ value is evaluated for each model and shown in the plots.}
\end{figure*}

\begin{figure*}
  \begin{minipage}{0.48\hsize}
  \includegraphics[width=8.6cm,pagebox=cropbox]{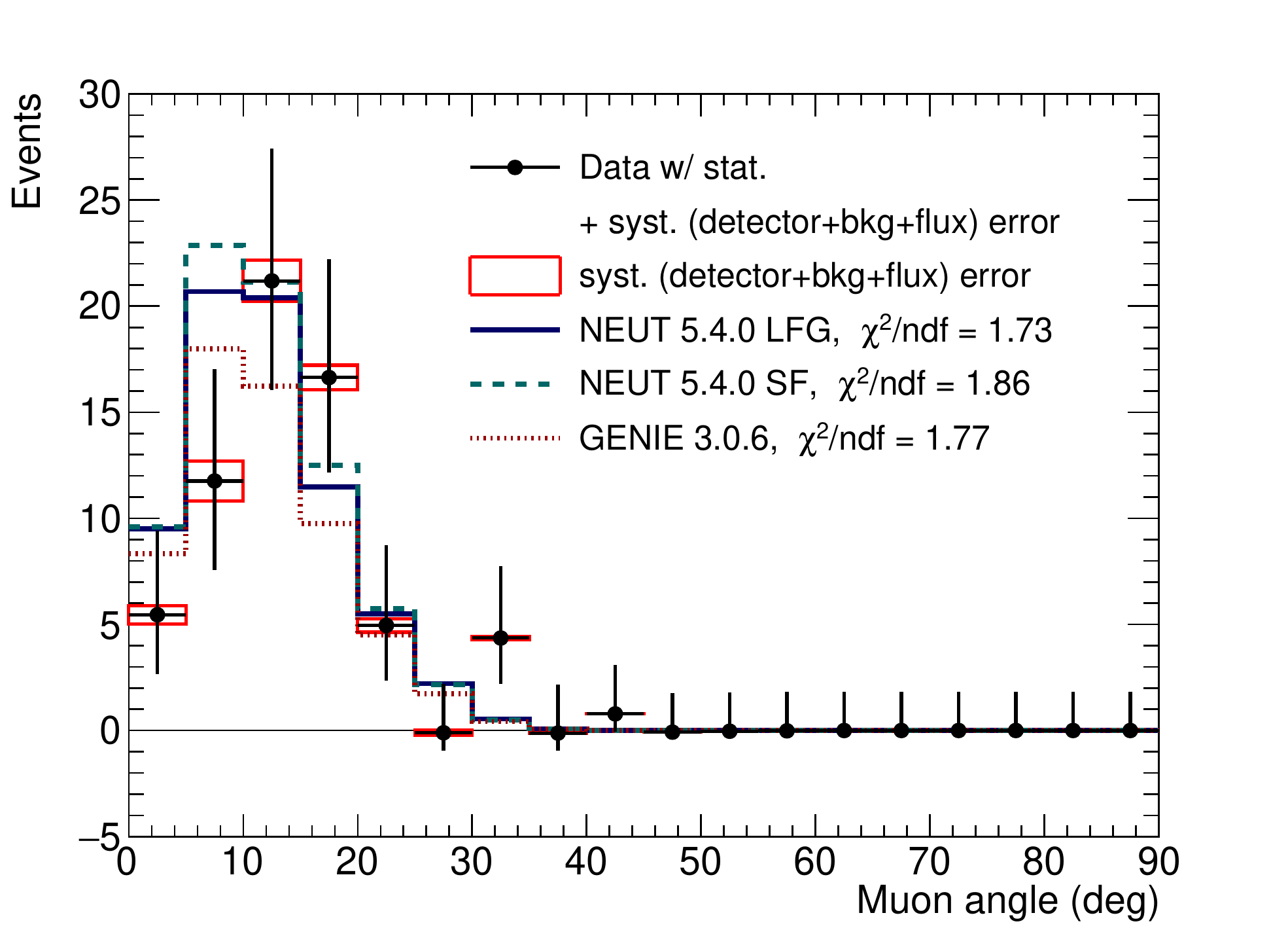}
  \end{minipage}
  \begin{minipage}{0.48\hsize}
    \includegraphics[width=8.6cm,pagebox=cropbox]{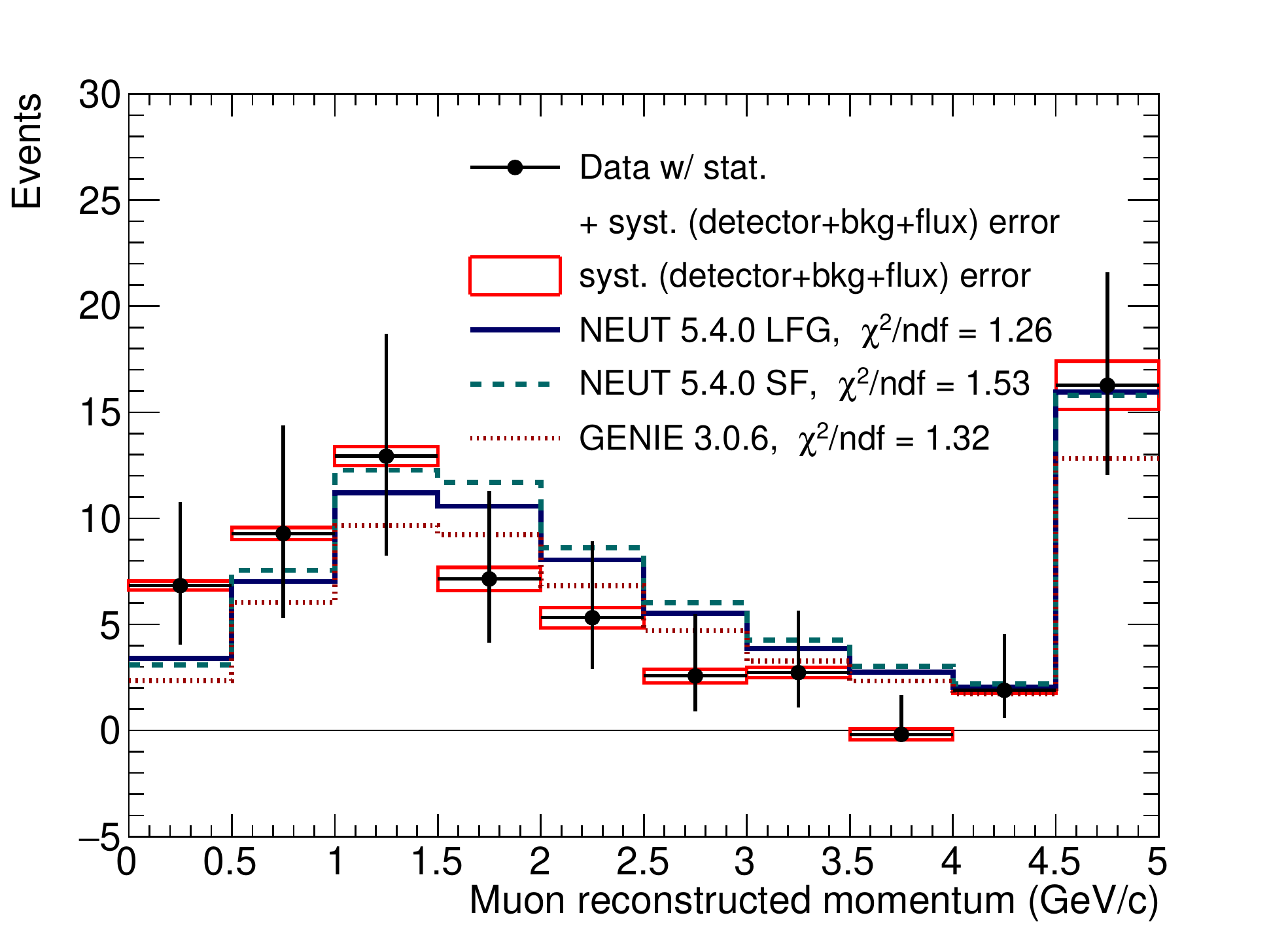}
  \end{minipage}
    \begin{minipage}{0.48\hsize}
  \includegraphics[width=8.6cm,pagebox=cropbox]{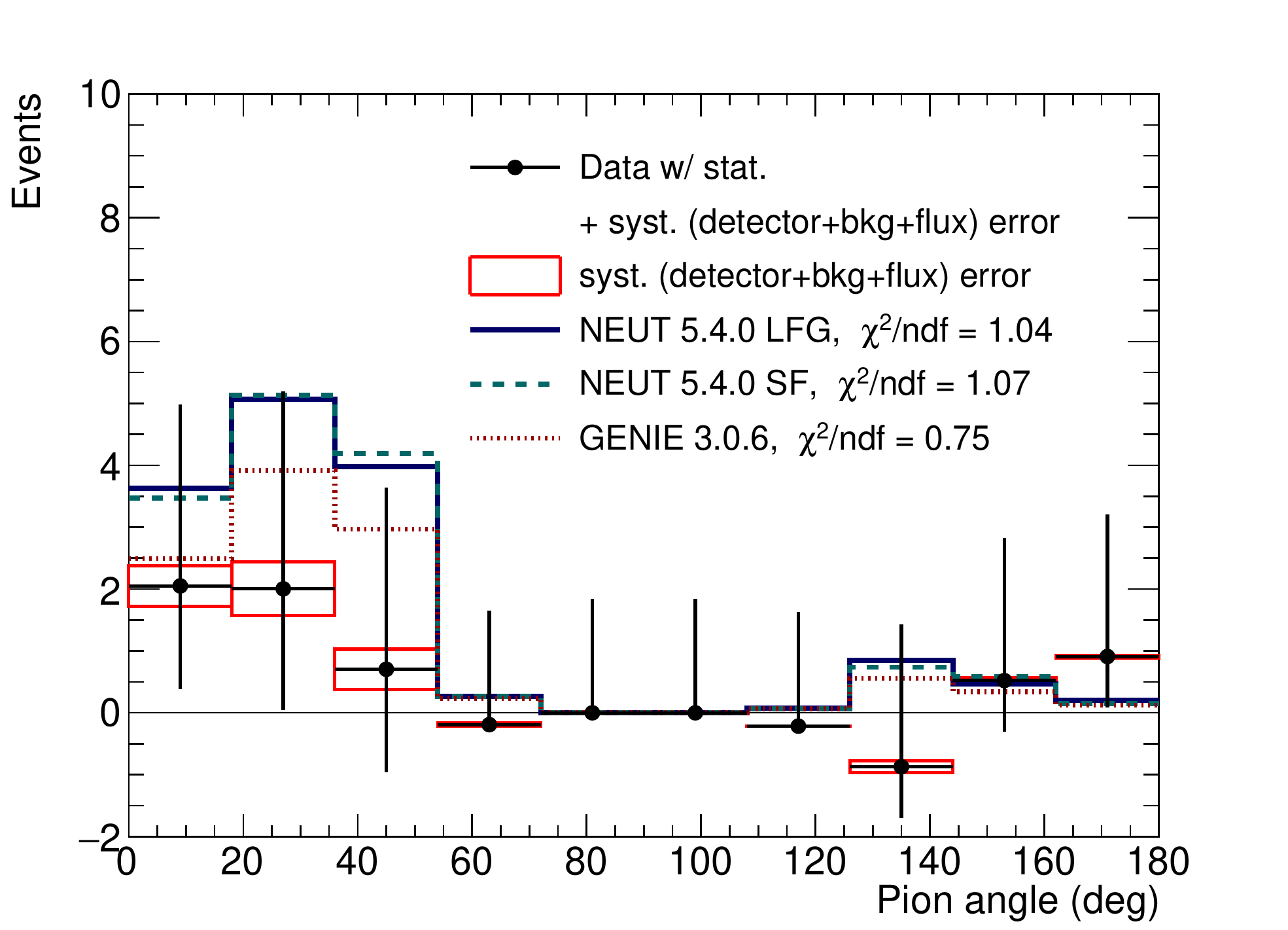}
  \end{minipage}
    \begin{minipage}{0.48\hsize}
  \includegraphics[width=8.6cm,pagebox=cropbox]{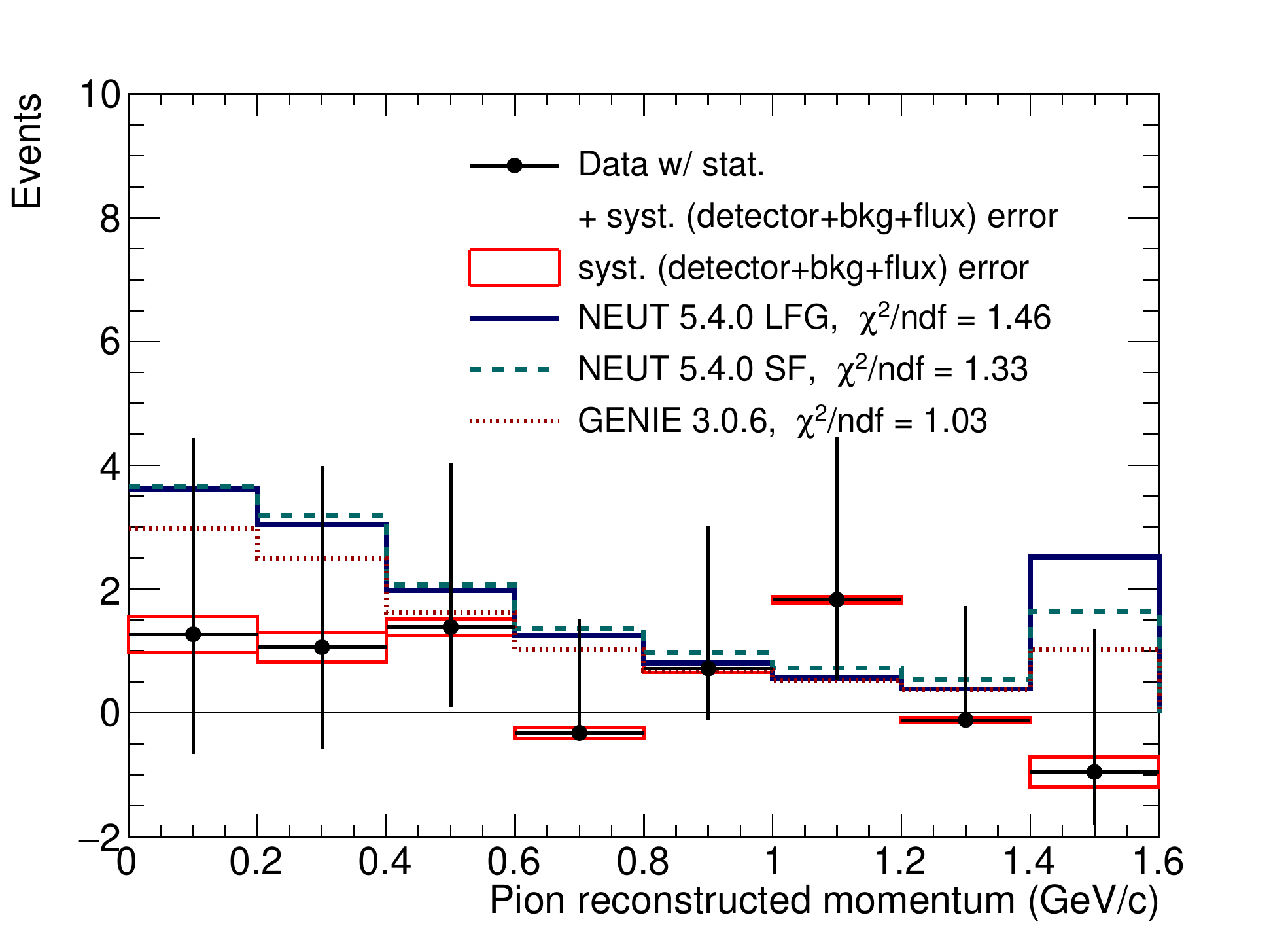}
  \end{minipage}
  \begin{minipage}{0.48\hsize}
    \includegraphics[width=8.6cm,pagebox=cropbox]{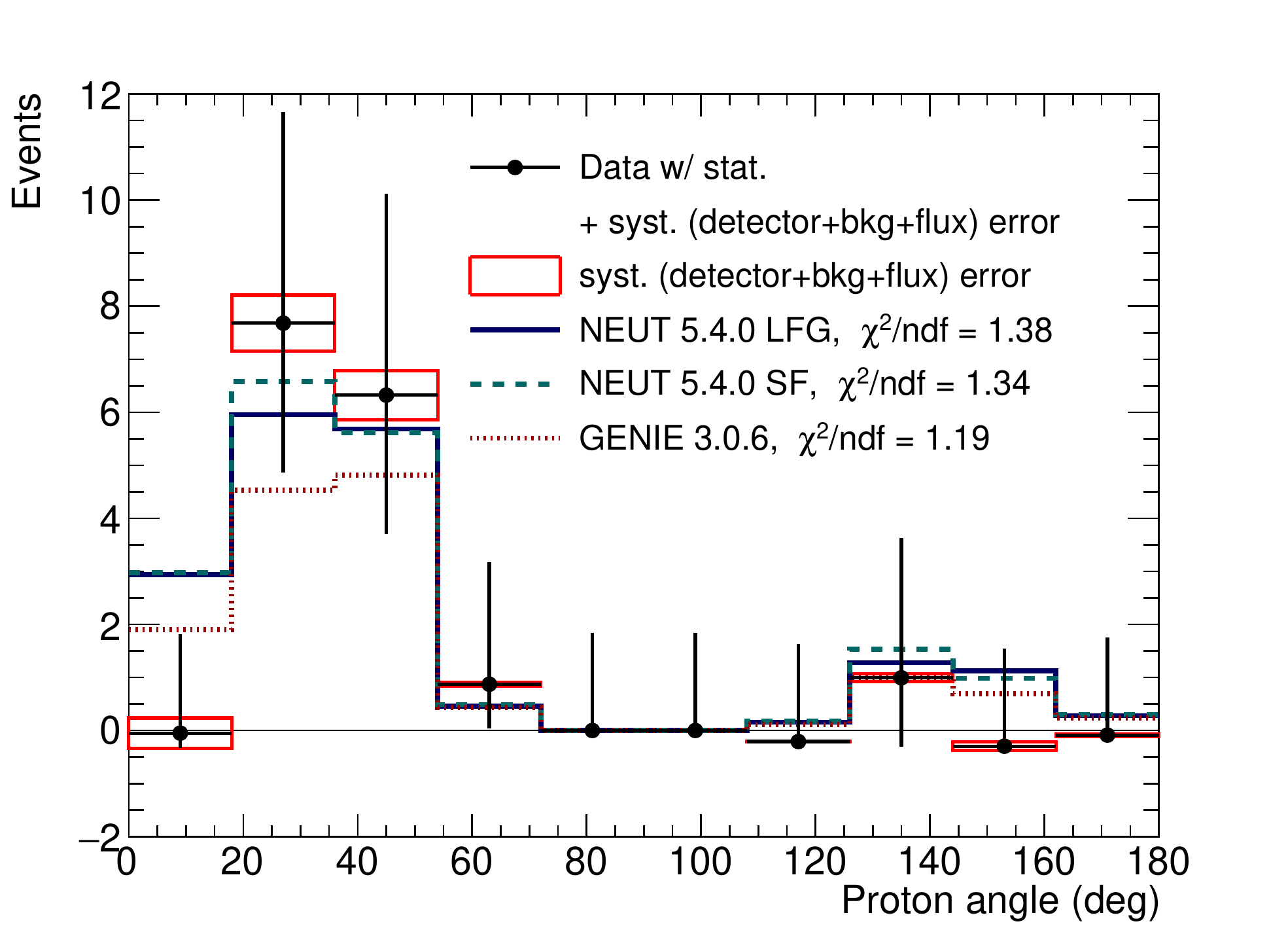}
  \end{minipage}
  \begin{minipage}{0.48\hsize}
    \includegraphics[width=8.6cm,pagebox=cropbox]{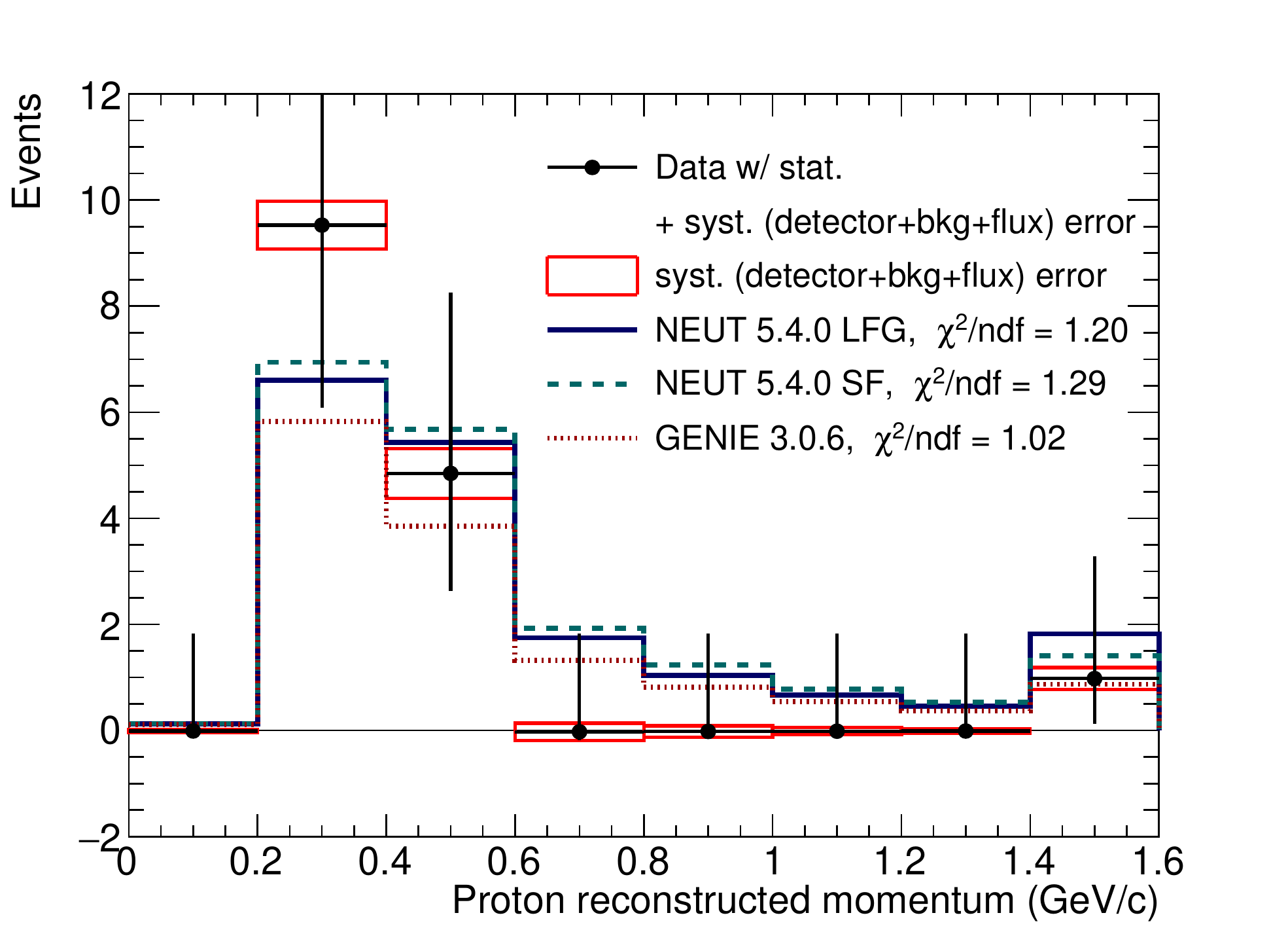}
  \end{minipage}
  \caption{\label{fig:results_model}Distributions of muon, pion, and proton kinematics in comparison with NEUT LFG (nominal), NEUT SF, and GENIE predictions. Backgrounds are subtracted from the data using the nominal prediction. The left column shows angle distributions, while the right column shows momentum distributions. Though the angular resolution for all particles is sufficiently small compared to the bin width in the angle plots, the momentum resolution is typically larger than the momentum binning, especially for high momentum muons. In the right-most bins of the momentum distributions, all the events with momenta above 5\,GeV/$c$ for muons and 1.6\,GeV/$c$ for pions and protons are contained. The flux, detector response, and background estimation uncertainties are included in the error bars on the data points. Reduced $\chi^2$ value is evaluated for each model and shown in the plots.}
\end{figure*}

Finally, an alternative model of NEUT using SF, and another generator, GENIE~\cite{andreopoulos2010,andreopoulos2015}, are studied for comparisons with the nominal model of NEUT using LFG. Figures~\ref{fig:ntrk_model} and \ref{fig:results_model} show the results. The interaction models used in the nominal MC simulation is summarized in Tab.~\ref{tab:intmdl}. Interaction models used in the alternative model of NEUT using SF is almost the same as those in Tab.~\ref{tab:intmdl}, but the nuclear model is changed to SF, and $M\mathrm{_A^{QE}}$ is set to 1.21\,GeV/$c^2$. GENIE v3.0.6 with $\rm{G18\_10b\_02\_11a}$ tuning is used as an alternative generator. In GENIE, different axial mass values are used, and the Berger--Sehgal model~\cite{berger2007} is used for the single pion production. For the FSI simulation, GENIE hN cascade model~\cite{dytman2009} is employed. Reduced $\chi^2$ values are evaluated by a log-likelihood method assuming a Poisson distribution in each bin, and summarized in each plot in Figs.~\ref{fig:ntrk_model} and \ref{fig:results_model}. Only the statistical errors are used for the evaluation. With more statistics, our measurement can discriminate the models.

\vspace{0.5cm}
\section{Conclusion}
\vspace{0.5cm}
The first results of the NINJA pilot run using a water-target emulsion detector are reported in this paper. Multiplicity, angle, and momentum distributions of the outgoing muons, charged pions, and protons from neutrino-water interactions are reported. Protons from neutrino-water interactions are measured with a 200\,MeV/$c$ threshold for the first time. Although the statistical uncertainty is large, we found that the current neutrino interaction models predict the kinematics distributions well within the measurement uncertainty, including for low momentum protons down to 200\,MeV/$c$. In addition, we found that there is a tendency to overestimate the number of charged-pions in the MC simulation in the measurements of pion kinematics. The muon distributions show slightly higher angle and lower momentum shape than the MC prediction. The related data shown in this paper can be found in~\cite{datarelease}.

The first physics run of the NINJA experiment concluded its beam exposure with the neutrino mode beam in early 2020. We expect 15 times more neutrino interactions, thus the statistical uncertainty will be as small as the current systematic uncertainty. In the current analysis, relatively large systematic uncertainty is applied to the measurements of pions and protons due to the uncertainty of the PID performance. This uncertainty can be reduced through further understanding of our detector response, and the size of the total uncertainty will be similar to that of the muon measurements. Then, we will measure the differential cross-section with about 10\% uncertainty. We aim to characterize the nature of 2p2h interactions, which has the largest uncertainty in the current model. Moreover, differential cross-section measurements with respect to the number of protons with measurement of kinematics correlations allow us to gain more insights to nuclear effects. The results of this pilot run clearly demonstrate the capability of the emulsion detector to achieve this goal.

\clearpage
\begin{acknowledgments}
We thank the T2K collaboration for their great support in conducting this experiment as well as the J-PARC staff for their superb accelerator performance. We would also like to acknowledge the T2K INGRID group for their stable operation and for providing data as well as the T2K neutrino beam group for providing a high-quality beam and helping us with the beam MC simulation. We also express our gratitude to K. Kuwabara, K. Ohzeki, and T. Yoshida for their valuable insights and advice based on their experience and expertise in nuclear emulsion. We also acknowledge the support of Ministry of Education, Culture, Sports, Science and Technology in Japan (MEXT) and Japan Society for the Promotion of Science (JSPS) through various grants (JSPS KAKENHI Grant Number 25105006, 25707019, 26287049, 26105516, 17H02888, 18H03701, 18H05537, 18H05541, and 17J02714).

\end{acknowledgments}

%

\end{document}